\documentclass[aps,prx,twocolumn,showpacs,psfig,superscriptaddress,longbibliography]{revtex4-1}
\usepackage{times}
\usepackage{graphicx}
\usepackage{float}
\usepackage{latexsym,amsmath,amssymb,bm,euscript}
\usepackage{color}
\usepackage{subfigure}
\usepackage{epstopdf}
\usepackage[colorlinks=true,linkcolor=blue,citecolor=blue]{hyperref}
\usepackage{hyperref}
\usepackage{soul}
\usepackage{ulem}
\usepackage{mathrsfs}
\usepackage{amsmath}
\usepackage{dsfont}
\usepackage{natbib}

\usepackage{xspace}

\newcommand{\Tr}[1]{\mathrm{Tr\left[#1\right]}}
\newcommand{\order}[1]{\mathcal{O}{\left(#1\right)}}
\newcommand{\orderb}[1]{\mathcal{O}{\bigl(#1\bigr)}}

\newcommand{\Sec}[1]{Sec.~\ref{#1}}
\newcommand{\App}[1]{Appendix~\ref{#1}}
\newcommand{\Eq}[1]{Eq.~\eqref{#1}} 
\newcommand{\Eqs}[1]{Eqs.~\eqref{#1}}
\newcommand{\EQ}[1]{Equation~\eqref{#1}}
\newcommand{\Fig}[1]{Fig.~\ref{#1}}
\newcommand{\FIG}[1]{Figure~\ref{#1}}

\def\Dstar{D^\ast}
\def\Sztot{S^z_{\mathrm{tot}}}

\def\taui{\tau_0}
\def\Dstari{\Dstar_0}
\def\Di{D_0}

\def\Tc{\ensuremath{T_c}\xspace}
\def\TcS{\ensuremath{\Tc^{S}}\xspace}
\def\TcX{\ensuremath{\Tc^{\mathrm{X}}}\xspace}

\def\Ncut{\mathcal{N}_{\rm{c}}}

\newcommand\numberthis{\addtocounter{equation}{1}\tag{\theequation}}

\newcommand{\aw}[1]{{\color[rgb]{.9,.5,.2}{#1}}}

\begin{document}

\title{Exponential Thermal Tensor Network Approach for Quantum Lattice Models}

\author{Bin-Bin Chen}
\thanks{These two authors contributed equally.}
\affiliation{Department of Physics, Key Laboratory of Micro-Nano Measurement-Manipulation and Physics (Ministry of Education), Beihang University, Beijing 100191, China}

\author{Lei Chen}
\thanks{These two authors contributed equally.}
\affiliation{Department of Physics, Key Laboratory of Micro-Nano Measurement-Manipulation and Physics (Ministry of Education), Beihang University, Beijing 100191, China}

\author{Ziyu Chen}
\affiliation{Department of Physics, Key Laboratory of Micro-Nano Measurement-Manipulation and Physics (Ministry of Education), Beihang University, Beijing 100191, China}

\author{Wei Li}
\email{w.li@buaa.edu.cn}
\affiliation{Department of Physics, Key Laboratory of Micro-Nano Measurement-Manipulation and Physics (Ministry of Education), Beihang University, Beijing 100191, China}
\affiliation{International Research Institute of Multidisciplinary Science, Beihang University, Beijing 100191, China}

\author{Andreas Weichselbaum}
\email{weichselbaum@bnl.gov}
\affiliation{Department of Condensed Matter Physics and Materials
Science, Brookhaven National Laboratory, Upton, NY 11973-5000, USA}
\affiliation{Physics Department, Arnold Sommerfeld Center for Theoretical Physics,
and Center for NanoScience, Ludwig-Maximilians-Universit\"at,
Theresienstrasse 37, 80333 Munich, Germany}

\begin{abstract}

 {We speed up thermal simulations of quantum many-body systems
in both one- (1D) and two-dimensional (2D) models in an exponential way by iteratively
projecting the thermal density matrix $\hat{\rho}=e^{-\beta \hat{H}}$ onto itself.
We refer to this scheme of doubling $\beta$ in each step
of the imaginary time evolution as the exponential tensor
renormalization group (XTRG).
This approach is in stark contrast to conventional
Trotter-Suzuki-type methods which evolve 
$\hat{\rho}$ on a linear  {quasi-continuous} grid in inverse temperature $\beta \equiv 1/T$.} 
 {As an aside, the large steps in XTRG allow one to swiftly 
jump across finite-temperature phase transitions, i.e., without
the need to resolve each singularly expensive phase transition point
right away, e.g., when interested in low-energy behavior.}
 {A fine temperature resolution can be obtained, nevertheless,
by using interleaved temperature grids.}
In general, XTRG can reach low temperatures
exponentially fast, and thus not only saves computational time but also merits better
accuracy due to significantly fewer truncation steps.  For similar
reasons, we also find that the series expansion thermal tensor
network (SETTN) approach benefits in both efficiency and precision,
from the logarithmic temperature scale setup.  
We work in an (effective) 1D setting
exploiting matrix product operators (MPOs)
which allows us to fully and uniquely implement non-Abelian
and Abelian symmetries to greatly enhance numerical performance.
We use our XTRG machinery to explore the thermal properties
of Heisenberg models on  {1D chains and 2D square and triangular lattices 
down to low temperatures approaching ground state properties}.
The entanglement properties, as well as the renormalization group flow of entanglement
spectra in MPOs, are discussed, where logarithmic entropies (approximately $\ln{\beta}$) 
are shown in both spin chains and square lattice models with gapless
towers of states. We also reveal that XTRG can be employed to
accurately simulate the Heisenberg XXZ model on the square
lattice which undergoes a thermal phase transition. We determine
its critical temperature based on thermal physical observables, as
well as entanglement measures.
 {Overall, we demonstrate that XTRG provides an elegant,
versatile, and highly competitive approach to explore thermal
properties, including finite
temperature thermal phase transitions as well as the different
ordering tendencies at various temperature scales for frustrated
systems.}

\end{abstract}

\date{\today}


\maketitle

\section{Introduction} 

Efficient simulations of interacting quantum many-body systems
are crucial for a better understanding of correlated materials.
In particular, accurate computation of thermodynamic quantities
including magnetization, heat capacity, magnetic susceptibility,
etc., enables a direct comparison to experiments and helps to
identify relevant microscopic models. The exotic quantum matter
includes Luttinger liquids in one (1D)
\cite{Xiang.j+:2017:CopperNitrate, Lake.b+:2013:Multispinon}
and spin liquid materials in two dimensions (2D)
\cite{Balents.l:2010:Spinliquid, Nasu2015:Kitaev,
Shen.y+:2016:Spinliquid, Do.s+:2017:Kitaev,
Yamashita.s+:2017:Spinliquid, Kelly.z.a+:2016:Spinliquid}.
Besides, the exploration and understanding of the rich and diverse
behavior of quantum many-body physics at different energy or,
equivalently, temperature scales are interesting from a theoretical
perspective.  One example  {are} thermal states near 1D quantum
critical point  {which} show universal entropy in the
partition function due to emergent conformal symmetry
\cite{Tu.h:2017:Klein,Tang.w+:2017:CFT,UniEntropy-2017} in the
low-energy regime. Another prominent example is the thermal
fractionalization in the honeycomb Kitaev spin liquid (KSL) at
finite temperature \cite{Nasu2015:Kitaev}, which has been
experimentally observed \cite{Do2017} in proximate KSL material
$\alpha$-RuCl$_3$ \cite{Plumb14,Banerjee2016}.

At a first glance, the simulation of thermal many-body states seems
a task more than challenging.  There exist exponentially many
excited states in the energy spectrum, many of which possess
volume-law entanglement and deny any efficient representation in
classical computers. However, it turns out that the ensemble density
operator, say $e^{-\beta H}$ with $\beta\equiv 1/T$ being the
inverse temperature, can be efficiently expressed and manipulated in
terms of thermal tensor network (TTN) states. The matrix product
operator (MPO) is a very natural TTN for describing 1D quantum
systems at finite temperature [i.e., (1+1)D], due to the
``entanglement" area law in thermal states of both gapped and
gapless systems with local interactions. Intuitively, thermal
fluctuation effectively ``opens" an excitation gap and introduces a
finite correlation length in mixed states, rendering an area law in
terms of total correlation \cite{Eisert.j+:2010:AreaLaws} (as well
as operator space entanglement \cite{Marko.z+:2008:Complexity}).
However, it was estimated that the required MPO bond dimension has
an upper bound scaling as $D \sim e^{\beta}$
\cite{Hastings.m.b:2006:Gapped} which still seems to pose a severe
barrier towards obtaining low-$T$ properties.

Nevertheless, on the other hand, various renormalization group
algorithms have been proposed to accurately compute thermodynamics
in the (1+1)D problems, practically even down to very low
temperatures.  These methods include the transfer matrix
renormalization group (TMRG) \cite{Bursill.r.j+:1996:DMRG,
Wang.x+:1997:TMRG, Xiang.t:1998:Thermodynamics} and
finite-temperature DMRG \cite{Feiguin.a.e+:2005:ftDMRG} which are
based on traditional density matrix renormalization group (DMRG),
tensor network algorithms such as the linearized tensor
renormalization group (LTRG) \cite{Li.w+:2011:LTRG,
Ran.s+:2012:Super-orthogonalization, Dong.y+:2017:BiLTRG}, and
variational projected entangled pair operator (PEPO) method in
(2+1)D \cite{Czarnik.p+:2012:PEPS, Czarnik.p+:2015:PEPS,
Czarnik.p+:2016:TNR}. Besides, a combination of finite-temperature
DMRG and Monte Carlo samplings called minimally entangled typical
thermal states (METTS) was proposed \cite{White.s.r:2009:METTS,
Stoudemire.e.m+:2010:METTS} and recently generalized to (2+1)D
\cite{Bruognolo.b+:2017:MPS}.  The success of these algorithm in
(1+1)D, and partly in (2+1)D, strongly suggests that $D$ does not
necessarily scale exponentially with $\beta$.

\begin{figure}[tb]
\includegraphics[width=0.95\linewidth]{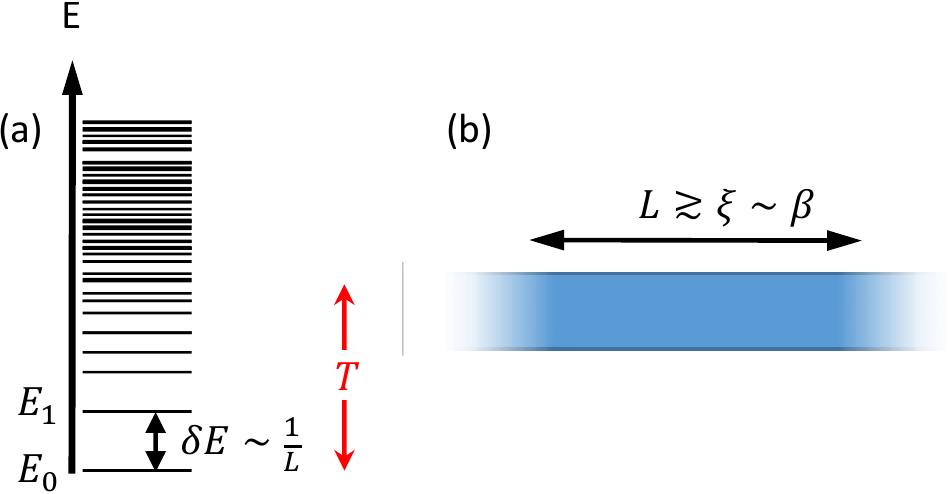}
\caption{(Color online) 
   (a) Finite-size spectra $E_s$ with low-energy
   level-spacing $\delta E \sim 1/L$. By requiring
   $T>\delta E$ for thermal averaging, this suggests
   $L\gtrsim\beta$, and therefore a thermal correlation
   length $\xi \sim \beta$.
   (b) A large or infinite system has an effective
   thermal cutoff $\xi\sim \beta$ in system length
   when measuring local observables. Therefore,
   provided  $\xi \lesssim L$ {\it finite} systems can 
   be used, as a very good approximation,
   to simulate thermal properties in the thermodynamic limit.}
\label{Fig:FS-spec}
\end{figure}

To estimate the computational cost in thermal simulations, one can
introduce a formal entanglement entropy in the TTN, e.g., in the MPO
representation of the mixed state density matrix, as introduced in
Refs.~\cite{Prosen.t+:2007:Entropy, Marko.z+:2008:Complexity}.
It has been revealed that this MPO entanglement saturates for gapped
systems and scales logarithmically (as $\frac{c}{3} \ln{\beta}$) for
quantum critical spin chains \cite{Prosen.t+:2007:Entropy,
Marko.z+:2008:Complexity}.  Very recently, two independent
works \cite{Barthel.t:2017:FiniteT,Dubail17}, deployed  {conformal field 
theory (CFT)}
arguments to show on general grounds that the Renyi entropy of
thermal states of effective 1D systems scales as $S_E^{(n)} \sim
\frac{c}{6} (1+\frac{1}{n}) \ln{\beta}$. In the limit $n\to1$ this
implies that also the von Neumann entropy scales like $S_E \sim
\frac{c}{3} \ln{\beta}$ for thermal states in 1D on general
grounds.

Intuitively, this scaling can be understood simply by considering
finite-size spectra with many-body low-energy level spacing $\delta
E \sim 1/L$, as schematically depicted in \Fig{Fig:FS-spec}.
In order to sample a thermal average, it must hold that at the very
least $T\gtrsim \delta E$, or equivalently $L\gtrsim \beta$ [in
other words, finite $\beta$ introduces an effective cutoff $\xi\sim
\beta$ of system size; see \Fig{Fig:FS-spec}(b)].  Now 
 {given that low-energy states violate a strict area law via
logarithmic corrections}, one has the block entropy $S_E \simeq \tfrac{c}{6}
\ln L + \mathrm{const}$ \cite{Bazavov17} for individual low-energy
states using open boundary conditions (OBC).  By choosing $L=a
\beta$ with fixed constant $a \gtrsim 1$, and by going from
individual low-energy pure states $|s\rangle$ to a thermal state
with weights $\rho_s$, i.e., $|s\rangle \to \sum_s \rho_s |s\rangle
\langle s|$, the block entropy for the outer product $|s\rangle
\langle s|$ acquires another factor of $2$.
Thermal averaging does not change this scaling due to the
subadditivity of the von Neumann entropy [see \App{App:SE}], given
the constraint $L=a \beta$ with fixed $a$.  By thermal averaging
over a similar set of low-energy states, the MPO block entropy at
the center of the system saturates by further increasing $L \gg a
\beta$ at a finite value, i.e., is cut off by \begin{eqnarray}
S_E[\rho(\beta)] \sim \tfrac{c}{3}\log\beta+\mathrm{const}
\label{eq:SE:scaling} \end{eqnarray} and, importantly, becomes
independent of $L$.  This block entanglement entropy of the thermal
state scales similar versus $\beta$ to the block entropy of a ground
state calculation versus $L$ using periodic boundary conditions
(PBC).

The above intuitive argument fits the holographic picture in terms
of thermal multiscale entanglement renormalization ansatz (MERA)
\cite{TNR-MERA}, where the minimal surface (of half the system) in
thermal MERA, as well as the corresponding (bipartite) entanglement
entropy, is argued to be proportional to $\ln{\beta}$
\cite{Swingle-holography}.

Furthermore, the argument of translating the scaling of the entropy
in $\ln L$ to that of $\ln \beta$ is completely consistent with the
notion within CFT that $\beta$ and $L$ are equivalent directions
connected via a modular $\mathcal{S}$ transformation.  This has
direct consequences for conformal TTN framework in (1+1)D, i.e.,
with one spatial (horizontal) and one imaginary time (vertical)
axis.  The horizontal transfer matrix $e^{-\tau H}$ across different
temperatures has the ground state of $H$ as its dominant eigenvector
which thus contains logarithmic entanglement.  For thermodynamics of
an infinite-size quantum chain ($L\gg \beta \gg1$), we therefore
expect that the vertical transfer matrix (across different
real-space sites) also has a dominant eigenvector  {with entanglement 
entropy $S_E\sim\ln(\beta)$}. In
addition, however, by definition of the partition function, it
acquires {\it intrinsic} PBC in the direction of temperature, which
therefore doubles the prefactor in entanglement entropy scaling, in
agreement with the earlier arguments.

This logarithmic growth of entropy versus $\beta$ provides a tight
upper bound in efficient thermal simulations
\cite{Barthel.t:2017:FiniteT}.  This, together with the constant
entanglement for gapped systems, suggests that the bond dimension
$D$ only needs to scale algebraically (constantly) as $\beta$
increases for critical (non-critical) quantum chains, respectively.

\begin{figure}[tbp]
\includegraphics[angle=0,width=1\linewidth]{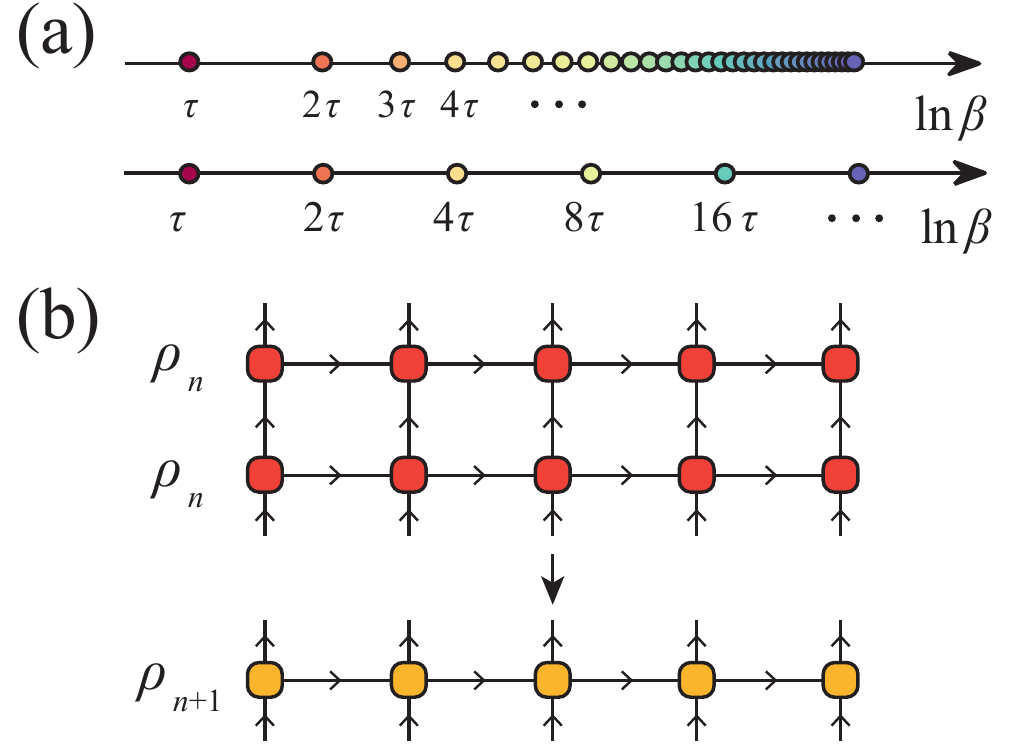}
\caption{(Color online)
(a) Linear versus logarithmic temperature scale employed in
   thermal simulations.
(b) A single step in XTRG evolution by
   projecting MPO $\rho_n$ (at $\beta=2^n \tau$) to itself.
   Following common notation, tensor networks are graphically
   depicted by blocks (i.e., tensors) connected by lines which are
   to be contracted.  Here vertical lines indicate physical state
   spaces, whereas horizontal lines indicate virtual or bond state
   spaces.  The exploitation of symmetries, quite generally,
   mandates directed lines, hence each line carries an arrow.
}
\label{Fig:Diagram}
\end{figure}

Conversely, it directly follows from the above logarithmic scaling
of the entanglement entropy that $\beta$ needs to change
significantly on a {\it relative} and not an absolute scale, in
order to see sizeable effect on the entanglement entropy. This
suggests for simulations that the numerical grid in $\beta$ should
be {\it logarithmically} discretized. In particular, as depicted in
\Fig{Fig:Diagram} by doubling $\beta \to 2\beta$, one can therefore
design an exponentially faster cooling procedure, in contrast to
current standard simulation techniques which linearly evolve the
full density matrix \cite{Bursill.r.j+:1996:DMRG,Wang.x+:1997:TMRG,
Xiang.t:1998:Thermodynamics,Feiguin.a.e+:2005:ftDMRG,Li.w+:2011:LTRG,
Ran.s+:2012:Super-orthogonalization,Dong.y+:2017:BiLTRG} or the
typical sampling states
\cite{White.s.r:2009:METTS,Stoudemire.e.m+:2010:METTS,
Bruognolo.b+:2017:MPS} in imaginary time.
We note that essentially a similar, even though much more involved
strategy was pursued in Refs.
\onlinecite{Czarnik.p+:2015:PEPS,Czarnik.p+:2016:TNR, HOTRG}.  Their
approach was based on a dimensional reduction via a nested
contraction of linear Trotter gates, followed by a variational
optimization of coarse-graining transformations
\cite{Levin.m+:2007:TRG}.  In contrast, our approach does not rely
on Trotter gates, and hence is straightforwardly applicable to
arbitrary Hamiltonians (1D and 2D). Overall, it represents an
extremely simple yet also very efficient approach.

In this work, inspired by the logarithmic MPO entanglement entropy,
we propose a one-way exponential tensor renormalization group (XTRG)
scheme along imaginary time.  Interestingly, this allows to draw
parallels to the concept of energy scales in the Numerical
Renormalization Group (NRG) \cite{Wilson75,Bulla08, Wb12_FDM}. There
also, with every new iteration the energy scale is reduced by a
factor, typically $\gtrsim \sqrt{2}$. Consequently, this also zooms
into the low-energy regime in an exponential fashion, while dealing
only with a very manageable linear number of iterations.

We benchmark our results with conventional linear evolution schemes.
The results show that, by following the entanglement structure and
exploiting the logarithmic temperature scale, one can obtain more
accurate results with less cost. By implementing non-Abelian
symmetries in the MPO, we can even simulate 2D clusters down to low
temperatures with high precision, and investigate thermodynamics and
related entanglement properties.

The model systems considered here are (anisotropic)
spin-half Heisenberg XXZ models 
\begin{eqnarray}
   H=J \sum_{\langle i,j \rangle} \bigl(S_i^x S_j^x  + S_i^y S_j^y + \Delta S_i^z S_j^z\bigr)
\label{eq:HeisenbergH}
\end{eqnarray}
both, in 1D Heisenberg chains ($\Delta=1$) of length $L$, as
well as in the 2D square lattice ($\Delta=1$ and $5$) for
systems of width $W$ and length $L$, and thus with a total of $N=W
L$ sites, using open (OBC) as well as periodic (PBC) boundary conditions.
We only include nearest
neighbor couplings as indicated by the sum $\langle \cdot,\cdot
\rangle$. 
For the purpose of benchmarking, we also consider
the spin-half XY-chain with $J_z=0$, i.e.,
\begin{eqnarray}
   H=J \sum_{\langle i,j \rangle}
   \bigl( S_i^x S_{j}^x + S_i^y S_{j}^y \bigr)
\label{eq:HeisenbergXY}
\end{eqnarray}
as this can be mapped to a fermionic tight-binding chain.
The XXZ model in Eq.~(\ref{eq:HeisenbergH}) possesses an U(1)
symmetry, which restores a larger SU(2) symmetry when $\Delta=1$.
Symmetries, whether non-abelian or abelian, are fully exploited,
throughout.
We also set $J:=1$ as the unit of energy, unless specified
otherwise. Furthermore we use units $k_B=\hbar=1$.

The rest part of the paper is organized as follows. In
\Sec{Sec:MPOD}, we introduce the XTRG scheme with symmetries
implemented, as well as an improved series-expansion thermal tensor
network (SETTN) method \cite{Chen.b+:2017:SETTN}  {based on a 
pointwise Taylor expansion algorithm that also exploits the logarithmic 
temperature scale}. 
The performances of these methods in the simulations of both
1D and 2D quantum many-body system, are presented and compared in
\Sec{Sec:Bench}.  In \Sec{Sec:TTNE}, the entanglement properties of
MPO are investigated, where logarithmic entanglement entropies
versus $\beta$ in the Heisenberg chain and also square lattice
models are discussed. XTRG is also employed to study the
finite-temperature phase transition of 2D Heisenberg XXZ model,
where we demonstrate that XTRG can accurately pinpoint the critical
temperature.
 
\section{Symmetric Thermal Tensor Networks in Logarithmic
Temperature Scale} \label{Sec:MPOD}

By construction, a thermal density matrix $\rho=e^{-\beta H}$
is a scalar operator, and thus shares exactly the same symmetries as
the Hamiltonian. For example, symmetries are preserved for
Trotter-Suzuki type TTNs \cite{Li.w+:2011:LTRG,
Ran.s+:2012:Super-orthogonalization}, where every local tensor
(storing Boltzmann weights) is symmetric.  Similarly, in the
series-expansion TTNs, it is also clear that arbitrary $H^n$'s have
exactly the same symmetry as $H$ (any unitary symmetry
transformation that leaves $H$ intact also leaves $H^n$ intact), and
so does the resulting tensor network representation of
$\rho(\beta)$. 

Concepts such as spontaneous symmetry breaking apply to individual
low-energy (eigen)states, but not to a thermal state.  Hence the
full exploitation of all symmetries of the Hamiltonian, abelian and
non-abelian alike, are very natural in XTRG. For finite systems with
open boundary, in particular, XTRG shares the same benefits in
efficiency as DMRG in quasi-1D. There is a notable difference,
however: as long as the thermal correlation length is clearly
smaller than the system size under consideration, local thermal
properties in the center of the system can be regarded as in the
thermodynamic limit.

\subsection{Symmetric Matrix Product Operator}

The explicit implementation of non-Abelian symmetries has been
regarded as a standard technique in ground state DMRG simulations
($T=0$) \cite{McCulloch.i.p+:2002:NonAbelian}, which has many
important applications including exploring quantum spin liquids in
frustrated quantum magnets \cite{Depenbrock.s+:2012:SpinLiquid,
Gong.s+:2014:Plaquette}, and is also shown to be useful in
METTS-type thermal simulations \cite{Bruognolo.b+:2015:SYMETTS,
Binder.m+:2017:SYMETTS}. However, the implementations of non-Abelian
symmetries in MPO for finite-temperature simulations are still
absent. Here by virtue of the flexible and versatile QSpace
framework \cite{Weichselbaum.a:2012:QSpace}, we implement
non-Abelian SU(2), as well as Abelian U(1), symmetry in the MPO
algorithm and thus realize a very efficient thermal renormalization
group (RG) algorithm that can also be applied to 2D problems.

In our MPO-based thermal algorithm, we start by constructing an
SU(2) invariant MPO representation of $H$.  As this involves reduced
matrix elements in the spirit of Wigner-Eckart theorem
\cite{Weichselbaum.a:2012:QSpace}, we refer to this as the reduced
MPO in contrast to the full MPO  {when not exploiting non-Abelian 
symmetries}.  By switching from a state-based to a
multiplet-based description, we can reduce the overall bond
dimension from $D$ states to $\Dstar <D$ multiplets.
The reduced MPO representation of $H$ can be constructed by automata
method \cite{Crosswhite.g.m+:2008:Automata, Frowis.f+:2010:Tensor,
Pirvu.b+:2010:MPO} for 1D Hamiltonians and MPO sum-and-compress
scheme \cite{Hubig.c+:2017:MPO} for more complicated 2D lattice
models. For a Heisenberg chain with nearest-neighbor interactions, a
full MPO requires $D_H=5$ bond states. As  {these correspond} to two
singlets and one triplet, i.e., $\underbar{1}^2 \oplus
\underbar{3}^1$ where $\underbar{d}^n$ specifies $n$ multiplets of
dimension $d$ each, the reduced SU(2) invariant MPO only involves
$\Dstar _H=3$ multiplets.  For the Heisenberg model on a 2D square
lattice, we map a system of width $W$ to a 1D snake-like chain with
``long-range" interactions (up to distance $2W-1$).  Then e.g.,
using OBC, the full MPO requires $D_H = 3 W + 2$ bond states, while
the reduced SU(2) invariant MPO has a significantly more compact
representation with only $\Dstar_H = W + 2$ multiplets
($\underbar{1}^2 \oplus \underbar{3}^{W}$).  More details on the
symmetric MPO representation of total Hamiltonian can be found in
\App{App:SYMPO}.

The computational cost in a tensor network algorithm typically
scales with some power $\order{D^m}$ aside other factors concerning
number of sites etc., where for the MPO structure of this paper we
encounter $m=[3,\ldots,6]$.  By exploiting non-abelian symmetries,
the computational cost can be effectively reduced to
$\order{(\Dstar)^m}$ which leads to a gain in numerical efficiency
by $\orderb{\bigr( \tfrac{\Dstar}{D}\bigl)^m}$.  For a single SU(2)
symmetry, it roughly holds, on average, $D/\Dstar  {\sim 3\ldots4}$ for
spin-1/2 systems. Note also that multiplet dimensions are typically
somewhat larger in thermal MPO as compared to matrix product ground
states which renders us even greater numerical gain of symmetry
implementation.  The underlying reason for this is that an MPO has
two physical indices associated with the same site. Therefore their
direct product already also leads to an enlarged effective local
spin.  With this in mind, for the sake of readability, we will
generally quote estimates in numerical efficiency in terms of $D$
since after all,  $\order{D^m} = \orderb{(\Dstar)^m}$ with the
overall scale factor $(\Dstar/D)^m \ll 1$ absorbed into the
definition of $\order{\ast}$.

\subsection{Exponential Tensor Renormalization Group}
\label{SubSec:XTRG}

For one-dimensional critical systems, the entanglement entropy in
the MPO of a thermal state diverges only logarithmically in $\beta$.
Therefore to see a sizeable effect in the properties of a thermal
state, $\beta$ must change significantly on a relative and not an
absolute scale. E.g., a change $\beta \to a\beta  \to a^2\beta \to
\ldots$ with some constant $a>1$ will change the entanglement by
linear increments. This strongly suggests to scale $\beta$ on a
logarithmic and not on a linear scale.

We can take fully advantage of this by a novel approach, which we
refer to as exponential tensor renormalization group (XTRG), to
simulate quantum many-body systems at finite temperatures, with high
efficiency and accuracy. We start by preparing an MPO of the
(unnormalized) thermal state $\rho(\tau)=e^{-\tau H}$ at
exponentially small $\tau$, i.e., at very high temperature. Then we
can proceed to cool down the system exponentially by multiplying the
thermal state with itself, 
\begin{eqnarray}
  \rho_0 \equiv \rho(\tau)
  \to \rho_1 \equiv \rho(\tau) \cdot \rho(\tau) = \rho(2\tau)
  \text{.}
\label{eq:rhoinit}
\end{eqnarray}
Feeding the last MPO iteratively into the next step, with $\tau_{n}
\equiv 2^n\tau$ and therefore $\taui \equiv \tau$, we obtain,
\begin{eqnarray}
   \rho_n \equiv \rho(\tau_n) = \rho_{n-1} \cdot \rho_{n-1}
\text{ .}\label{eq:rho:scale2}
\end{eqnarray}
This directly implies an exponential acceleration to reach low
temperatures.

Importantly, in the present XTRG scheme we can easily start from
exponentially small $\tau$.  For example, for $\tau J = 10^{-3}$
with $J:=1$ the largest local energy scale here given by the
Heisenberg coupling strength, we can use an efficient series
expansion scheme [cf.  \Sec{SubSec:SETTN}].  For $\tau$ as small as
$10^{-6}$ even simplest lowest-order linear expansion of $e^{-\tau
H}$ can suffice, which extremely simplifies initialization even for
longer-ranged Hamiltonians  {which become cumbersome for 
Trotter-like decompositions}, 
or for 2D Hamiltonians in the effective 1D-MPO
setup.  In the latter setup, with minor modifications, the MPO of
the Hamiltonian already encodes the essential structure of the
thermal state using the {\it same} bond dimension. A detailed
comparison of different initialization strategies, including the
Trotter-Suzuki decomposition, SETTN, and simple linear
initialization for small $\taui$ is provided in \App{App:Init}.

Given an MPO representation for $\rho(\tau_n=\beta/2)$, we can
compute the (unnormalized) thermal state at temperature $T=1/\beta$
via $\rho(\beta) = \rho(\tau_n)^\dagger \rho(\tau_n)$, i.e., by
contracting $\rho(\tau_n)$ with its conjugate.  This guarantees
positivity of the thermal state $\rho(\beta)$ even in the presence
of truncation of the MPO for $\rho(\tau_n)$.  Furthermore, we can
also compute the partition function at $\beta=2\tau_n$ via
$\mathcal{Z}(\beta) = \Tr{ \rho(\tau_n)^\dagger \rho(\tau_n) }$, and
thus gain another factor of two to reach lower temperatures.  The
latter can be simply obtained by computing the Frobenius norm
squared of $\rho(\tau_n)$.  Not incidentally, many of the features
above are directly related with common procedures within the setup
of a purified thermal state.  \cite{Bursill.r.j+:1996:DMRG,
Wang.x+:1997:TMRG,Li.w+:2011:LTRG, Xiang.t:1998:Thermodynamics,
Feiguin.a.e+:2005:ftDMRG, Dong.y+:2017:BiLTRG, Czarnik.p+:2015:PEPS,
Czarnik.p+:2016:TNR, Schwarz17}.

In case the grid of inverse temperature values is too sparse,
intermediate values can be easily obtained by shifting the initial
value of
\begin{subequations}
\label{eq:zshift}
\begin{eqnarray}
   \taui \to \taui \cdot 2^{z} \qquad \text{with }
   z \in [0,1)
\text{ ,}\label{eq:zshift:1}
\end{eqnarray}
a procedure that is entirely analogous to $z$-shifts within the NRG.
In order to obtain a uniform logarithmic grid over $n_z$ shifts, one
may simply choose
\begin{eqnarray}
z_i = \tfrac{i}{n_z} \qquad \text{with } i=0,\ldots,n_z-1
\text{ .}\label{eq:zshift:2}
\end{eqnarray}
\end{subequations}
Different ``$z$-shifts'' can be computed completely independently
from each other and can therefore be efficiently parallelized.
Truncation errors are still kept minimal by moving to large $\beta$
as quickly as possible in an accurate manner.  Alternatively, one
can obtain intermediate values of $\beta$ also by computing
$\rho_{n,n'} \equiv \rho_{n-1} \cdot \rho_{n'-1}$ for various $n'
\le n$.

\subsection{Series Expansion Thermal Tensor Networks}
\label{SubSec:SETTN}

\begin{figure}[tbp]
\includegraphics[angle=0,width=1\linewidth]{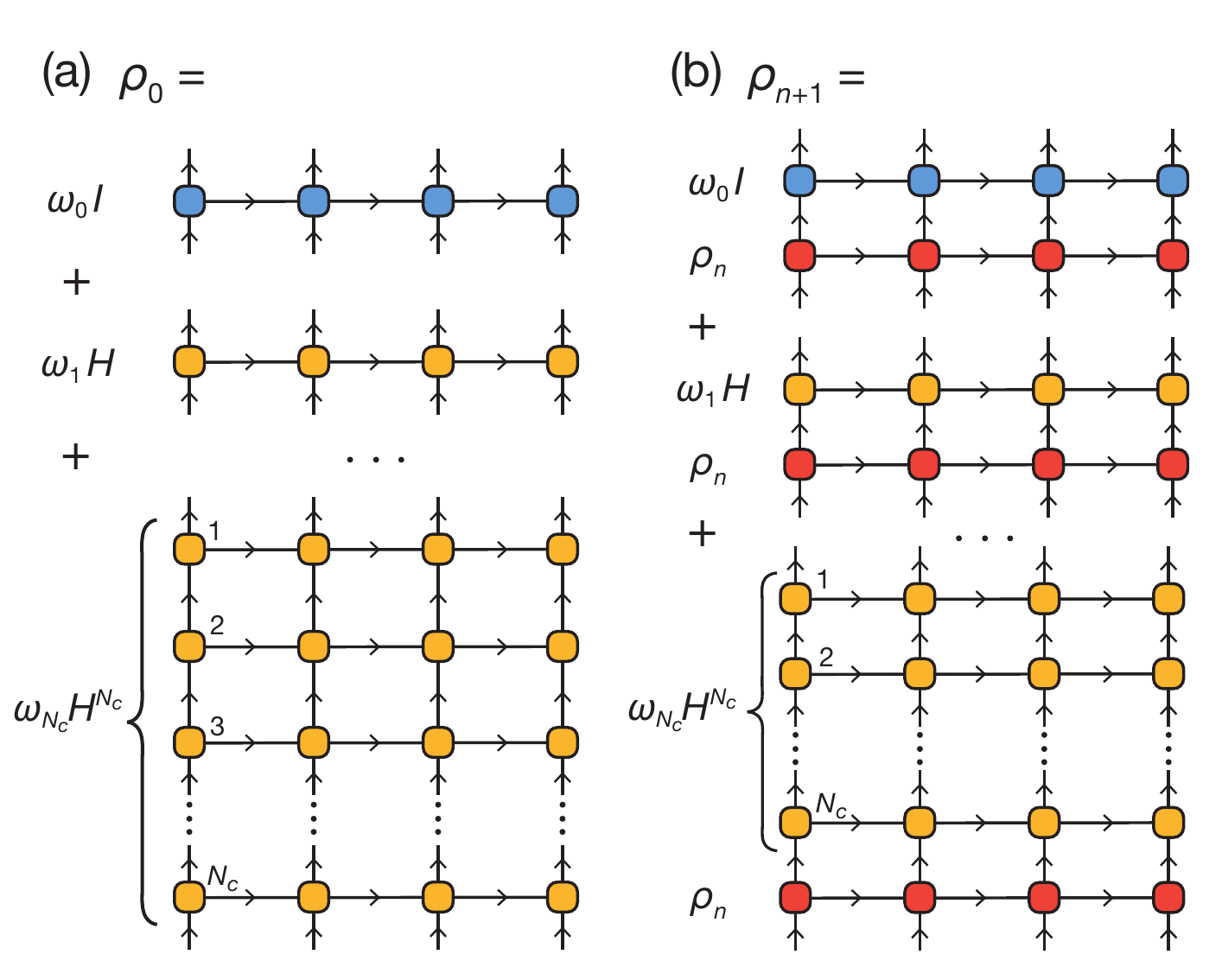}
\caption{(Color online)
   (a) SETTN initialization of $\rho_0$ (with $\beta=\taui$)
   using Maclaurin expansion with coefficients $\omega_k
   \equiv (-\taui)^k / k!$ [see \Eq{Eq:RhoTau}].  (b)
   Pointwise SETTN algorithm exploiting the logarithmic
   $\beta$ scale, here with coefficients $\omega_k \equiv
   (\tau_n - \tau_{n+1})^k / k!$ [see \Eq{Eq:Tay}].
}
\label{Fig:SETTNAlg}
\end{figure}

Series-expansion thermal tensor network (SETTN) is a
``continuous-time" RG approach for the accurate simulation of
quantum lattice models at finite temperature
\cite{Chen.b+:2017:SETTN}. By exploiting the series expansion of
density matrix in Eq.~(\ref{Eq:RhoTau}), SETTN is essentially free
of discretization errors, making it distinct from previous
Trotter-Suzuki type RG methods including TMRG
\cite{Wang.x+:1997:TMRG, Xiang.t:1998:Thermodynamics},
finite-temperature DMRG \cite{Feiguin.a.e+:2005:ftDMRG}, LTRG
\cite{Li.w+:2011:LTRG, Ran.s+:2012:Super-orthogonalization,
Dong.y+:2017:BiLTRG}, and METTS \cite{White.s.r:2009:METTS}, etc.
The efficient MPO representations of $H^n$ is the key for the
algorithm to work, and both OBC and PBC chain systems can be equally
well dealt with in SETTN (here, specifically, we simply use one
long-range bond for the simulation of PBC).  Being free of Trotter
errors, SETTN has better controllable and uniformly higher accuracy,
compared to conventional thermal RG methods. 

To initialize $\rho_0$ for small $\taui$, a series expansion yields
[cf. Fig.~\ref{Fig:SETTNAlg}(a)]
\begin{subequations}
  \label{Eq:SETTN}
\begin{equation}
\rho_0 \equiv \rho(\taui) \simeq 
\sum_{k=0}^{\Ncut}
  \tfrac{(-\taui)^k}{k!} H^k
\text{ .}\label{Eq:RhoTau}
\end{equation} 
The required cutoff order of the expansion is $\Ncut \sim N \taui$ ,
i.e., proportional to the total number of sites $N$. In practice,
$\Ncut$ is determined automatically by only allowing a negligibly
small expansion error ($<10^{-15}$).  Therefore for sufficiently
small $\taui$, the initialization of $\rho(\taui)$ above is
well-controlled and accurate, typically resulting in $\Ncut\lesssim
10$.

The high-temperature Maclaurin expansion in \Eq{Eq:RhoTau} can be
employed not only in the intialization stage, but also for
simulating low-temperature thermal states, as shown in Ref.
\onlinecite{Chen.b+:2017:SETTN}.  Despite its competitive
performance, this method still leaves room for further improvement.
Since \Eq{Eq:RhoTau} expands $\rho$ around the infinitely high
temperature, i.e., $\beta=0$, the power series in $H^n$ involves
large $\Ncut \propto N \beta$ for large system size $N$ and low
temperature $1/\beta$.  The precision of  SETTN is limited by the
truncation in $H^n$ [see \Sec{Sec:trunc}], which generally increases
as $n$ increases \cite{Chen.b+:2017:SETTN}.  In this sense, a
pointwise Taylor expansion can help reduce the expansion order
$\Ncut$ and improve the accuracy, i.e.,
\begin{equation}
   \rho(\tau_{n+1}) \simeq \Bigl(
   \sum_{k=0}^{\Ncut}
   \tfrac{(\tau_n - \tau_{n+1})^k}{k!} H^k \Bigr) \ 
   \underset{\equiv \rho_n}{\underbrace{e^{-\tau_n H}}} 
\text{ .}\label{Eq:Tay}
\end{equation}
\end{subequations}
\EQ{Eq:Tay} expands the density operator around an arbitrary
but fixed $\tau_n$.  For generality, the initialization in
\Eq{Eq:RhoTau} may be viewed as iteration $n=-1$ having
$\rho_{-1} = \mathbb{I}$ and $\tau_{-1} =0$.

Now given the density operator $\rho(\taui)$ obtained by
initialization via \Eq{Eq:RhoTau}, $\rho(a\taui)$ with $a>1$ can be
obtained via Taylor expansion around $\rho(\taui)$. In particular,
this also hold for $a=2$ which thus may serve as a complimentary
scheme to the XTRG above.
For example, alternative to the plain doubling scheme above, SETTN
may be employed to cool down the system and obtain the MPO form of
density operators at the inverse temperature grid $\tau_n$.  Given
$\rho_n$, the MPO representation of $\rho_{n+1}$ can be expanded as
in \Eq{Eq:Tay}.  Each term in the summation there can be obtained
iteratively by projecting $H$ onto $(H^{n-1} \rho_n)$ and
compressing the product.
For the overall sum then [cf. \Fig{Fig:SETTNAlg}(b)] we also employ
variational optimization (see \App{App:VariComp}) to finally arrive
at the MPO representation of $\rho_{n+1} \equiv \rho(\tau_{n+1})$.  

By repeating this procedure, we also can follow the XTRG protocol to
cool down the  {system} along the inverse temperature grid $\tau_n
= 2^n \taui$.  In contrast to the plain doubling scheme in XTRG,
however, in case of SETTN the step size $\delta\tau \equiv
\tau_{n+1}-\tau_{n}$ can be chosen continuously.  In this sense,
SETTN is more flexible as it permits the flexible exploration of
thermal properties in the immediate vicinity of temperature $\tau_n$
with only modest cost.

Note that for this {\it improved} SETTN, as we will refer to it,
using an exponentially increasing $\tau_n$ series does not acquire
exponential acceleration as XTRG does, since in the case of SETTN
one still needs to perform projection and compression operations
$\Ncut \propto \beta N$ times.  Nevertheless, from the point of view
of SETTN, the exponential $\tau_n$ series is computationally
preferable to, say, a linear $\tau_n$ series as expansion points,
since the former can reduce expansion overhead and thus save
computational time, in practice, without losing any accuracy (see
\App{App:LvsX} for a detailed comparison).

\subsection{MPO compression and numerical cost}
\label{Sec:trunc}

In SETTN we start with a reduced SU(2) invariant MPO for $H$. Then
we iteratively apply the projections $H$ onto $(H^{k-1} \rho_n)$ to
obtain $H^k \rho_n$, with \Eq{Eq:RhoTau} represented by $\rho_{n=-1}
= \mathbb{I}$.  These projections need to be combined with a
compression algorithm to reduce numerical cost in a controlled
manner.  In the present context, however, truncation by discarded
weight is dangerous since small weights for small $\taui$ can affect
the accuracy for large $\tau_n$.  Hence we truncate by number of
multiplets, throughout.  For this, we introduce the control
parameters $\Dstar_{n,k}$ which stand for the maximum number of
multiplets $\Dstar$ to be kept in the $k$-th iterative term when
computing $\rho_{n}$.  For simplicity, we set this parameter
constant, i.e., $\Dstar_{n} \equiv \Dstar_{n,k}$, which also stands
for the bond dimension of the target state $\rho_{n}$. Furthermore,
we choose constant $\Dstar \equiv \Dstar_{n>0}$ but, for the sake
of the analysis, may use a different value for $\Dstari$ for the
initialization in \Eq{Eq:RhoTau} if specified.
 
For an extremely small $\taui$ (say, as small as $10^{-4}$ to
$10^{-8}$), the initialization of $\rho(\taui) =e^{-\taui H}$ can be
simplified to lowest-order, i.e., linear expansion $\rho(\taui)
\simeq 1-\taui H$. Having $\mathcal{N}_c=1$ in \Eq{Eq:RhoTau}, the
result shares the same bond-dimension $\Dstari = \Dstar_H$ as $H$
itself.
In constrast, when expanding around finite $\tau_n$ as in
\Eq{Eq:Tay}, the bond dimension $\Dstar$ in $ \rho(\tau_n)$
typically needs to grow significantly, and therefore is fixed to
some specified $\Dstar \gg \Dstari$.

The compression of the SETTN projections above can be achieved
either by a singular value decomposition (SVD) technique quite
similar to that in Ref.~\cite{Chen.b+:2017:SETTN}, apart from the
fact that the MPOs here have SU(2) symmetry, or by a variational
optimization which can greatly improve numerical efficiency (see
\App{App:VariComp} for more details on related MPO compression
techniques).  Within SETTN, the cost of either compression scheme
scales like $\order{D^3}$.  We tested both and found comparable
numerical accuracy.
Finally, we variationally add up the MPOs for $H^k \rho_n$ with
coefficients as in \Eqs{Eq:SETTN} to obtain $\rho_{n+1}$ (cf.
\App{sec:compress:sum}).

In contrast, XTRG projects $\rho_n$ onto itself in
\Eq{eq:rho:scale2}. So {\it both} MPOs involved have large bond
dimension $\Dstar \gg  \Dstar_{0}, \Dstar_H$.  Here a direct SVD
compression is numerically costly, $\order{D^6}$, to be compared to
$\order{D^3}$ for SETTN.  For the XTRG iteration in
\Eq{eq:rho:scale2} we therefore constrain ourselves to a variational
compression which scales like $\order{D^4}$ [see
\App{sec:compress:prod}].

\begin{figure}[tbp]
\includegraphics[angle=0,width=1\linewidth]{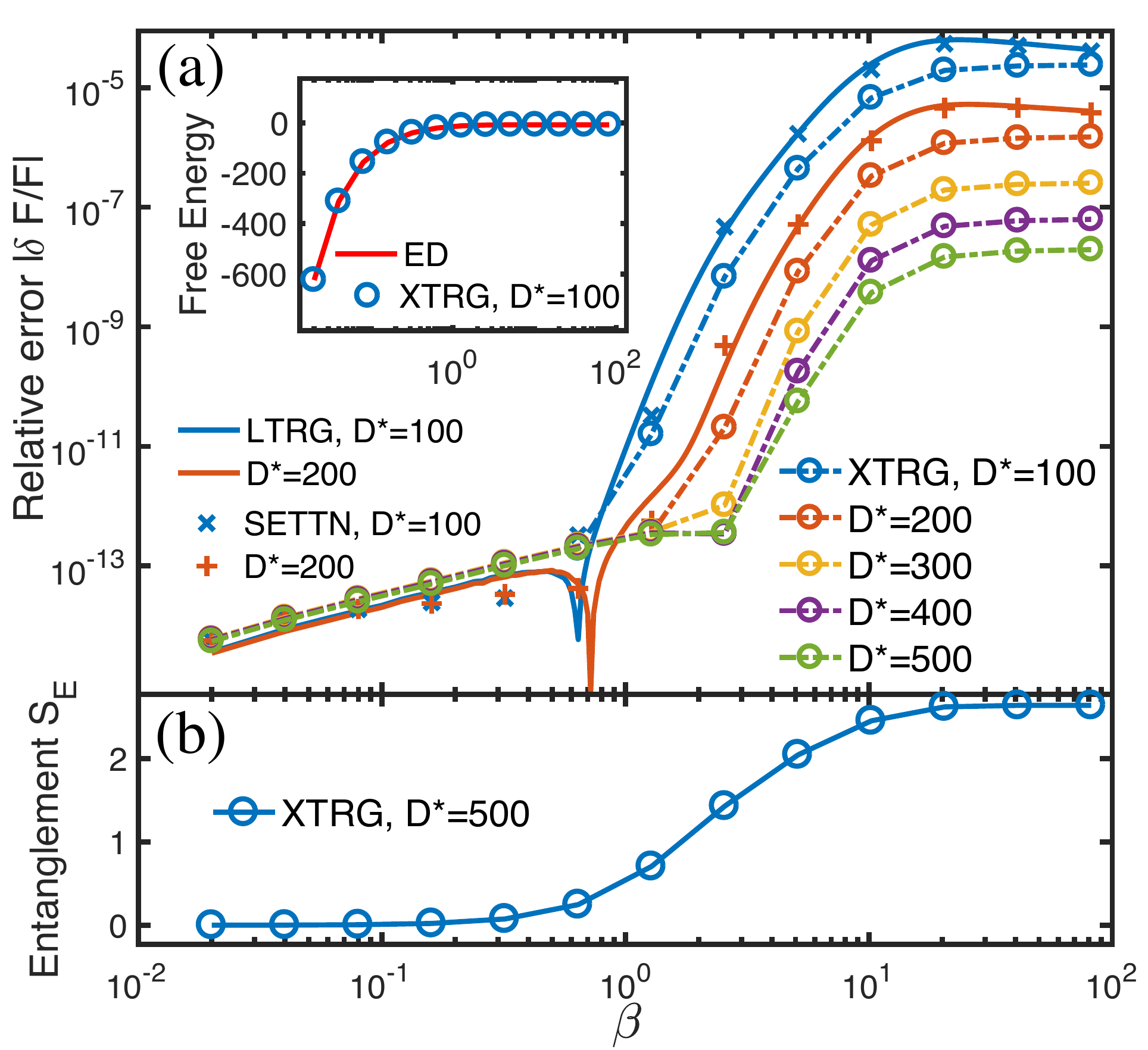}
\caption{(Color online)
   (a) Relative errors of free energy of an $L=18$ spin-1/2
   Heisenberg chain (PBC), calculated by TTN algorithms
   including LTRG,  XTRG, and SETTN relative to ED, with
   SU(2) symmetry implemented, throughout. Data for different
   methods share the same color for the same $\Dstar$.  Here
   $\rho(\taui=0.01)$ was initialized using SETTN.  (b)
   Bipartite entanglement entropy $S_E$ in the middle of the
   MPO, which increases  {monotonically}.  The time costs of the
   LTRG calculations are  8 (10) times as long as SETTN and
   180 (113) times of XTRG, in $\Dstar=100$ $(200)$
   calculations, respectively.
}
\label{Fig:FeErr}
\end{figure}

More explicitly, we summarize in Tab.~\ref{Tab:Cost}
the time costs of the three algorithms involved in the current
discussions.
\begin{table}[tb]
\caption{ 
   Time complexity of thermal tensor network
   methods for a lattice of
   length $L$ and width $W$,
   i.e. a total of $N\equiv W L$ sites,
   assuming $\beta>1$
   (by default, $\beta$ and $\tau$ are 
   in units of $1/J$).}
\begin{tabular}{cccc}
    \toprule
    Methods & XTRG & SETTN  & LTRG \\
    \hline
    complexity &  $\order{\ln(\beta)N D^4}$ & $\order{\beta N^2 D^3 D_H}$ & $\orderb{\frac{\beta}{\tau} N D^3 W}$ \\       
    relative cost $q$ \footnote{XTRG cost is set as the time unit.} & 1 &
    $\order{\tfrac{\beta N}{D \ln{\beta}} D_H}$ &
    $\order{\frac{\beta/ \tau}{D \ln{\beta}} W}$ \\    
    \hline\hline
\end{tabular}
\label{Tab:Cost}
\end{table}
The numerical cost of SETTN for an entire run up to inverse
temperature $\beta$ scales as $\order{\beta N^2 D^3 D_H}$ assuming
$\mathcal{N}_c \propto \beta N$ for large $\beta,N$ with
$\beta=\tau_n$ in \Eqs{Eq:SETTN}, whereas XTRG scales as
$\order{\ln(\beta)N D^4}$.  We can thus estimate the relative run
time of SETTN over XTRG as $q_\mathrm{S}\equiv \tfrac{\beta
N}{D\ln{\beta}} D_H$.
For practical simulations as in Figs.~\ref{Fig:FeErr} and
\ref{Fig:FeErrXY}, we find that XTRG calculations are faster than
SETTN by more than one order of magnitude. In 1D critical systems,
since the required bond dimension scales as $D\sim e^S \sim
\beta^{\lambda}$ ($\lambda \lesssim 1$ for $c=1$ CFTs, say, spin-1/2
Heisenberg chain, see Ref. \onlinecite{Barthel.t:2017:FiniteT}).
Thus, $q_\mathrm{S} \gtrsim \tfrac{N}{\ln{\beta}} D_H$, with
$q_\mathrm{S} \gg 1$ for $N$ large and $\beta>1$ (in units of
$1/J$). Similarly, also Trotter-Suzuki type linear thermal RG
methods, like the finite-temperature DMRG
\cite{Feiguin.a.e+:2005:ftDMRG} and LTRG \cite{Li.w+:2011:LTRG,
Dong.y+:2017:BiLTRG}, with scaling $\orderb{\frac{\beta}{\tau} N
D^3}$ (last column of Tab.~\ref{Tab:Cost} with $W=1$), are
typically much slower by a factor $q_\mathrm{L} \simeq
\frac{1}{\tau \ln{\beta}} \gg 1$ as compared to XTRG. 

It is also revealing to compare the efficiency of XTRG with
currently most efficient scheme in 2D systems, i.e., Trotter-Suzuki
decomposition plus swap gates \cite{Bruognolo.b+:2017:MPS}.  The
numerical (time) cost of the latter scheme scales like
$\orderb{\frac{\beta}{\tau} N D^3 W}$, where the additional factor
$W$ stems from the number of required swap gates which is
proportional to the width $W$.  For 2D, however, typically $D
\gg \beta$. Therefore the relative cost of XTRG scales like
$q_\mathrm{2D} \equiv \frac{\tau D}{W \beta} \ln{\beta}$,
resulting in $q_\mathrm{2D}\sim\order{1}$ [e.g., with
$W=8, \tau=0.05, \beta=50$, and $D^* \sim 2000$ (correspondingly
$D\simeq 8\times10^3 \sim 10^4$) in SU(2) simulations, or
$D\sim2000$ in U(1) calculations, one obtains $q_\mathrm{2D} \sim
0.98$].  Nevertheless, XTRG is still clearly advantageous over
Trotter gates due to the far fewer truncation steps involved.
Besides, XTRG can be simply and efficiently parallelized based on
$z$-shifts [cf.~\Eq{eq:zshift}].
For thermal simulations that are dominated by Trotter error and
swap gates in LTRG schemes to reach low temperatures, and not
necessarily by the truncation error due to entanglement growth, XTRG
may be crucial to reach the lowest temperature scales e.g. in
systems with more than one well separated physical energy scales.
As a very interesting example, in the Kitaev honeycomb model
the gauge flux excitation peak in the specific heat curve locates at
very low temperature, to be referred to as $T_l$,
compared to the high-temperature peak at $T_h$.
To see the $T_l$ peak in the specific heat, one needs to cool
down the system till extremely low temperatures ($T_l/T_h 
 {\lesssim 10^{-2}}$, 
depending on the coupling constants and boundary conditions,
\cite{Nasu2015:Kitaev}) 
where the exponential acceleration in XTRG can play a very
important role. A similar scenario is observed in the 2D triangular
Heisenberg lattice which we will discuss in more detail below.

\section{Benchmark calculations: 1D and 2D Heisenberg models at finite temperature}
\label{Sec:Bench}

 {In this section we benchmark our XTRG starting from well understood
models such as the 1D Heisenberg chain [\Sec{sec:HChain}] and the
2D square Heisenberg lattice (SLH) [\Sec{Sec:2DHeisenberg}].
We then proceed towards the much less understood and thus much less trivial
finite temperature behaviors of the triangular lattice Heisenberg (TLH) model [\Sec{sec:frustrated}] where quite generally, the strong local frustration
represents a huge challenge to numerical simulations.}

\subsection{ {Thermodynamic quantities}}

\begin{figure*}
  \centering
  \subfigure{
    \label{fig:subfig:a} 
    \includegraphics[width=0.305\linewidth]{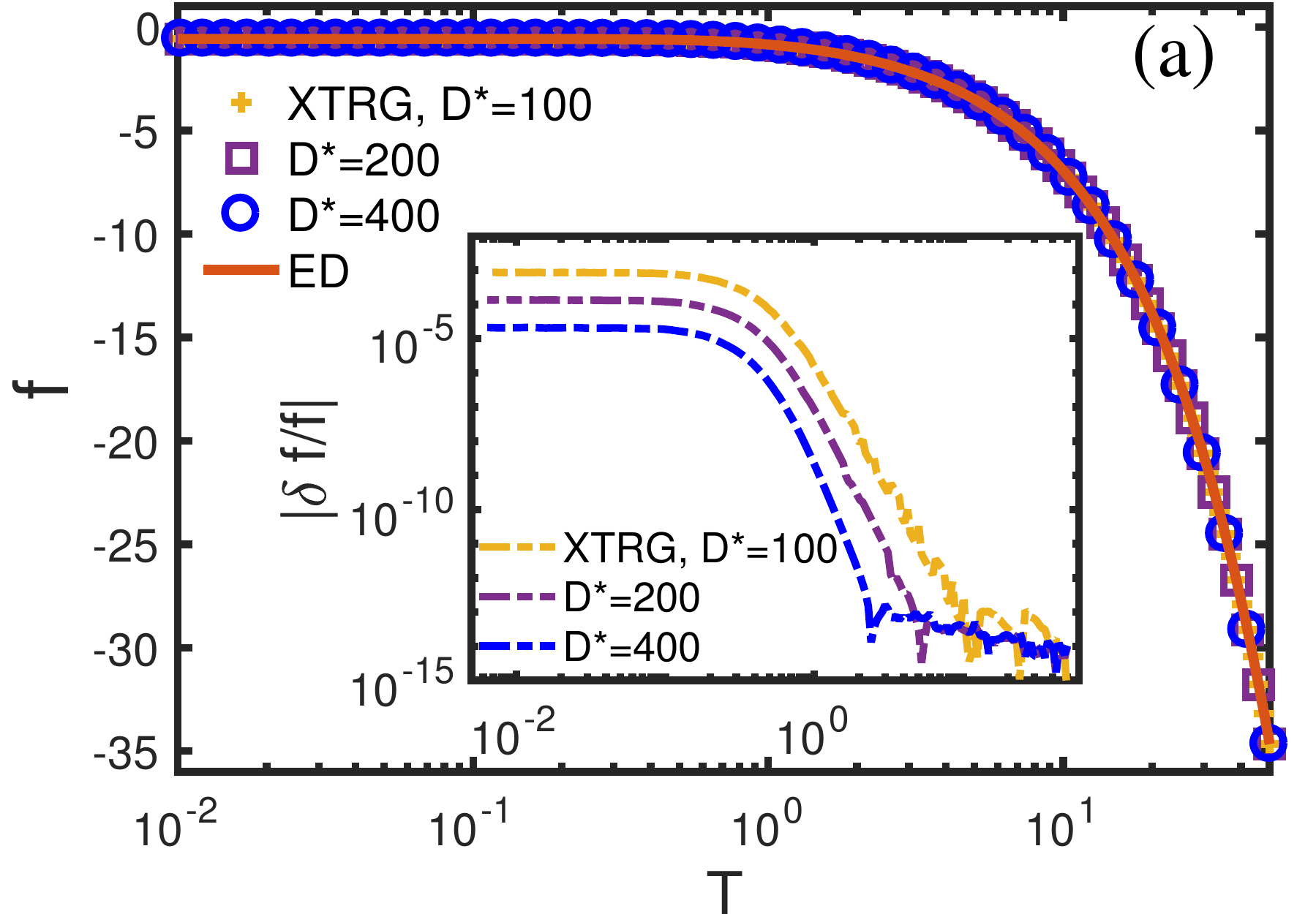}}
  \subfigure{
    \label{fig:subfig:b} 
    \includegraphics[width=0.305\linewidth]{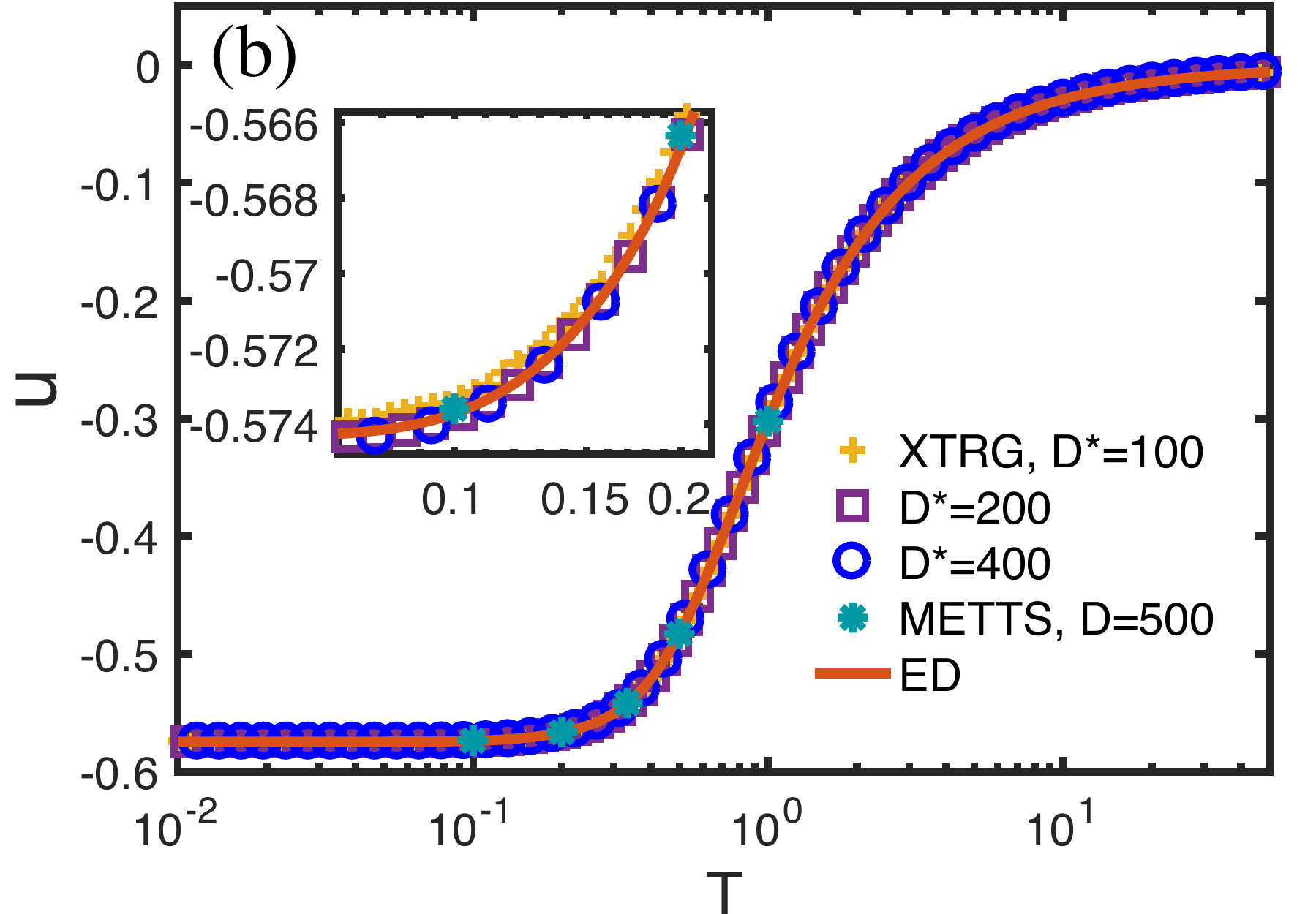}}
   \subfigure{
    \label{fig:subfig:b} 
    \includegraphics[width=0.305\linewidth]{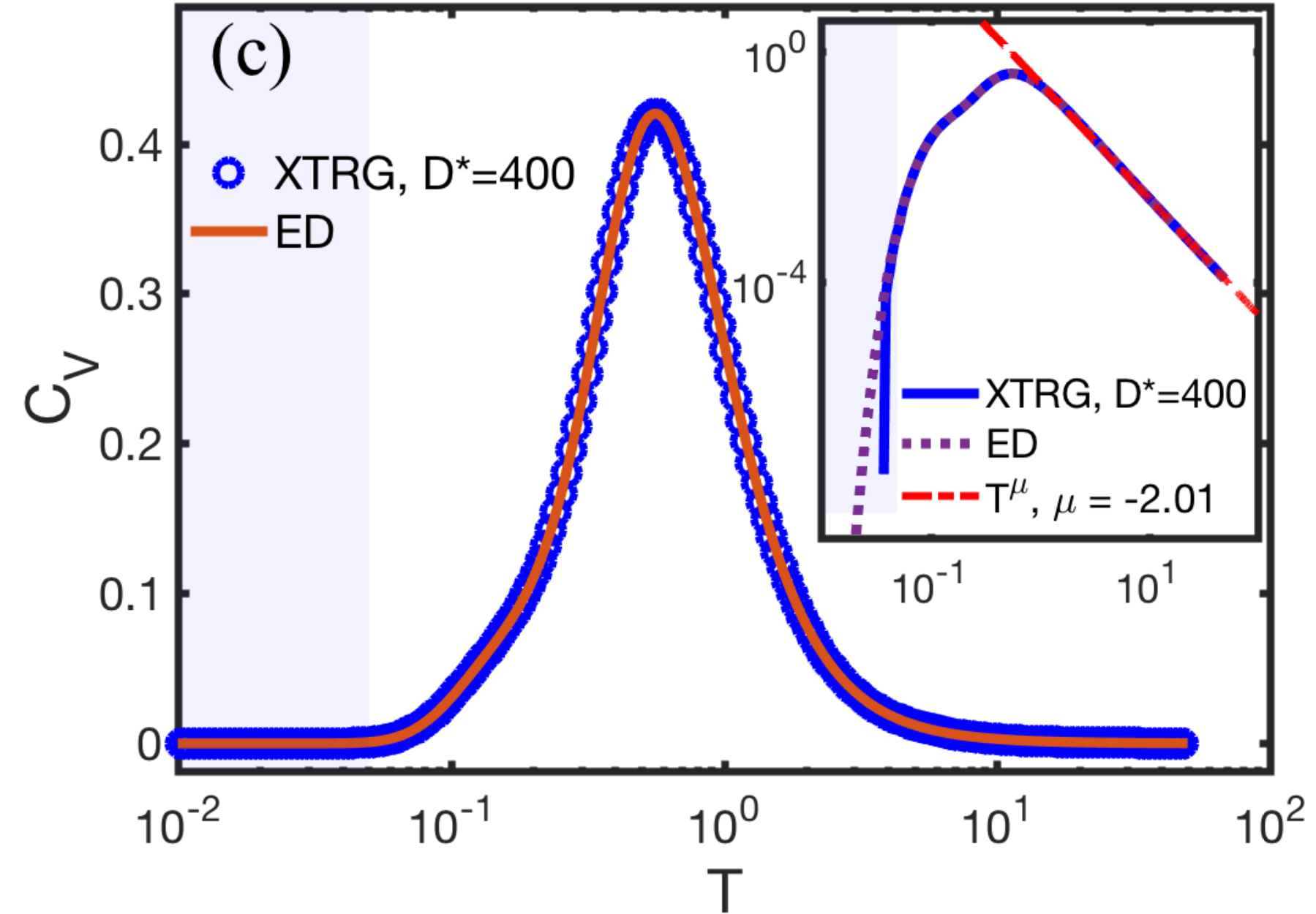}}
   \subfigure{
    \label{fig:subfig:b} 
    \includegraphics[width=0.305\linewidth]{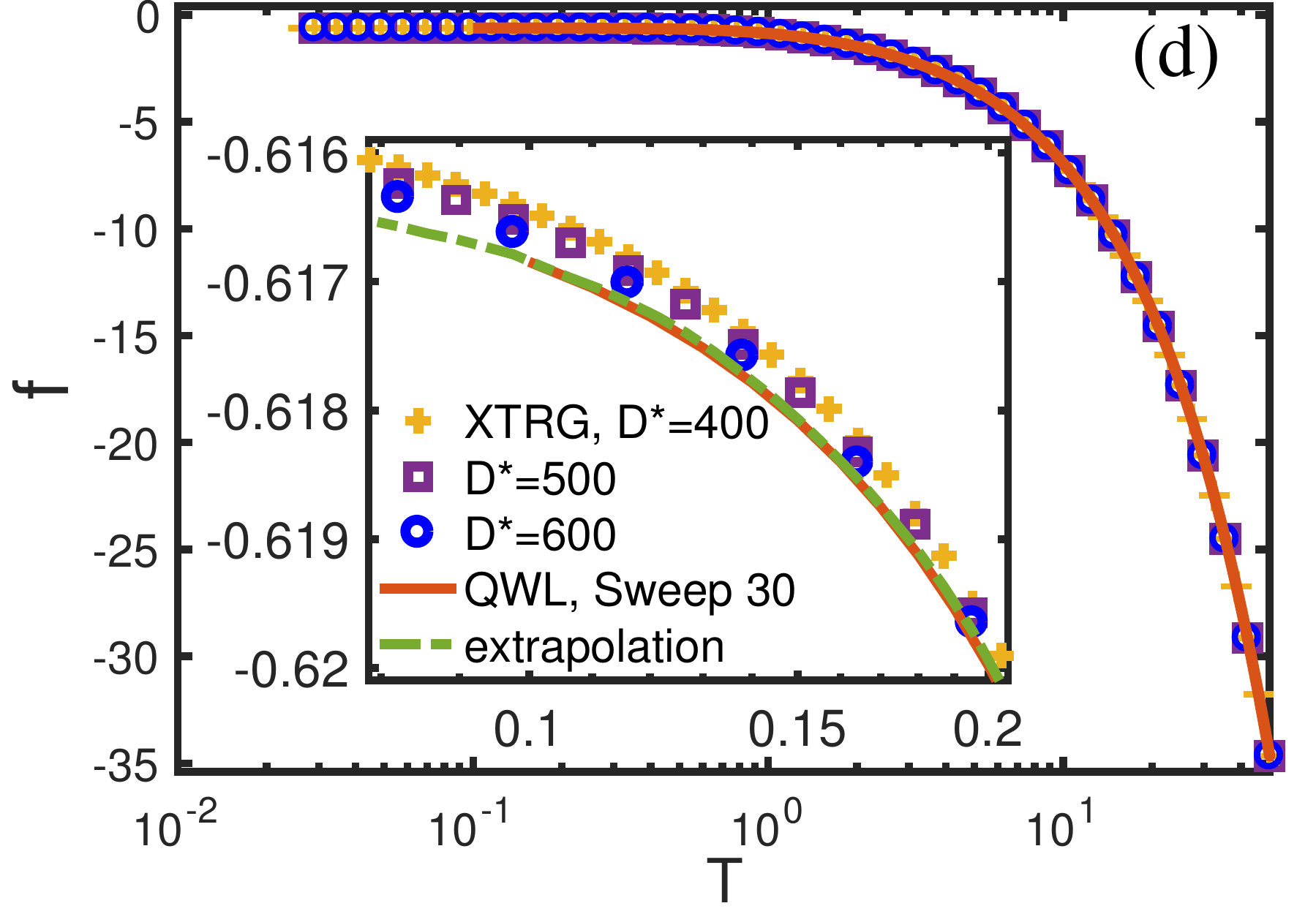}}
   \subfigure{
    \label{fig:subfig:b} 
    \includegraphics[width=0.305\linewidth]{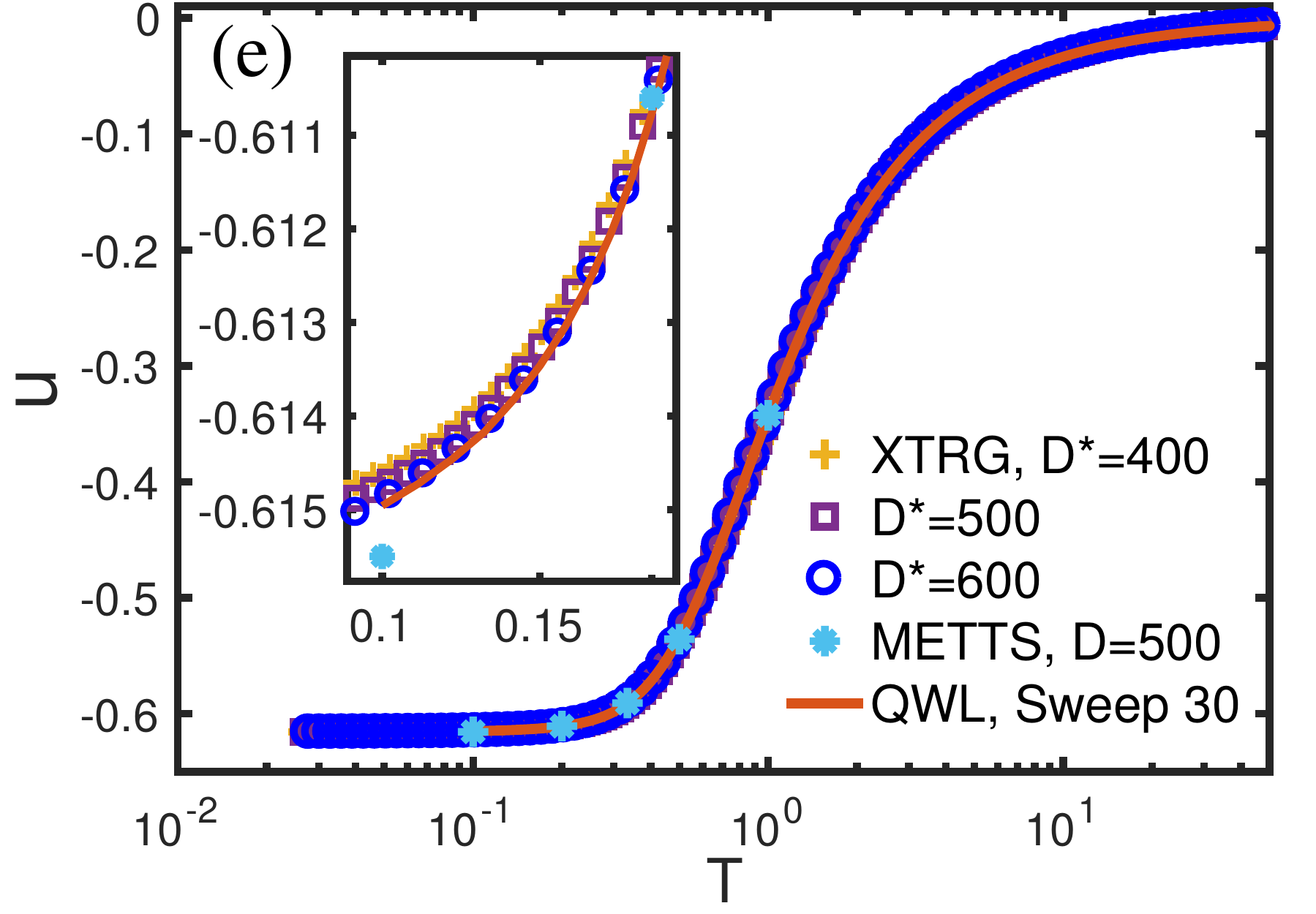}}
   \subfigure{
    \label{fig:subfig:b} 
    \includegraphics[width=0.305\linewidth]{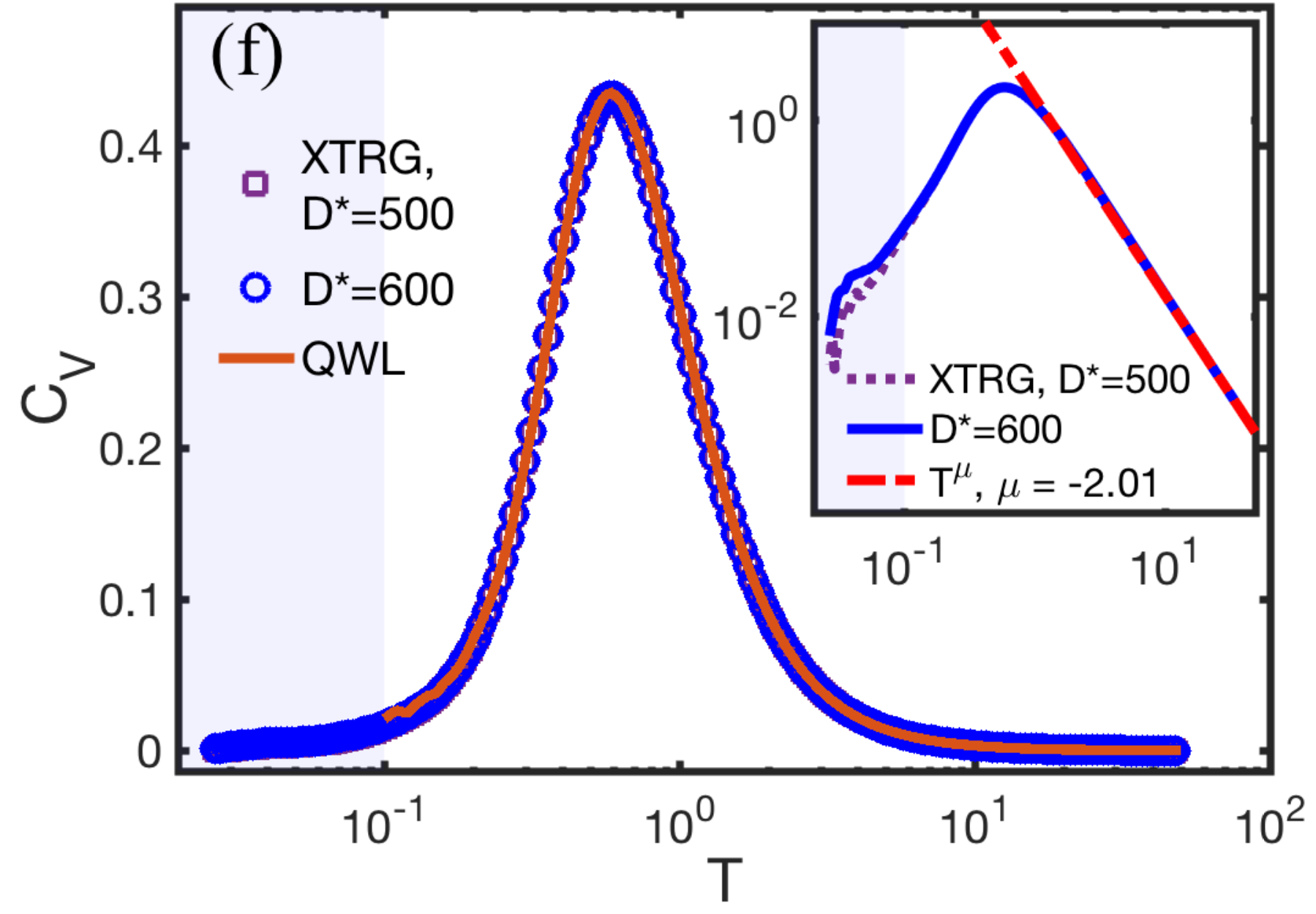}}
\caption{(Color online)
   Free energy $f$, internal energy $u$, and specific heat $c_V$ of
    {the isotropic} Heisenberg square lattice
   on an $L=4, W=4$ (upper panels a-c) and
   $L=16, W=5$ (lower panels d-f) square lattice.  The XTRG results
   are in very good agreement with those of ED for $4\times4$
   lattice and quantum Wang-Landau (QWL, expansion order 1200, sweep
   number 30, down to $T/J=0.1$) for the $16\times5$ lattice.  With
   $\Dstar$ the number of retained bond multiplets in the SU(2) MPO
   and $D$ the corresponding number of individual U(1) states, data
    {is} shown for $\Dstar$ ($D$) equal to $\Dstar=400$ (1462),
   $\Dstar=500$ (1832), and $\Dstar=600$ (2400), resulting in a
   maximal ratio $D/\Dstar\simeq 4$.  The maximal truncation errors,
   i.e., at largest $\beta$, are $\delta\rho \sim 1.4 \times
   10^{-5}$ for $4\times4$ lattice at $\Dstar=400$, and
   $\delta\rho\sim 1.9 \times 10^{-4}$ for $16\times5$ lattice at
   $\Dstar=600$. These truncation errors directly scale with the
   relative error in the partition function, and thus the free
   energy.  Extrapolating $1/\Dstar \to 0$ by a quadratic polynomial
   based on $\Dstar\sim300 - 600$ data results in perfect agreement
   (green dashed line) with QWL data in (d). The low temperature
   regions in (c,f) are shaded where we have limited accuracy of
   $c_V$.  Insets in (c,f) fit the high-$T$ specific heat with
   $T^{\mu}$, resulting in $\mu=-2.01$ in both cases (based on last
   eight points, i.e., $T>30$).
}
\label{Fig:FeSq}
\end{figure*}

 {Here we briefly summarize the thermodynamic quantities
that will be computed and analyzed in detail below.}
An equilibrium thermal state is described by the partition function
$\mathcal{Z}(\beta) \equiv \Tr{e^{-\beta H}} \equiv
\Tr{\rho(\beta)}$.  Typical interesting thermodynamic quantities,
which constitute important tasks for the TTN algorithms to compute,
including
\begin{subequations}
\label{Eq:ThermalQuan}
\begin{align}
& f \equiv{\tfrac{F}{N}} = -\tfrac{1}{N \beta} \ln \mathcal{Z}(\beta)
&\! &\text{free energy} \\
& u \equiv{\tfrac{E}{N}} 
  = \tfrac{\partial (\beta f)}{\partial \beta}
= \tfrac{1}{N} \tfrac{\Tr{H\cdot \rho(\beta)}}{\mathcal{Z}(\beta)}
&\!  &\text{(internal) energy}
\label{eq:energy:0}
 \\
& c_V =   
  \tfrac{\partial u}{\partial T}
 =  - \beta^2 \tfrac{\partial u}{\partial \beta}
&\! &\text{specific heat}
\label{eq:Cv:0}
\end{align}
\end{subequations}
etc. The computation of  {the} free energy $f$ and energy density $u$, are
straightforward, where $H$ and $\rho(\beta)$ are expressed as MPO,
and the calculations amount to efficient contractions of tensor
networks consisted of these MPOs.  The linear derivatives in $\beta$
for the specific heat as well as the energy density, however, are
not very natural for XTRG, which obtains the thermal data on a
uniform logarithmic $\beta$ grid.  Therefore it is more suitable to
use $\tfrac{\partial}{\partial\beta} =
\tfrac{\partial}{\beta\partial(\ln\beta)}$, i.e.,
\begin{subequations}
\label{Eq:ThermalQuan:X}
\begin{eqnarray}
u &=& \tfrac{1}{\beta} \tfrac{\partial (\beta f)}{\partial \ln{\beta}}
\label{eq:energy}
\\
c_V &=& -\beta\tfrac{\partial u}{\partial \ln\beta}
\text{ ,}\label{eq:Cv}
\end{eqnarray}
\end{subequations}
instead. This is also more stable numerically for small temperatures
since, the quotient of numerical differences is divided by $T$ for
the specific heat in \Eq{eq:Cv} and not by $T^2$ as in \Eq{eq:Cv:0},
and is multiplied by $T$ for the internal energy in \Eq{eq:energy}.
This formula is used to compute the specific heat in
Figs.~\ref{Fig:CvFit}, \ref{Fig:FeSq}.  In order to reduce numerical
differential errors, independent calculations with slightly
different initial $\tau$ values are run in parallel, e.g., using
$n_z=16$ in \Eq{eq:zshift:2} which produces interleaved data points
with $\delta z = 1/16$, i.e., $\delta \ln{\beta} =\delta z \ln 2
\simeq 0.0433$.

Finally, we note that the LTRG approach adopted in this work, e.g.,
for the data in Figs.~\ref{Fig:FeErr} and \ref{Fig:FeErrXY} below,
has been streamlined with the remainder of the TTN procedures used
in this work.  It differs from the original LTRG algorithm in Refs.
\cite{Li.w+:2011:LTRG, Dong.y+:2017:BiLTRG} in that it successively
projects the MPO for $\rho(\taui)$ to the density operator
$\rho(\beta)$ to increase $\beta$ linearly. Since the
Trotter-Suzuki decomposition is not involved in the procedure, it
is thus free of Trotter error.

\subsection{Heisenberg chain \label{sec:HChain}}

 \begin{figure}[tbp]
\includegraphics[angle=0,width=1\linewidth]{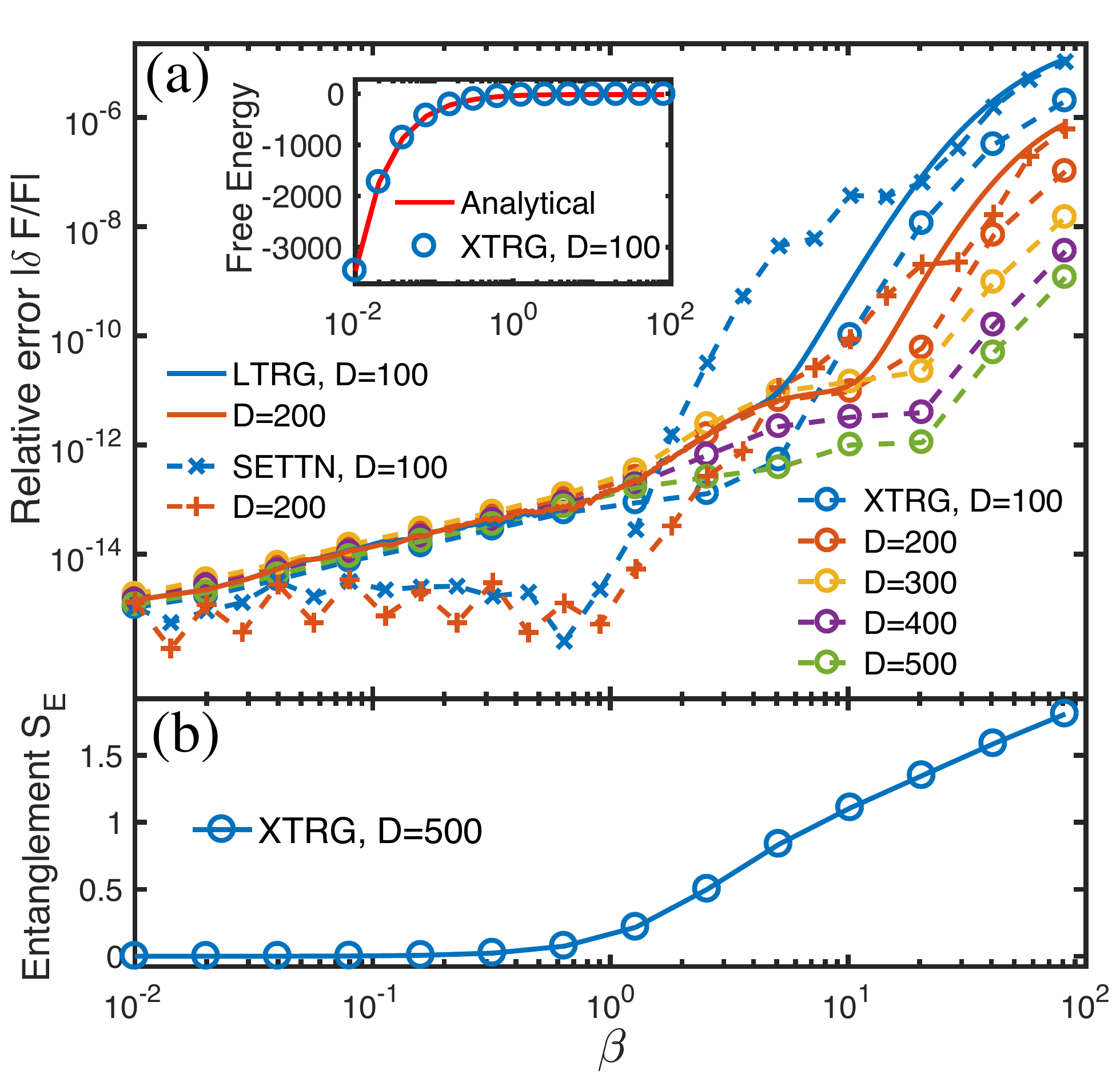}
\caption{(Color online)
   (a) Relative errors of free energy
   in an $L=50$ spin-1/2 XY chain with OBC,
   calculated by TTN algorithms including LTRG,  XTRG, 
   and SETTN, with U(1) symmetry encoded,
   relative to the analytical solution.
   Similar presentation as in \Fig{Fig:FeErr}, otherwise.
   Here $\rho(\taui=0.01)$ was again also initialized
   using SETTN.
   (b) Entanglement entropy in the middle of the system. 
   Overall, time costs of LTRG are 7 (7.4) times that
   of the SETTN run, and 363 (183) times that of XTRG, 
   for bond dimension $D=100$ $(200)$,
   respectively.
}
\label{Fig:FeErrXY}
\end{figure}

Firstly, we benchmark XTRG results with conventional linear
evolution in Fig.~\ref{Fig:FeErr}, where an $18$-site spin-1/2
Heisenberg (PBC) chain [cf.  \Eq{eq:HeisenbergH}] is calculated up
to $\beta \simeq 82$.  From Fig.~\ref{Fig:FeErr}(a) it is clear that
the accuracy in the low-$T$ regime gets continuously improved as
$\Dstar$ increases in XTRG. Starting from a fixed $\tau=0.01$, XTRG
reaches a precision as good as $10^{-8}$ also at the lowest
temperatures for $\Dstar=500$ as compared to exact diagonalization
(ED) data.  By keeping the same bond dimensions, LTRG and SETTN have
almost the same accuracy in all temperature regimes.  When compared
to XTRG, they are of similar accuracies only at high temperatures
($\beta \lesssim 1$), but are clearly less accurate in the low-$T$
region where truncation errors dominate. These remarkable results
suggest that as XTRG targets the low-$T$ properties much faster than
LTRG as well as SETTN, due to the (much) fewer truncation steps and
its algorithmically much simpler setup, it also gains better
results.

The MPO entanglement, as defined in \Sec{Sec:TTNE} and measured in
the center of the chain, is plotted in \Fig{Fig:FeErr}(b), which
offers a quantitative estimate of computational complexity.  The
entanglement data  {suggests} that truncation errors start to develop
for $\beta\gtrsim 1$ and the simulation errors stop increasing due
to the convergence of entanglement for $\beta \gtrsim 10$. This is
also clearly reflected in the overall error in physical quantities
such as the free energy in \Fig{Fig:FeErr}(a).

Besides the Heisenberg chain, we also benchmark XTRG, LTRG and SETTN
for an XY chain [cf. \Eq{eq:HeisenbergXY}] with size $L=50$, where
analytical solutions are available (\App{App:XY-Chain}).  As shown
in Fig. \ref{Fig:FeErrXY}, again XTRG gets better results than LTRG
and SETTN, and the accuracy in the low-temperature regime also
improves continuously as we increase bond dimensions $D$.  This
simulation on longer XY chain again confirms that increasing
$\beta$ exponentially fast not only improves the efficiency but
also gains in accuracy.

\begin{figure}[tbp]
\includegraphics[angle=0,width=1\linewidth]{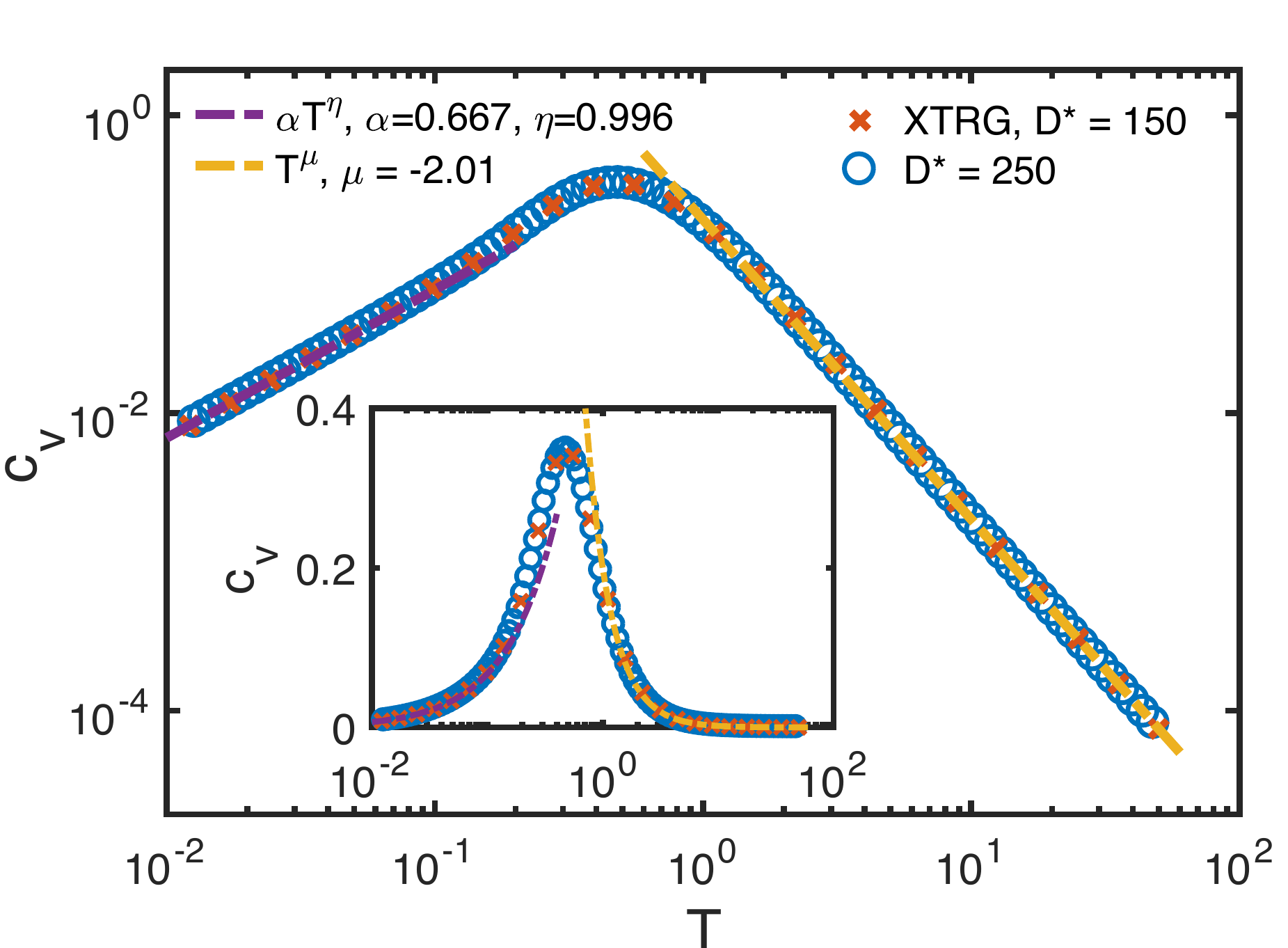}
\caption{(Color online)
   Specific heat of a Heisenberg chain of length $L=300$,
   with up to $\Dstar=250$ multiplets retained.  For low
   temperatures $T\ll 1,$ a universal linear behavior versus
   T is observed, i.e., $c_V = \alpha T^{\eta}$ with fitted
   exponent $\eta=0.996$ and slope $\alpha\simeq 2/3$, in the
   regime $T\leq0.025$. Exploiting the fact $\alpha =
   \frac{\pi c}{3 v}$ ($v=\pi/2$ for spin-1/2 Heisenberg
   chain), we extract the central charge $c\simeq1$. For
   large $T$, the specific heat shows a universal $1/T^2$
   temperature dependence (the fit shown was performed for
   $T>15$).  The inset shows the same data on a linear
   vertical scale.
}
\label{Fig:CvFit}
\end{figure}

Besides the free energy, we also calculate the specific heat of a
spin-1/2 Heisenberg chain of length $L=300$, utilizing the XTRG
algorithm with bond dimension up to $\Dstar=250$. As shown in
\Fig{Fig:CvFit}, in the low-$T$ region, the specific heat of the
system shows a universal linear relation versus temperature, as
indicated by the polynomial fitting (purple dashed line) with $\eta
\simeq 1$.  In addition, the fitted slope is also in perfect
agreement with the well-known value $2/3$ from CFT prediction
\cite{Affleck.I:1986:UniversalTerm}, from which we extract the
central charge $c\simeq1$.  On the other hand, in the high-$T$
regime, the specific heat is also universal, and decays as $1/T^2$
(a polynomial fit in the log-log scale, depicted by the yellow
dashed line yields an exponent $\mu = -2.01 $).  This exponent can
be confirmed by a high-temperature expansion up to the second order,
which approximates the energy as
\begin{eqnarray*} E{=}\tfrac{\mathrm{Tr} (e^{-\tau H}
H)}{\mathcal{Z}(\tau)} {\simeq} \tfrac{1}{\mathcal{Z}^0} \Bigl[
\mathrm{Tr} H - \mathrm{Tr} (H^2) \tau + \tfrac{(\mathrm{Tr}
H)^2}{\mathcal{Z}^0} \tau + \order{\tau^2} \Bigr] \end{eqnarray*}
where $\mathcal{Z}^0=\rm{Tr} (\mathbb{I})$, with the high-$T$ limit
$c_V \sim 1/T^2$.

\subsection{Square lattice Heisenberg model}
\label{Sec:2DHeisenberg}

Symmetric TTN methods, including the XTRG and the SETTN, can be
conveniently employed to calculate 2D systems, with minor
adaptations. We map the 2D clusters into a 1D snake shape, and
prepare the MPO representation of this Hamiltonian (with
``long-range" interactions) as elaborated in \App{App:SYMPO}.  Other
than that, one follows exactly the same line as in 1D simulations
and can represent the density matrix of the 2D systems accurately in
terms of MPO.  

Here we perform calculations on 2D clusters and benchmark the
calculations with ED for small (OBC) systems ($4\times4$) in
Figs.~\ref{Fig:FeSq}(a-c) and quantum Monte Carlo (QMC) for larger
systems ($16\times5$) in Figs.~\ref{Fig:FeSq}(d-f).  With
non-Abelian symmetries implemented in the highly efficient XTRG
algorithms, we obtain high quality data till quite low temperatures,
which was not accessible before by other thermal RG algorithms.

In Figs.~\ref{Fig:FeSq}(a-c), we show the free energy, energy and
specific heat results of a 4$\times$4 square lattice Heisenberg (SLH) model.
Very nice agreement between XTRG and ED data is observed in all
three plots.  As seen in the inset of Fig.~\ref{Fig:FeSq}(a), the
relative accuracy is quite high, i.e., $10^{-4}$ ($10^{-5}$) for
$\Dstar=200$ $(400)$ at low temperatures ($T\leq 0.05$).  The energy
density shown in \Fig{Fig:FeSq}(b)  {is} obtained by taking
derivatives of interleaved XTRG free energy data [cf.
\Eq{eq:energy}].  The error in the energy density is small even down
to $T\leq0.05$, as seen in the inset of \Fig{Fig:FeSq}(b) by
zooming in the low-$T$ region. XTRG data  {differs} from the ED
results in the fourth digit for $\Dstar=200$, and is bounded by
numerical differentiation error for $\Dstar=400$.  In
\Fig{Fig:FeSq}(c), we show our results of the specific heat, which
was calculated by taking derivatives of energy data as in
\Eq{eq:Cv}.  Inset plots $c_V$ on a log-log scale, from which we
again observe an algebraic behavior ($1/T^2$) at high temperatures.
In the low-$T$ region, it shows a very rapid (exponential) decay
versus the temperature. This can even be observed in the dashed
region of \Fig{Fig:FeSq}(c), although there the XTRG data  {departs}
from ED results due to lack of accuracy.

For a 16$\times$5 SLH which is far beyond the
scope of ED calculations, we compare our XTRG results to those of
quantum Wang-Landau (QWL) simulations \cite{Bauer.b+:2011:ALPS} in
Figs. \ref{Fig:FeSq}(d,e,f).  We run the calculation down to
$T=0.025$.  For the smallest temperature $T=0.1$ for which with have
well-converged QWL reference data at comparable numerical cost, the
error in the $\Dstar=600$ data for the free energy is $\sim 2
\times10^{-4}$.  Since the truncation errors is generally larger on
$16\times5$ lattice, we extrapolate the free energy in
\Fig{Fig:FeSq}(d) to $1/\Dstar\to 0$ and observe a perfect
agreementwith QWL data, with the error further reduced by about an
order of magnitude.
Besides the free energy, in Figs.~\ref{Fig:FeSq}(e-f) we also show
that the energy density $u$ as well as specific heat $c_V$ all have
very good accuracy.  In the high-$T$ region, $c_V$ again shows a
$1/T^2$ relation, as shown in the inset of Fig.~\ref{Fig:FeSq}(f)
and as already discussed with \Fig{Fig:CvFit}. In the very low
temperatures [dashed region of \Fig{Fig:FeSq}(f)], though, our
accuracy becomes somewhat limited as the derivative to obtain $c_V$
develops minor wiggles at the lowest temperatures, seen on the
log-scale in the inset. However, the upturn does not appear to be
due to numerical inaccuracies, but appears to be physical, in the
sense that, similar to \Fig{Fig:FeSq}(c), it is a precursor before
the finite-size spectral gap sets in.

In Figs.~\ref{Fig:FeSq}(b, e), for comparison we also included METTS
data exploiting $U(1)$ symmetry only \cite{Bruognolo.b+:2017:MPS}.
The METTS results also show good agreement with our other methods in
both cases, apart from the fact that the METTS energy data is not
strictly variational, i.e., could be even lower than the (quasi)
exact value.  As for the $16\times5$ plot in Fig.~\ref{Fig:FeSq}(e),
note that $T=0.1$ is currently the typical lowest temperatures that
2D METTS simulations can reach \cite{Bruognolo.b+:2017:MPS} 
 {at computational resources that are comparable to XTRG}. 
However, the current 2D
METTS involves many swap gates and needs at least a few hundreds of
samples.  In contrast, our XTRG method, with SU(2) symmetry
implemented, is much more efficient and can reach much lower
temperatures with great accuracy in 2D.

\begin{figure*}[tbp]
\includegraphics[angle=0,width=1\linewidth]{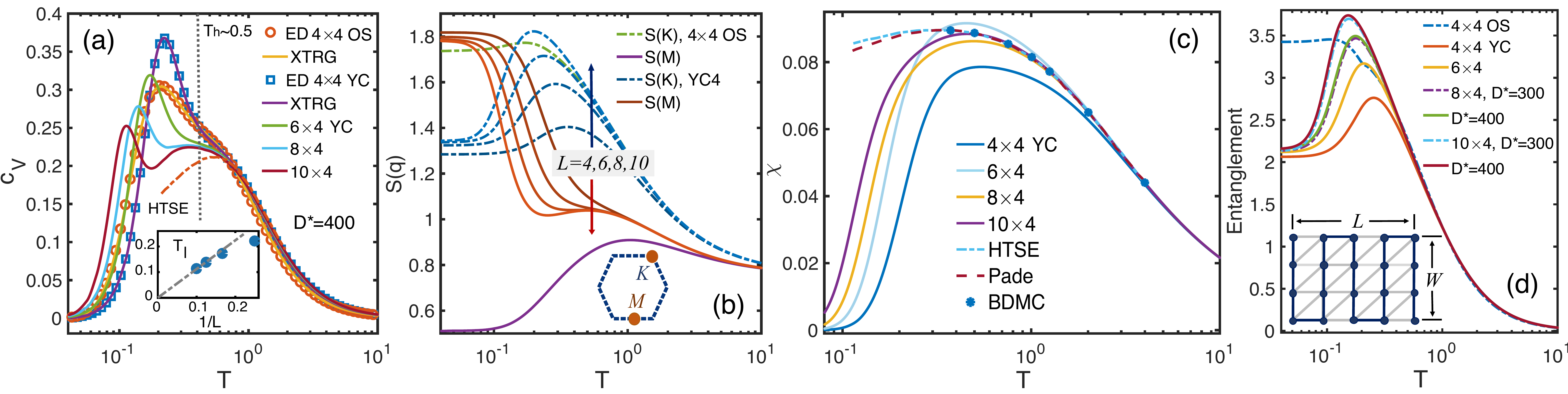}
\caption{(Color online)  {Thermodynamics of triangular lattice Heisenberg model defined on two $W=4$ geometries, OS with OBC on both directions and Y cylinder (YC) with various length $L$, i.e., PBC along the vertical direction [see inset in (d)
for our specific 
choice of the YC geometry, as well as the TLH lattice layout in \Fig{Fig:L8W4}(f) 
below].
The presented quantities include (a) specific heat $c_V$,
(b) static structure factor 
$S(q)$ at $q=K$ and $q=M$ 
at the boundary of first Brillouin zone (see inset),
(c) uniform magnetic susceptibility $\chi$, and (d) thermal entanglement vs. temperatures.
Inset in (a) depicts the temperature $T_l$ of the peak in $c_V$
at lower temperature vs. inverse length $1/L$, with the dashed line a guide for eyes.
Note that the high temperature scale in $c_V$ stays at $T_h \sim 0.5$.
The HTSE and Pade data  {is} taken from Refs.~\cite{Elstner-1993,Zheng-2005}, and BDMC from Ref.~\cite{Kulagin2013}.
}}
\label{Fig:TLH}
\end{figure*}

\subsection{ {Triangular lattice Heisenberg model}}
\label{sec:frustrated}

 {Since Anderson's famous conjecture on resonating valence bond (RVB) state
in the frustrated triangular lattice Heisenberg (TLH) model \cite{Anderson1973,ANDERSON1987}, TLH continues 
to intrigue people until today \cite{Elstner-1993,Bernu1994,Bernu01, PhysRevLett.82.3899,White07, Zheng-2005,Zheng-2006,Zheng2006PRL, Kulagin2013,PhysRevLett.120.207203,Alicea06}. 
Although the ground state turns out to be a 120$^{\circ}$ ordered magnetic state \cite{Bernu1994,PhysRevLett.82.3899,White07}, finite-temperature properties of TLH appear to be anomalous, in that they cannot be described by renormalized classical behavior, as one would expect for systems with a magnetically ordered ground state,
and that thermal data  {``extrapolates"} to a disordered state \cite{Elstner-1993,Zheng2006PRL,Zheng-2006,Kulagin2013,Alicea06}.
TLH materials have been realized experimentally \cite{Shirata2012,Zhou2012,Susuki2013,Cui2018}. Therefore the theoretical understanding of its thermodynamic properties becomes more pressing. So far, however, low-temperature simulations of TLH were hindered by lack
of sufficiently powerful numerical approaches.} 

\begin{figure*}[tbp]
\includegraphics[angle=0,width=0.95\linewidth]{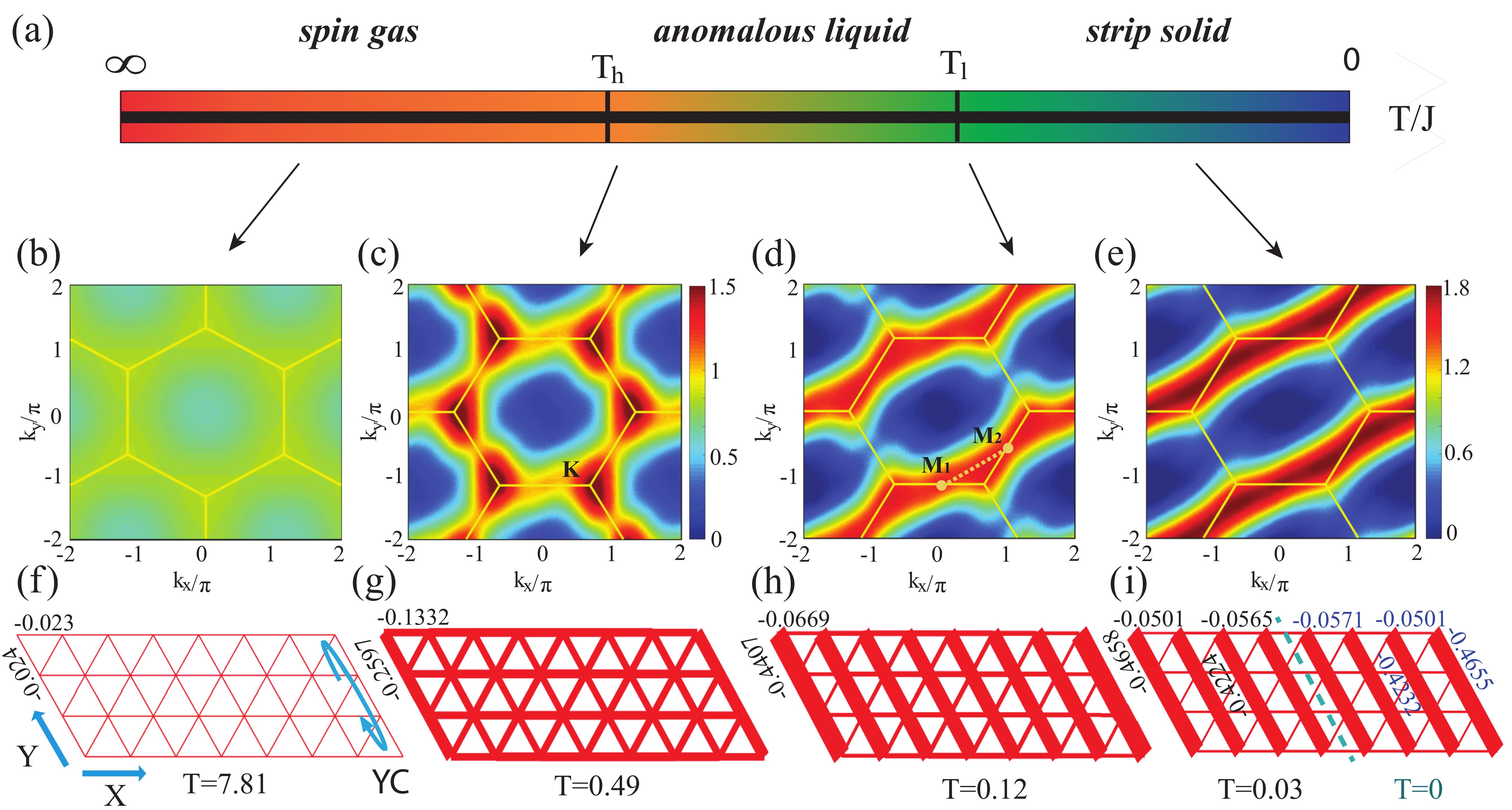}
\caption{(Color online)
 {(a) Finite-temperature phase diagram of TLH on a YC4 lattice, which consists of high-$T$ ``gas", intermediate-$T$ liquid, and low-$T$ solid states. (b-e) show the static structure factor at four representative temperature points (left to right, from high temperature to low), and (f-i) exhibit corresponding bond textures at finite-$T$, and the values are the bond energies to whose absolute value the thickness of nearest neighbor bonds are proportional. In (i) we also show the bond energy distribution at $T=0$ calculated by DMRG, where the agreement between XTRG results (left half) at low-$T$ and DMRG data (right half) is apparent, and one can recognize the strip solid structure quite distinctly. 
Dashed lines are guide for eyes, representing in (e) the line connecting two inequivalent $M$ points (labelled as $M_1$ and $M_2$) with enhanced intensity. In (f) we label the X and Y directions, as well as the way we wrap the lattice into a Y cylinder.}
}
\label{Fig:L8W4}
\end{figure*}

 {Besides non-frustated systems such as the SLH, XTRG can also be applied to frustrated magnets like TLH. In Fig.~\ref{Fig:TLH}, we present exemplary XTRG results of TLH on width $W=4$ systems [see inset of Fig.~\ref{Fig:TLH}(d)], including open strips (OS) and Y cylinders (YC). 
To be specific, we consider YC4 $\equiv$ YC(W=4) geometry with various aspect ratio $L/W=1 \sim 2.5$, as well as a $4\times4$ OS for comparison.The YC geometry we adopt in practice is shown in the inset of Fig.~\ref{Fig:TLH}(d), and also in Figs.~\ref{Fig:L8W4}(f-i) after a proper transformation to restore the corresponding triangular lattice geometry. Our realization is essentially equivalent to conventional YC in previous DMRG studies \cite{Hu2015} at $T=0$.
We cool down the system from high temperatures to as low as $T/J=0.03$, and $D^*=400$ multiplets are kept in the following calculations, to insure convergence vs. bond dimensions.}

 {Although YC4 shows clear finite-size effects in the ground state at $T=0$, 
nevertheless, we can observe how correlations gradually develop as we lower temperatures,
with relevant finite-$T$ physics in the thermodynamic limit at larger $T$
down to temperatures where finite-size effects set in.
In Fig.~\ref{Fig:TLH}(a) we plot the specific heat of a $4\times4$ OS and various YC4 lattices. By comparing Fig.~\ref{Fig:TLH}(a) to corresponding $c_V$  in Fig.~\ref{Fig:FeSq}(c) of the SLH, we observe quite distinct features in the TLH case.
Already on the $4\times4$ lattice, either OS or YC,
the specific heat $c_V$ 
exhibits a peak, whose location will be denoted by $T_l$, 
and a shoulder structure at a higher temperature $T_h \sim 0.5 \sim J$.
For YC4 lattices with increasing length $L=6,8,10$,
these features in $c_V$ develop into
a pronounced two-peak structure where the
plateau-like feature for $L=4$ develops into a broad peak
located around the same stable value of $T_h$.
Note that $T_h$ is also about the same temperature scale
where the data from high-temperature expansion (HTSE)
shows a round peak \cite{Elstner-1993}.
In stark contrast, the peak at $T_l$ moves towards lower temperatures.
Its value 
scales like $T_l  \sim 1/L$  as seen in inset of Fig.~\ref{Fig:TLH}(a) and therefore is clearly linked to the finite system size.
Overall, the stark qualitative difference
between the data for TLH in Fig.~\ref{Fig:TLH}(a)
as compared to SLH in Fig.~\ref{Fig:FeSq}(c) may
be ascribed to finite-temperature effects of magnetic frustration.
}

 {
To gain a better understanding, we analyze the static structure factor
\begin{equation}
S(q)=\sum_{j}  e^{-i q \cdot r_{0j}} 
\, \langle {\mathbf{S}}_{0}  \cdot \bold{S}_j \rangle 
\end{equation}
where $r_{0j} \equiv r_j - r_0$ with $r_j$ the lattice location of site $j$.
In practical calculations, site $0$ 
is fixed in the system center, while $j$ runs over the whole lattice.
Therefore, by inversion symmetry of the TLH, $S(q)$ is a real number.
Note that also the structure factor above can be conveniently and efficiently
obtained via the expectation value of single $q$-dependent MPO
of bond dimension $D_S^*=2$.
}

 {The static structure factor for the THL is analyzed at specific values
of $q$ in \Fig{Fig:TLH}(b). However, before discussing these in detail,
let us look at the static structure factor over the entire Brillouin zone,
as presented in \Fig{Fig:L8W4} for a YC4 system at $L=8$
along with the bond energy texture, at various temperatures (including $T=0$, obtained by DMRG). 
In agreement with the previous discussion,
Figure~\ref{Fig:L8W4}(a) indicates the existence of three regions, separated
by the two temperature scales $T_l$ and $T_h$ 
as introduced with \Fig{Fig:TLH}(a). For $T\gg T_h$, the system is in a ``spin gas" paramagnetic phase, with featureless structure factor and bond texture in Figs.~\ref{Fig:L8W4}(b) and (f), respectively. In Fig.~\ref{Fig:L8W4}(c) we can observe that the intensities of $S(q)$ at $q=K \equiv \pm \frac{2\pi}{3}(1, \sqrt{3})$, $\pm \frac{2\pi}{3}(-1, \sqrt{3})$, and $\pm \frac{2\pi}{3}(2, 0)$ become prominent in the intermediate region $T_l< T<T_h$, representing the development of strong 120$^{\circ}$-ordering correlations.}

 {In addition to that, there also emerges an extended region with considerable intensity around $q=M \equiv \pm 2\pi (0, \frac{1}{\sqrt{3}})$, $\pm \pi (1, -\frac{1}{\sqrt{3}})$, and $\pm \pi (1, \frac{1}{\sqrt{3}})$, near which roton-like excitations were reported both theorectically \cite{Zheng-2005,Zheng-2006,Zheng2006PRL,Alicea06} and experimentally \cite{Ito2017}. The (anomalous) 
enhancement of $S(q)$ around these $M$ points
for temperatures $T\sim T_h$, which can be more clearly 
observed still in Fig.~\ref{Fig:TLH}(b), may thus 
be related with roton excitations. By association with $T_h$, these
are also linked to 
the round peak/shoulder in the specific heat  $c_V$. The corresponding bond texture in Fig.~\ref{Fig:L8W4}(g) reveals much stronger spin-spin correlation than those in paramagnetic phase, while it still maintains a uniform pattern that respects the TLH lattice symmetry, like a liquid state. Thus we dub the finite $T$ region between $T_l$ and $T_h$ anomalous quantum liquid phase.}

 {As $T$ is lowered further down to less than $T_l\sim0.14$,
the system in \Fig{Fig:L8W4} undergoes 
a rapid crossover.
Due to the finite YC circumference, tightly bound RVB rings of length $W$ form
as can be clearly observed 
in the bond textures in \Fig{Fig:L8W4}(h,i).
There emerges a vertical strip pattern around the cylindrical circumference at $T\lesssim T_l$ [Fig.~\ref{Fig:L8W4}(h)], which becomes strongly dominating for lower temperatures [Fig.~\ref{Fig:L8W4}(i)]. Note that the low-temperature ($T/J=0.03$) bond texture [left half of Fig.~\ref{Fig:L8W4}(h)]
is in perfect agreement with $T=0$ DMRG results [right half in Fig.~\ref{Fig:L8W4}(h), glued together at dashed center line]. The agreement between bond energies is better than 0.1\% for most bonds, indicating the XTRG already effectively reached the $T=0$ regime.
This is further reflected in the average energy per site $e_0=-0.53027(1)$, for given
YC4 system at $L=8$ and $T/J=0.03$, which is in perfect agreement with, \aw{i.e. just above} the DMRG ground state energy $e_0=-0.53034$ [with $D=4000$ U(1) states retained].
Similarly, also the structure factor is in very good agreement, e.g., with $S(M)= 1.82$ at $T=0$ DMRG to be compared to $S(M)=1.79$ at $T/J=0.03$ XTRG.}

 {As a consequence of the tightly bound rings around the
circumference of the YC4 cylinders, the correlations
become significantly enhanced along a line connecting 
two of the three initially equivalent $M$ points in the  {Brillouin zone}
[dashed line in \Fig{Fig:L8W4}(d)].
This strongly competes the triangular magnetization associated with the $K$ points
[Fig.~\ref{Fig:L8W4}(d)], such that the structure factor $S(M)$ at these points 
eventually dominates over $S(K)$ [\Fig{Fig:L8W4}(e), and also \Fig{Fig:TLH}(b)]. 
This gives rise to the low-$T$ peak of the specific heat curve in Fig.~\ref{Fig:TLH}(a). 
From Fig.~\ref{Fig:TLH}(b), it can be observed that almost exactly at $T_l$, 
the two correlations cross and $S(M)$ becomes larger than $S(K)$ for $T<T_l$.
Here the symmetry breaking across the initially equivalent $M$ points
is due to the finite-size cylindrical structure.
}

 {The results of magnetic susceptibility $\chi$ of various YC4 lattices are shown in Fig.~\ref{Fig:TLH}(c). It can be observed that $W=4$ data can produce results in agreement with the HTSE \cite{Elstner-1993}, bold diagrammatic Monte Carlo (BDMC) \cite{Kulagin2013}, as well as Pade approximation results \cite{Zheng-2005}, down to $T\sim 0.4$ (for YC $10\times4$), despite a very limited circumference. This observation is quite surprising, which suggests a very small finite-size effect and correlation length $\xi$ $(\lesssim 1)$ for $T/J \geq 0.4$, in accordance to previous studies \cite{Elstner-1993}. }

 {In Fig.~\ref{Fig:TLH}(d), we plot the bipartite entanglement entropy $S_E$ in the purified state. Since $S_E$ differs from bond to bond, we show there the maximal value amongst all bonds in the effective
1D chain structure adopted in the calculation [inset in Fig.~\ref{Fig:TLH}(d)]. It can be clearly seen in Fig.~\ref{Fig:TLH}(d) that this low-temperature entanglement reaches a relatively small value around $S_E=2.16$. DMRG simulation shows that the corresponding entanglement $S_E \simeq 1.1$ which, as expected, is roughly one-half of the entanglement in low-$T$ mixed state.
}

 {The tightly bound stripes in the low-$T$ regime in given
YC4 system lead to a strongly reduced entanglement entropy
when cutting the system exactly in between the stripes,
i.e. vertically the inset in \Fig{Fig:TLH}(d), where
the entanglement is reduced e.g. down to $S_E\sim 0.9$ at $T=0.03$
}.
 {Therefore at low temperatures $T<T_l$ 
the system is close to a direct product of tightly bound uniform 1D rings
around the circumference of the YC4 cylinder.
These four-site rings favor a well-known RVB ground state $| \rm{RVB} \rangle = \frac{1}{2\sqrt{3}}(\phi_{1 2} \phi_{3 4}  - \phi_{1 4} \phi_{2 3})$,
where $\phi_{i j} \equiv \vert \uparrow_i \downarrow_j - \downarrow_i \uparrow_j \rangle$ constitutes a valence bond, and the lowest excited state is separated by a significant energy gap $\Delta \sim J$.
}
 {These tightly bound rings are disrupted when opening
the boundary. Therefore in the OS geometry,
e.g., in the $4\times4$ data in \Fig{Fig:TLH}(b),
$S(K)$ remains strong, and never crosses with $S(M)$
at  $M=\pm 2\pi (0, \frac{1}{\sqrt{3}})$ as $T$ is lowered.
However, for the $4\times4$ OS, the dominant weight of the structure
factor turns out to be located at one out of three types of 
N\'eel antiferromagnetic order,
i.e., $S(M)$ with $M= \pm(\pi, -\pi/\sqrt{3})$ [data not shown in \Fig{Fig:TLH}]. The entanglement also increases  {monotonically} on the OS lattice, until it saturates due to finite system size [Fig.~\ref{Fig:TLH}(d)]}.

 {Recently, BDMC has been employed to explore thermal properties of TLH, and it was found that the ``extrapolated" ground state is disordered, via a particular quantum-to-classical mapping based on their best thermal data (with temperatures down to $T/J\simeq0.375$) \cite{Kulagin2013}. The anomalous thermodynamic behavior is in contradiction to the ground-state $120^{\circ}$ ordering, which reveals that the true low temperature regime has not been reached. 
To fully understand the finite-temperature anomaly and to resolve the above apparent contradiction, 
a more extensive XTRG survey of the TLH 
e.g. on wider cylinders down to low temperatures is required.
This is beyond the scope of this paper, and thus 
will be reported elsewhere \cite{TrianLatt2018}.
For the purpose of this paper, nevertheless, we have
demonstrated that XTRG provides a highly competitive approach
that allows one to tackle complex and challenging problems.
}

\section{Entanglement in Thermal Tensor Networks}
\label{Sec:TTNE}

\subsection{Thermal Entanglement Renormalization Group Flow}

The entanglement measure in a thermal state is more complicated as
compared to a ground state due to the interplay of classical
correlation and quantum entanglement. Among various definitions, we
take a very natural and most relevant measure of the entanglement in
practise, i.e., entanglement in the  normalized ``superstate"
\cite{Zwolak.m+:2004:Superoperator}
\begin{eqnarray}
   |\Psi(\beta) \rangle \equiv
   \tfrac{1}{\sqrt{\mathcal{Z}(\beta)}} \,
   \underset{\equiv |\tilde{\Psi}(\beta) \rangle}
   {\underbrace{|e^{-\frac{\beta}{2} H}\rangle}}
\label{eq:purif}
\end{eqnarray}
which vectorizes the MPO for $e^{-\beta H/2}$.  In other
words, the MPO is simply transformed into a matrix product state
(MPS) with doubled local state spaces.  Then the partition function
is equivalent to the overlap of the unnormalized superstate
$\mathcal{Z} (\beta) = \langle \tilde{\Psi}(\beta) |
\tilde{\Psi}(\beta) \rangle$, whereas $\langle \Psi(\beta) |
\Psi(\beta) \rangle=1$.
Note that this definition, is a specific (and most natural) choice
of purification \cite{Verstraete.f+:2004:2DRenormalization,
Feiguin.a.e+:2005:ftDMRG}, which in some other context is also
called the thermofield double (TFD) state $|\Psi(\beta) \rangle =
\frac{1}{\sqrt{\mathcal{Z(\beta)}}} e^{-\frac{\beta}{2} E_n}
|n,\bar{n}\rangle$, where $E_n$ is the eigen-energy of eigenstate
$|n \rangle$ and its duplicate $|\bar{n} \rangle$ in the auxiliary
state space \cite{Takahashi.y+:1996:TFD, Maldacena.j:2003:ads,
Kallin.a.b+:2011:Entanglement, Schwarz17}.  Here, for simplicity of
notation, by the entanglement entropy or the entanglement spectrum
(ES) of the MPO or the thermal state, we refer to precisely these
quantities obtained from the underlying TFD, or equivalently, the
purified and normalized state in \Eq{eq:purif}.  Specifically, the
entanglement spectrum is given by the eigenspectrum of
$\mathcal{H}_\mathrm{ES} \equiv -\ln \mathcal{R}$ where
$\mathcal{R}$ is the `super'-density-matrix of the purified thermal
state $|\Psi(\beta) \rangle$ as in \Eq{eq:purif}.
 
Note that the MPO entanglement analyzed here is not directly related
to the entanglement of purification, which is defined as the minimal
value amongst various purification schemes
\cite{Terhal.b.m+:2002:Entanglement, Hauschild17}.  Nevertheless,
through the optimal truncation via orthogonal state spaces in the
XTRG (and also LTRG), one is simultaneously optimizing the
super-state overlap (i.e., partition function), as well as this MPO
(TFD) entanglement.  Therefore, this MPO entanglement, as well as
some other measures, like the mutual information, quantifies the
resources required to perform efficient thermal simulations  {and, thus,} have
attracted recent interest \cite{Nguyen.p+:2017:Entanglement,
Takayanagi.t+:2017:Holographic, Barthel.t:2017:FiniteT}. 

\begin{figure}[tbp]
\includegraphics[angle=0,width=1\linewidth]{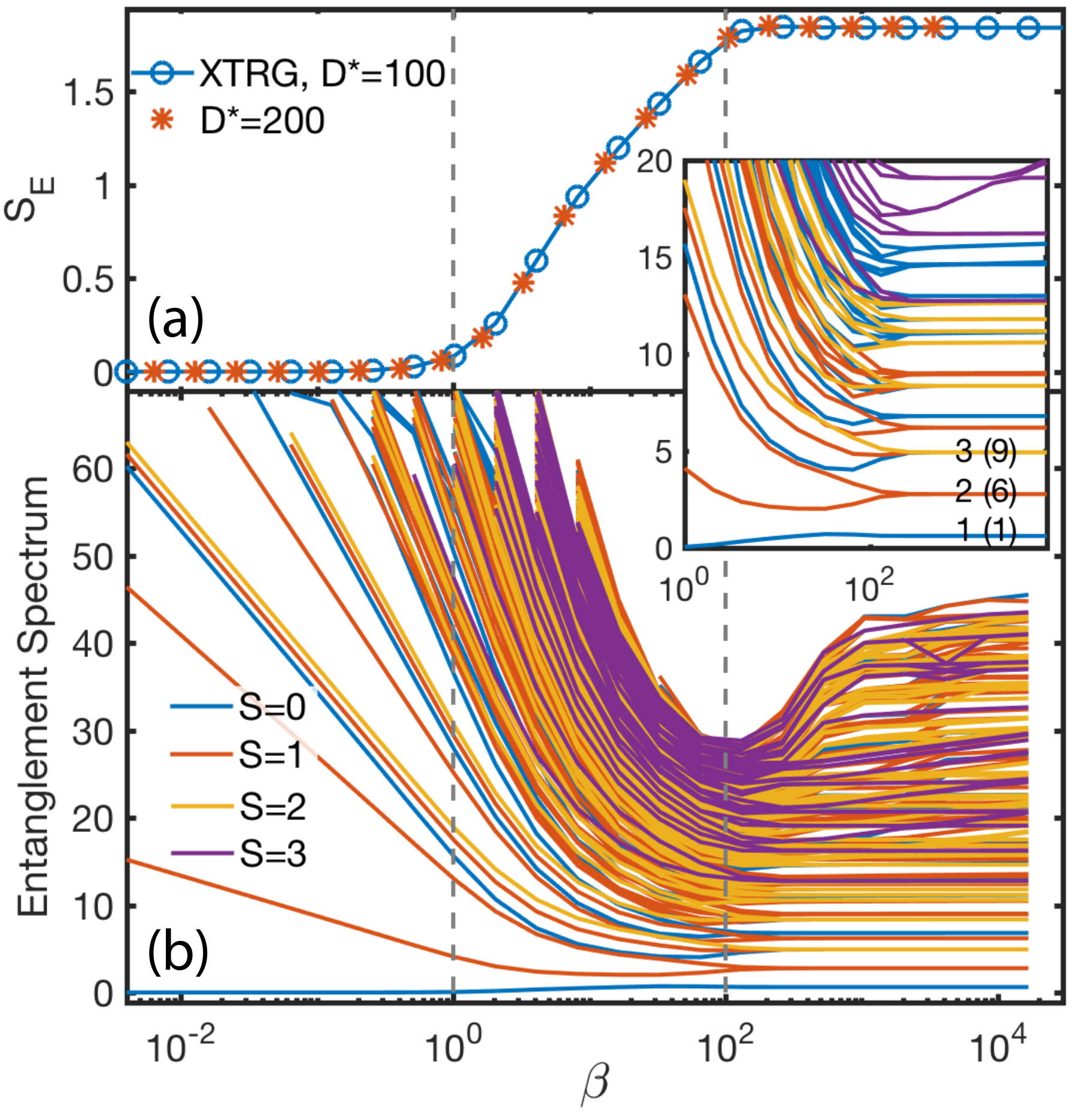}
\caption{(Color online)
   (a) Bipartite MPO entanglement entropy $S_E$ across the
   center of the system of a spin-1/2 Heisenberg chain
   (length $L=100$, OBC).  (b) Entanglement spectra obtained
   in the center of the system versus a wide range of
   temperatures presented as an RG flow in energy scales,
   where lines from the same symmetry sectors, i.e., $S=0 ,1,
   2, \ldots$ are plotted in the same color.  Vertical
   markers depict different temperature regimes (see text).
   Inset zooms in the region around $\beta=100$, and the
   labels $d^\ast$ ($d$) indicate the degeneracy in the RG
   fixed point ES in terms of individual multiplets (or
   states), respectively.
}
\label{Fig:EntRGFlow}
\end{figure}

We start by analyzing the entanglement spectra of the thermal state
for a spin-1/2 Heisenberg chain, from which we can also compute its
 entanglement block entropy $S_E$.  By lowering the temperatures,
 one generates a RG flow that directly reflects different physical
 regimes of the system at various temperatures, i.e., energy scales.
In Fig.~\ref{Fig:EntRGFlow}, we show the entanglement RG flow over a
very wide range of temperatures.  We can vary $\beta$ over 7 orders
of magnitude, which thus reaches far beyond Trotter-Suzuki type
calculations.
The RG flow reveals three distinct regimes, demarcated by vertical
dashed lines in \Fig{Fig:EntRGFlow}: (i) a low entanglement region
$\beta \lesssim  1$, (ii) an intermediate region $1 \lesssim \beta
\lesssim 100$ where entanglement rises quickly, and (iii) the
saturation region for $\beta \gtrsim100$ where the ES flows to a
fixed point, either converging to the ground state of a physically
gapped state in the thermodynamic limit, or resolving the gap of
finite size level spacing.

When approaching the low-energy fixed point ES, lines systematically
merge into groups with larger degeneracy as seen in the inset of
\Fig{Fig:EntRGFlow}. Given that the entropy already clearly
converges to a finite value at the lowest temperatures, this
suggests that the low-energy fixed-point spectrum must be related to
the tensor product space of two copies of the ground state (bra and
ket) which naturally results in systematically enlarged
degeneracies.

For the remainder of this section, we focus on the entanglement
entropy both in 1D chain and 2D lattice models over a wide range of
temperature scales.  Interesting logarithmic behaviors are observed,
which intimately relate to (gapless) low-energy excitations, yet
also suggest efficient computational complexity of thermal
simulations for the specific model systems considered.

\begin{figure}[tbp]
\includegraphics[angle=0,width=1\linewidth]{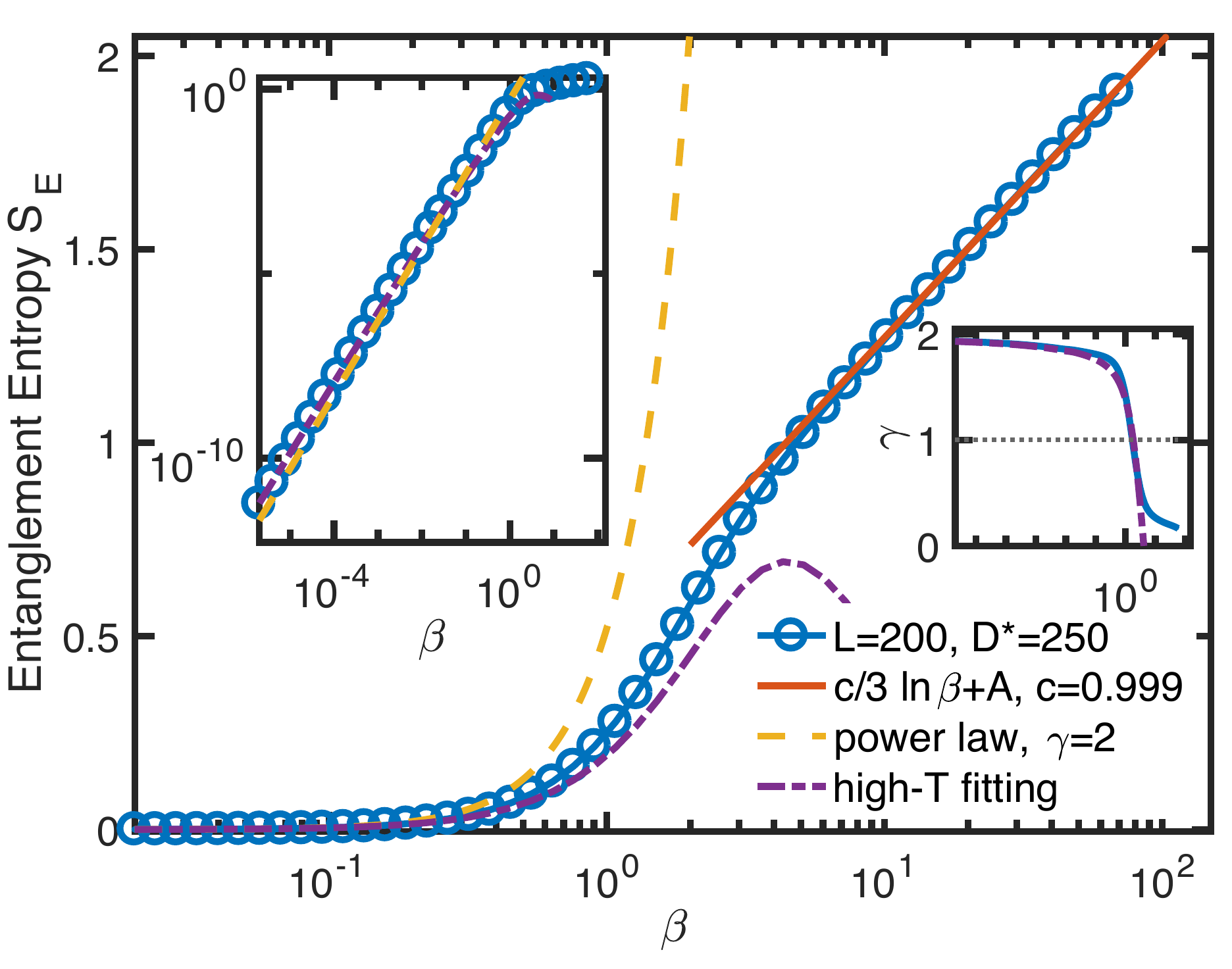}
\caption{(Color online)
   Bipartite MPO entanglement $S_E$ across the center of the
   system in an OBC Heisenberg chain of length $L=200$. At
   low temperatures ($\beta \geq 10$), $S_E$ diverges
   logarithmically.  The line in the main panel for large
   $\beta$ (small $T$) is a fitting to $S_E = c/3 \ln{\beta}
   + \mathrm{const}$, with central charge $c\simeq 0.999$
   determined by fitting the range $\beta > 12$.  Left inset
   shows the same data on a log-log plot, which emphasizes a
   power-law of entanglement versus $\beta$ for $\beta
   \lesssim 1$.  The purple dashed lines represent the fit in
   \Eq{eq:fit:SE} with $\alpha=0.05$, while the yellow dashed
   lines represent $\beta^{2}$ as guide to the eye.  Right
   inset shows the slope of the log-log data in the left
   inset, i.e., the power-law exponent $\gamma$ of the
   entanglement versus $\beta$.
}
\label{Fig:EntOpenChain}
\end{figure}

\subsection{Universal entanglement behavior in (1+1)D conformal
thermal states}

In Fig.~\ref{Fig:EntOpenChain} we plot the entanglement entropy in a
spin-1/2 Heisenberg chain.  By simulating twice the length ($L=200$)
as in \Fig{Fig:EntRGFlow}, we can observe a logarithmic divergence
in the low temperature region ($L \gtrsim \beta \gg 1$) The
logarithmic entropy was already observed in the past and related to
the computational complexity of finite temperature simulations
\cite{Prosen.t+:2007:Entropy,Marko.z+:2008:Complexity}.
More recently, this was further analyzed by numerical simulations on
Renyi entropy and also conformal field theory (CFT) analysis
\cite{Barthel.t:2017:FiniteT}.  Notably, the finite-temperature
entanglement calculation in \Fig{Fig:EntOpenChain} provides a
convenient and accurate way to extract the central charge $c$ of
CFT: without going into ground states calculation at $T=0$
\cite{Holzhey.c+:1994:RenormalizedEntropy, Calabrese09,Amico08}, one
can fit the MPO entanglement at finite temperatures. In
\Fig{Fig:EntOpenChain}, we observe that, by fitting $L=200$ data,
the estimate of central charge $c$ is already very accurate
($c=0.999 \simeq 1$).  For this we fit the data to the CFT
prediction $S_E=\tfrac{c}{3}\ln\beta + \mathrm{const}$
\cite{Barthel.t:2017:FiniteT}.  Importantly, by having the system
length sufficiently large, the physics of the thermal state in the
center of the system is effectively short-ranged by a thermal
correlation length. In this sense, the simulation of the central
charge in \Fig{Fig:EntOpenChain} does not yet see the finite open
boundary condition.  This is in stark contrast to the evaluation of
central charge using ground state properties.

Universal features of the entanglement property appear also at large
temperatures.  As seen in the left inset of \Fig{Fig:EntOpenChain},
the MPO entanglement shows a power-law behavior for $\beta \lesssim
1$.  The slope $\gamma$ on the log-log plot at very high
temperatures is analyzed in the right inset in
\Fig{Fig:EntOpenChain}, which suggests a power-law exponent $\gamma
\approx 2$ for $\beta \lesssim 1$. The growth in the entanglement
$S_E$, however, slows down strongly for $S_E\gtrsim1$, i.e., $\beta
\gtrsim 1$, where $\gamma$ drops significantly below 1 as seen in
the right inset of \Fig{Fig:EntOpenChain}.  The entanglement
behaviors at high temperatures can be understood from a lowest,
i.e., first-order expansion of density operator, $\rho(\tau) =
\mathbb{I} - \tau H$.  The singular value spectrum of this MPO is
given by the vector $s=[1,\alpha\tau]$, with $\alpha$ another
numerical vector. The resulting normalized ``density matrix'' of the
supervector in \Eq{eq:purif} has eigenvalues $r_i=s_i^2/\sum_{i'}
s_{i'}^2$ with lowest-order thermal contributions $\propto \tau^2$.
With the von Neumann entropy $S_E=-\sum_i r_i \ln(r_i)$ and

\begin{subequations} \label{eq:gamma}
\begin{eqnarray}
   \gamma(\tau) &\equiv& \tfrac{d \ln S}{d \ln\tau},
\end{eqnarray}
one obtains that
\begin{eqnarray}
   \gamma_0 &\equiv& \lim_{\tau\to 0^+} \gamma(\tau) =2.
\end{eqnarray}
\end{subequations}

For simplicity, one may consider a single value for the
vector $\alpha$, resulting in the two normalized weights
\begin{subequations}\label{eq:fit:SE}
\begin{eqnarray}
  (r_1,r_2) = \tfrac{1}{1+{\alpha}^2\tau^2} ( 1, {\alpha}^2 \tau^2 )
\end{eqnarray}
with von Neumann entropy
\begin{eqnarray}
  S(\tau; \alpha) = -\sum_{i=1}^2 r_i \ln r_i 
\,\text{.}
\end{eqnarray}
\end{subequations}
A subsequent one-parameter fitting of $S(\tau;\alpha)$ with respect
to $\alpha$ to the actual entanglement entropy $S_E$ for
$\tau=\beta\ll 1$ nicely reproduces the high-$T$ entanglement data,
as shown in the right inset in \Fig{Fig:EntOpenChain} for
$\alpha=0.05$.  The slope $\gamma$ decreases  {monotonically}, starting
from $\gamma=2$ and undergoing a sharp decrease around $\beta\sim1$.
Note, however, that the convergence towards the power-law exponent
of $\gamma=2$ for small $\tau$ is extremely slow, as also clearly
supported by the simple asymptotic analysis above. For example, for
$\tau=10^{-3}$, one only has $\gamma \simeq 1.94$.

Nevertheless, it follows from the generality of the above asymptotic
argument, that the exponent $\gamma=2$ for infinitesimal $\tau$ is
universal.  It should hold for any Hamiltonian, and therefore, in
particular, also in arbitrary dimensions.  Furthermore, given that
by construction, the exponent $\gamma=2$ only holds for $\beta\ll 1$
where $S_E\ll 1$, the growth of the entropy of the MPO may be
considered sublinear in this regime, in the sense that the entropy
grows slower than linear for infinitesimal $\tau$, having
$\lim_{\beta\to 0^+} \tfrac{d S_E}{d\beta} = 0$.

\begin{figure}[tbp]
\includegraphics[angle=0,width=1\linewidth]{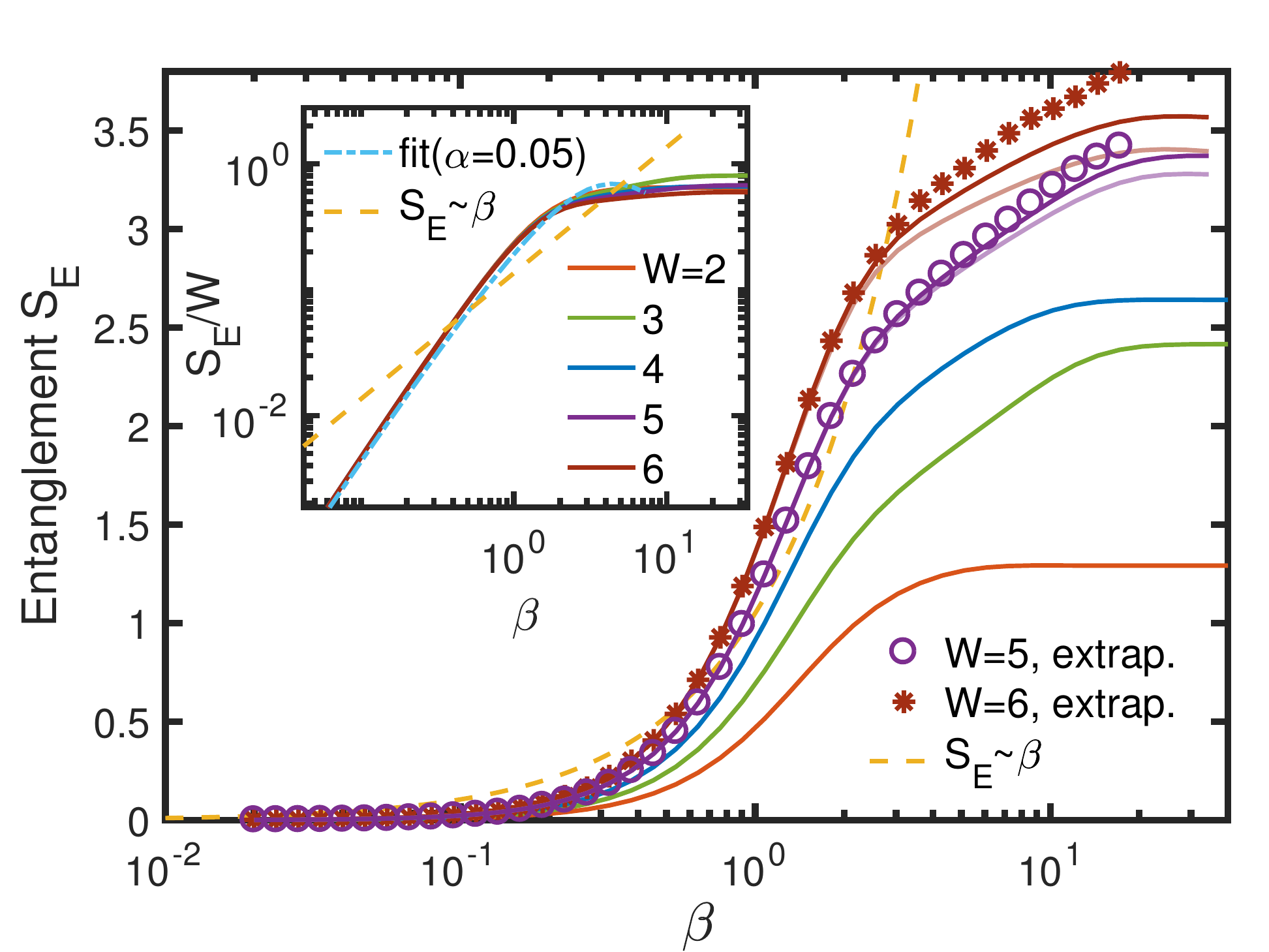}
\caption{(Color online)
   Bipartite MPO entanglement across the center of the system
   in the SLH, up to width $W=6$.
   The system length is fixed to $L=10$ for width
   $W=2,3,4,5$, and to $L=12$ for $W=6$.  For $W\leq 4$, the
   entanglement entropy in the center of the system is well
   converged by retaining $\Dstar=500$ bond multiplets.  For
   $W=5$ and $6$, we show data with $\Dstar (D) =300$ (1000)
   and $550$ (2000) (color matched lines) as well as
   extrapolated data (symbols) based on a quadratic
   polynomial extrapolation in $1/\Dstar \to 0$ that includes
   additional data between $\Dstar=300$ and $\Dstar=550$.
   Inset presents the entropy per leg ($S_E/W$) on a log-log
   scale, where the power-law increase for $\beta \ll 1$
   again clearly follows the large temperature fit in
   \Eq{eq:fit:SE} with approximate exponent $\gamma\simeq 2$,
   using $\alpha=0.05$.
   The dashed line shows
   $S_E\sim\beta$ as a guide to the eye for reference.
}
\label{Fig:EntSquare}
\end{figure}

\subsection{Logarithmic entanglement in thermal states of 2D
Heisenberg model} \label{sec:ent2D}

The low temperatures entanglement of the thermal state saturates for
gapped quantum chains and grows only polynomially for critical ones.
This directly implies excellent numerical efficiency in 1D quantum
systems, since the required bond dimension grows at most
polynomially with inverse temperature $\beta$
\cite{Prosen.t+:2007:Entropy,Marko.z+:2008:Complexity}, rather than
exponentially as originally estimated
\cite{Hastings.m.b:2006:Gapped}.
We take this as a motivation to explore the MPO entanglement of the
Heisenberg magnet on the square lattice, and take it also as an
indicator of computational complexity for the latter. 

In \Fig{Fig:EntSquare} we plot the entanglement property versus
inverse temperature $\beta$, for the SLH of various system sizes, ranging from width $W=2$ to
$W=6$.
As shown in the inset of \Fig{Fig:EntSquare}, in the high
temperature regime one still recovers the universal power-law
$\gamma\simeq 2$, c.f. \Eq{eq:gamma}.  Specifically, the data is
very well fitted by the function in \Eq{eq:fit:SE} with exactly the
same parameter $\alpha=0.05$ as in \Fig{Fig:EntOpenChain} for the 1D
quantum chain.

For the inset of \Fig{Fig:EntSquare}, since the entanglement entropy
satisfies area law, i.e., $S_E \propto W$, we divide the entropy
$S_E$ by the width $W$.  For high temperatures $\beta \lesssim 1$,
this collapses the data for different system widths on top of each
other, indeed, demonstrating universal area law in this regime.  For
intermediate and low temperatures, $\beta > 1$, deviations from the
strict scaling collapse of the area law can be observed.

In the main panel of \Fig{Fig:EntSquare}, the entropy $S_E$ changes
gradually into a logarithmic divergence versus $\ln{\beta}$ as $W$
increases for both, even and odd widths.  This suggests that
simulations are also efficient with increasing $\beta$ in the 2D
setting, while bearing in mind an additive constant term to the
entropy that is proportional to the width (note that in order to
satisfy area law, the entropy data for $2 \leq \beta \leq 20$ is
separated roughly by equal vertical offsets when incrementing the
width for $W\ge 3$).  Since the calculations of the entropy in the
system center are not fully converged for the wider systems, we also
extrapolate the width $W=5$ and $6$ systems in $1/\Dstar\to0$. This
actually further reinforces the regime of logarithmic increase of
$S_E$ vs. $\ln\beta$ for $2 \leq \beta \leq 20$.

A similar additive logarithmic scaling of the entanglement entropy
($\propto \ln{W}$) in 2D Heisenberg model has been also found
numerically via QMC calculations in the ground state
\cite{Kallin.a.b+:2011:Entanglement, Song.h.f+:2011:Entanglement,
Humeniuk.s+:2012:Entanglement, Kulchytskyy.b+:2015:Goldstone,
Luitz.d.j+:2015:Entanglement, Laflorencie.n+:2015:Entanglement}.
The coefficient of logarithmic correction was argued to be universal
\cite{Metlitski.m.a+:2011:Entanglement} (proportional to the number
of Goldstone modes in the system). In the 2D Heisenberg model here,
according ED and DMRG studies of the tower of states (ToS) in the
energy and the entanglement spectra \cite{Alba.v+:2013:Entanglement,
Kolley.f+:2013:Entanglement}, respectively, the relevant low-energy
ToS has characteristic level spacing that scales as $1/N$ with $N=W
L$ the total number of sites.  This is in contrast to spin wave
excitations (Goldstone modes) which have characteristic level
spacing that scales with inverse linear system size, i.e., $1/L$.

These $1/N$ ToS excitations are responsible for the logarithmic
entanglement at $T=0$, and possibly also relate to the $\ln{\beta}$
scaling of the entropy observed in the present study.  In
\Fig{Fig:EntSquare}, we restrict the length to be as small as
$L=10$, which suggests that the magnon excitations are gapped out at
temperatures as low as $T = 1/30$ -- 1/10. Therefore the relevant
energy scale are likely only ToS modes with energy level spacing
$\propto1/N \sim 1/50$, which is smaller than the temperatures in
the regime where logarithmic scaling $S_E\sim \ln \beta$ is
observed.

The logarithmic growth in entanglement in \Fig{Fig:EntSquare},
as well as the low-$T$ specific heat behavior, are quite remarkable.
As we will show shortly, they differ qualitatively from the
anisotropic case, $\Delta \neq 1$. In the isotropic case, the
entanglement curves  {do} not show any sign of singularity at any
finite $T$,  {which} suggests the absence of phase transition at $T
\neq 0$. The low-temperature specific heat $c_V$ gets significantly
enhanced when the width $W$ is increased from 4 to 5, but it still
grows  {monotonically with} increasing $\beta$,  {e.g., with no sign of 
a singularity at finite $\beta$}.  
This is, of course, in
complete agreement with the celebrated finite-temperature
Mermin-Wagner theorem \cite{Mermin-Wagner}.

\subsection{XTRG and thermal phase transitions in 2D}
\label{sec:thermal}

\begin{figure*}
\centering
\includegraphics[width=0.85\linewidth]{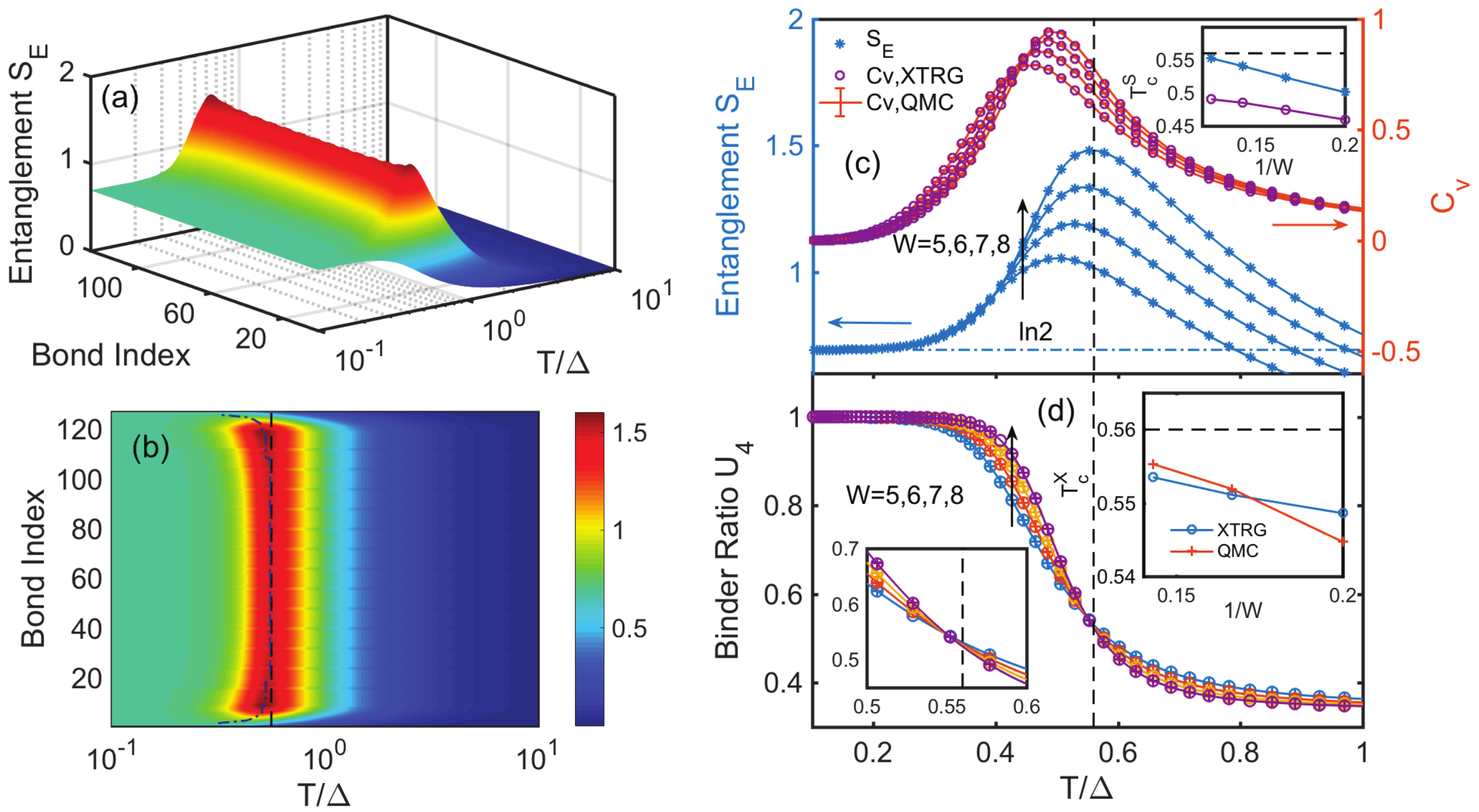}
\caption{(Color online) 
  Finite temperature phase transition in the XXZ
  Heisenberg model at $\Delta=5$, keeping $D=600$ states
  in U(1) XTRG.  (a) Entanglement landscape vs. temperatures
  $T/\Delta$ and bond indices on a $L=16,W=8$ open square
  lattice which exhibits pronounced peaks around the exact
  transition temperature $\Tc = 0.56$ \cite{Goettel12}
  [dashed black lines in (b,c,d)]. The dash-dotted line in
  (b) depicts the maximal entropy point for each bond,
  which converges to $\TcS \simeq 0.552$ in the central
  region of the system.
(c) Block entanglements $S_E$ at bond
  $[(N+W)/2]$ cutting across the center of the system (where $N=LW$
  and $[.]$ means the integer part), and also the specific heat
  $c_V$ with comparison to QMC data.  The data  {is} shown for various
  widths $W=5,6,7,8$ with fixed aspect ratio, using $L=2W$.  Both
  $S_E$ and $c_V$ curves show peaks near $\Tc$ (vertical dashed
  line).  The horizontal dashed line represents constant $\ln{2}$,
  manifesting the global $\mathbb{Z}_2$ symmetry which can be
  spontaneously broken at low temperatures if applying a small
  pinning field, otherwise.  The peak positions $\Tc^S$ of the $S_E$
  and $c_V$ are collected and shown versus $1/W$ in the inset.  The
  value of $\Tc^S$ from $S_E$ suffers smaller finite-size
  corrections than that obtained from the specific heat.
(d) Binder ratio $U_4$ for various system sizes $L=2W$.
  XTRG data  {is} plotted with open circles and looper
  QMC data with crosses with perfect agreement 
  between the two data sets. The cross point determines $\TcX$. 
  By zooming into the data in the lower left inset, we obtain the
  crossing points $\TcX(W)$ for pairs of consecutive system sizes
  $W$, as plotted versus $1/W$ in the upper right inset, again with
  good agreement between the XTRG and the QMC data.  From this we
  obtain $\TcX \simeq 0.554$ for our largest systems, which still
  trends towards $\Tc$ (horizontal dashed line) with increasing
  system size, and hence agrees with $\Tc$ to within an error bar
  of 1\%.
  }
\label{Fig:XXZTc}
\end{figure*}

Lastly, we analyze the 2D anisotropic XXZ model on
the square lattice, i.e., \Eq{eq:HeisenbergH} with $J=-1$, and
$\Delta \neq 1$. There exists a finite-temperature phase transition
at the critical temperature $\Tc$ towards a gapped low-energy
ferromagnetically ordered phase.  {Note that
while Trotter-like methods have to necessarily work
straight through a thermal phase transition point, which becomes
singularly hard already with the first one encountered,
our XTRG can {\it jump across} phase transition
points \cite{Czarnik.p+:2015:PEPS}
by reducing temperatures by a factor of $2$!}

In the following, we show that
XTRG can be employed to simulate such model with nonzero $\Tc$ with
high accuracy.  In particular, we determine the phase transition
point using various quantities including the block entanglement
entropy, specific heat, as well as the Binder ratio in spin
fluctuations.
We choose the same model parameters as in
Ref.~\onlinecite{Bruognolo.b+:2017:MPS}, i.e., $\Delta=5$, where the
critical temperature was estimated as $\Tc^\ast/\Delta = 0.56 \pm
0.01$. This was in agreement with the exact result $\Tc /\Delta =
0.56$ in Ref.~\cite{Goettel12} from QMC for much larger system
sizes, which we will also take as reference for our data below.

Our results are presented in \Fig{Fig:XXZTc}, where we also make
comparison to QMC data explicitly generated by the ALPS looper code
\cite{Bauer.b+:2011:ALPS}.
\FIG{Fig:XXZTc}(a) shows the landscape of the block entanglement
entropy across each bond of the MPO density matrices with decreasing
temperature. Because the low-temperatures phase is gapped, prominent
peaks are present around the critical temperature $\Tc=0.56$, with
only weak dependence on the snake-like serial ordering of our 2D
system, otherwise.  \FIG{Fig:XXZTc}(b) shows a top view of the same
data where the temperatures $\TcS$ of the maxima of the entropy data
$S_E$ (dashed dotted line) agrees very well with the exact QMC value
$\Tc$ (vertical dashed line). 

To see this more clearly, we also take a cut in the center of
the system [while shifted by half a column to avoid local minimal,
see caption of \Fig{Fig:XXZTc}(c)].  In \Fig{Fig:XXZTc}(c) we show
$S_E$ vs. temperature $T$ for various system widths $W$. From this
we observe that the peak positions $\TcS$ in $S_E$ vs. $T$ quickly
approaches $\Tc$ as $W$ increases.  Specifically, for $W=8$ we
already have $\TcS \simeq 0.552$ with an error  $|\TcS-\Tc | <
0.01$.
Similar calculations are also performed using cylindrical boundary
conditions, as show in \App{App:EntCyl} [\Fig{Fig:XXZCyl}], where we
also see a quite accurate agreement with $\Tc$ for rather small
$W$.

We also resort to other more standard thermal quantities
including specific heat $c_V$ [\Fig{Fig:XXZTc}(c)] and Binder ratio
[\Fig{Fig:XXZTc}(d)] to pinpoint the critical temperature.  Our
results for $c_V$ by XTRG are in perfect agreement with QMC
simulations, as shown in \Fig{Fig:XXZTc}(c).
The specific heat also exhibits a peak near $\Tc$. As is well known
from numerous QMC simulations, however, similar to other physical
observables such as the spin susceptibility, the specific heat
suffers significant finite-size corrections. So this often only
provides a first rough estimate for $\Tc$ in practical numerics,
where even the finite-size extrapolation to the thermodynamic limit
often has larger error bars still, as is also the case here [cf.
inset to \Fig{Fig:XXZTc}(c)].

The lowest order, finite size corrections in standard thermal
quantities, however, can be eliminated by taking ratios of
expectation values. In practice, a common way to pinpoint $\Tc$ more
precisely, is the Binder ratio \cite{Binder-1981a,Binder-1981b}
\begin{eqnarray}
   U_4 \equiv
   \frac{\langle (\Sztot)^2
   \rangle_{\beta}^2}{\langle (\Sztot)^4 \rangle_{\beta}}
\text{ ,}
\label{Eq:Binder}
\end{eqnarray}
where $\Sztot = \sum_i S_i^z$ is the total magnetization in
$z$-direction, with $\langle \Sztot \rangle =0$.  We calculate $U_4$
with both, XTRG and QMC the looper code, and find perfect agreement.
Note that the Binder ratio $U_4$ can be conveniently calculated
using an exact MPO representation for the total magnetization
(operator) $\Sztot = \sum_i S_i^z$ of bond dimension $D=2$. By
taking product of $\Sztot$ and performing compression (without any
essential truncations), one can obtain compact representations of
$(\Sztot)^2$ and $(\Sztot)^4$ with $D=3$ and $D=5$ MPOs,
respectively. With these MPOs one can evaluate the
finite-temperature expectation values $\langle \, . \,
\rangle_{\beta}$ of the two operators required for the Binder
ratio in Eq.~(\ref{Eq:Binder}).

From the data as in \Fig{Fig:XXZTc}(d), we collect the crossing
points $\TcX$ for two consecutive curves for widths $W$ and $W+1$.
The data is summarized in the inset to \Fig{Fig:XXZTc}(d). By
considering the data for the largest system to be the most accurate
given the aspect ratio $L/W=2$, we obtain $\TcX\simeq 0.554$ from
both the XTRG and QMC data.  By considering the minor trend still
with increasing system size, we find agreement between this $\TcX$
and $\Tc$ to within 1\% which thus serves as a very good estimate
for the thermodynamic limit.

From the above we conclude that XTRG can be used to determine $\Tc$
of thermal phase transition very accurately. Furthermore, the
maximum in the entanglement entropy $S_E$ itself can already provide
a good estimate for the critical temperature. Here, in particular,
it clearly outperforms conventional thermal quantities like the
specific heat $c_V$, as seen in the upper inset in
\Fig{Fig:XXZTc}(d). Nevertheless, since $S_E$ grows linearly with
$W$ for $T>\Tc$ but stays constant for $T<\Tc$ in given system (the
system nearly becomes a product state for $T\ll \Tc$), the peak
around \TcS is rather round.  This might lead to systematic offsets
in \TcS in the thermodynamic limit, but needs further studies.

\section{Conclusions and Outlook} 

Inspired by the logarithmic growth of entanglement of purified
(thermofield double) states, we propose an exponential speed up of
thermal simulation. This thermal tensor network algorithm employs an
MPO form of the density operator and proceeds via doubling of the
density matrix $\rho(\beta)$ along the imaginary time evolution.  We
show that this exponential tensor renormalization group (XTRG)
method gains both accuracy and efficiency, in thermal simulations of
the Heisenberg models. We also implement this idea of logarithmic
temperature setup in a pointwise series-expansion thermal tensor
network (SETTN) algorithm. Also there we get more efficient and
accurate results than previous Maclaurin SETTN
\cite{Chen.b+:2017:SETTN}.

We apply XTRG and SETTN to efficiently simulate thermal states of 1D
and 2D Heisenberg spin models, obtain accurately the thermodynamic
quantities including free energy, energy, and specific heat, and
study their low- and high-temperature behaviors. We have also
investigated the temperature dependence of entanglement properties
$S_E$ in the MPO, and observed logarithmic entropies $S_E\simeq a +
b\ln{\beta}$ with constants $a$ and $b$ at low-temperatures not only
in gapless quantum chains, but also in the SLH at fixed system size due to gapless ToS modes. 

We applied XTRG to a 2D Heisenberg models with a
thermal phase transition {, as well as the TLH with spin frustration}. 
The results demonstrate the efficiency of
the present algorithm,  {which is capable to show rich finite-$T$ physics, 
especially for those system with spin frustration}. 
 {It will be interesting to see how useful the
present algorithm can be in exploring more 2D challenging
systems,} such as the kagome and the
$J_1$--$J_2$ Heisenberg models on the square lattice, as well as
interacting fermionic systems.

With XTRG, however, we are not limited to high temperatures but can
simulate down to much lower temperatures than previously anticipated
\cite{Bruognolo.b+:2017:MPS}. The present MPO algorithms may be
improved in several directions, including combining them with METTS
samplings at low $T$, or linked-cluster expansion to reduce
finite-size effects, etc. Moreover, XTRG may also be
straightforwardly combined with efforts to reduce block entanglement
entropy further by operating with disentanglers on the auxiliary
state space in the purified scheme
\cite{Terhal.b.m+:2002:Entanglement, Hauschild17}, all of which
certainly deserves further exploration.

\begin{acknowledgments}

The authors thank Benedikt Bruognolo
for nicely providing the METTS data and for very constructive
discussions. B.-B. C. and W. L. would like to acknowledge helpful
discussions on related topics with Han Li and Yun-Jing Liu, and
A. W. also with Thomas Barthel.  {W. L. is indebted to Shou-Shu Gong for providing the DMRG data of YC4 TLH.} 
This work was supported by the
National Natural Science Foundation of China (Grant No. 11504014,  {11834014,} 
11474015, 61227902, 11774018) and the Beijing Key Discipline
Foundation of Condensed Matter Physics.  B.-B. C., L. C., and W. L. thank the
hospitality of Arnold Sommerfeld Center for Theoretical Physics,
University of Munich, where this work was finished.  A. W.
acknowledges support from the German Research Foundation (DFG)
WE4819/2-1 and  WE4819/3-1 until December 2017, and since 
then by US DOE under contract number DE-SC0012704.

\end{acknowledgments}

\appendix

\setcounter{figure}{0}    
\renewcommand\thefigure{A.\arabic{figure}}


\section{Entanglement entropy in thermal states}
\label{App:SE}

The block entropy in the center $\ell \sim L/2$ of an individual
low-energy pure state $|s\rangle$ scales like $S_E(|s\rangle) \simeq
\tfrac{c}{6} \log \ell + \mathrm{const}$ \cite{Calabrese09,Amico08}.
The MPO block entropy for the outer product $\hat{\rho}_s \equiv
|s\rangle \langle s|$ then acquires a factor of $2$, i.e.
$S_E(\hat{\rho}_s) \simeq \tfrac{c}{3} \log \ell + \mathrm{const}$.
Now by going to a thermal state $\hat{\rho} = \sum_s \rho_s |s
\rangle\langle s|$ with weights $\sum_s \rho_s=1$, this scaling
does not change.  More precisely, due to the subadditivity of
entanglement entropy, one obtains an upper bound $S_E(\hat{\rho})
\lesssim \tfrac{c}{3} \log \ell + \mathrm{const}$.  

More explicitly, the block entropy of the thermal density matrix
changes most for the {\it worst case} that its spectrum is altered
from $\varrho^{(s)}_i$ for some fixed $s$ to the set $\rho^s_{i}
\equiv \rho_s \varrho^{(s)}_i$.  The corresponding block entropy for
a cut across the center of the system has the upper bound
\begin{eqnarray}
   S_E(\hat{\rho}) & \lesssim & -\sum_{s,i}
      \rho^s_{i} \log \rho^s_{i} \notag \\
   &=&  \underset{\equiv S(\{\rho_s\})
     \simeq \mathrm{const}}{
     \underbrace{- \sum_{s}\rho_s \log \rho_s}}
   + \sum_{s} \rho_s \underset{ \equiv S_E(\hat{\rho}_s) 
      \simeq \tfrac{c}{3} \ln \ell + \mathrm{const} }{
      \underbrace{
      \Bigl( - \sum_{i}
      \varrho^{(s)}_i \log \varrho^{(s)}_i \Bigr)}} \notag  \\
   &\simeq&  \tfrac{c}{3} \ln \ell + \mathrm{const}
\text{ ,}\label{eq:SE:upperbound}
\end{eqnarray}
having used $\sum_s \rho_s=1$, as well as $\sum_i \varrho^{(s)}_i=1$
for all $s$.  Here $S(\{\rho_s\})$ is the entropy of the weight
distribution $\rho_s$.  

Now, \Eq{eq:SE:upperbound} provides an upper bound.  As argued with
\Fig{Fig:FS-spec}, in order to sample thermal averages, one requires
that the system size must be larger than the thermal correlation
length.  For physical systems whose finite size spectra have a
low-energy level spacing that scales like inverse system size, this
is achieved by choosing $L=a \beta$ with $a\gtrsim1$ sufficiently
large but constant.  Therefore one can substitute $\ell=L/2 \to
\beta$ in \Eq{eq:SE:upperbound}, resulting in the overall
block-entropy of the thermal state
\begin{eqnarray}
   S_E(\hat{\rho}) \simeq
   \tfrac{c}{3} \log \beta + \mathrm{const}
\text{ .}\label{eq:SE:beta}
\end{eqnarray}

The block-entropy of the thermal state thus saturates, since
the system length is effectively cut off by the thermal correlation
length $\xi$, resulting in $\ell \sim \min(L,\xi)$ in the
thermodynamic limit $L\to\infty$. 
See also \cite{Barthel.t:2017:FiniteT,Dubail17} for a more
rigorous derivation based on CFT arguments.  Furthermore note, that
even \Eq{eq:SE:beta} can be still considered an upper estimate,
since in the purification scheme, the thermal state allows one to
minimize block entanglement entropy by disentangling operations on
the auxilliary state space \cite{Terhal.b.m+:2002:Entanglement,
Hauschild17}.


\section{Symmetry invariant matrix product operator for Hamiltonians}
\label{App:SYMPO}

In this Appendix, we discuss our approach to the implementation of
both, abelian as well as non-abelian symmetries into the MPO
representation of a given Hamiltonian. Conceptually, non-abelian
symmetries proceed the same way as abelian symmetries, as we explain
below.  The actual implementation is based on the framework of the
tensor libary QSpace \cite{Weichselbaum.a:2012:QSpace} that can deal
with abelian and non-abelian symmetries such as SU($N$) or the
symplectic symmetry Sp($2N$) on a generic footing.
 
In order to emphasize the generality of the argument, we will
frequently use the notation $q$ for a label of a generic irreducible
representation (irep) of a given symmetry. Here, specifically, it
may either stand for the spin-projection $S^z$ or spin $S$ label in
the case of a U(1) [an SU(2)] symmetry, respectively. For this
reason, we also refer to $q=0$ only as the scalar representation
even if for U(1) symmetry all symmetries multiplets are actually
one-dimensional and in that sense, scalars.  Examples for scalar
operators are the full Hamiltonian, as well as all of its terms in
its sum including local 1-site terms. Vacuum states transform like a
scalar multiplet.

The dual representation $q^\ast$ of some given irep $q$ is defined
by the unique representation that allows to form a scalar, i.e.,
with Clebsch-Gordan coefficients (CGCs) $(q,q^\ast;0)$. These CGCs
when properly normalized define a unitary matrix $U^{[q]}$. This
will be referred to as $1j$-symbol by analogy e.g., to $3j$ symbols
for the SU(2) spin symmetry, with the difference, that here only a
single irep label is concerned.  While SU(2) symmetry is self-dual,
i.e., $q^\ast=q$, abelian U(1) symmetries are not, since one has
$q^\ast=-q$ such that $q+q^\ast = q'$ properly adds up to the
scalar representation $q'=0$.  For self-dual symmetries, such as
SU(2), however, one must be careful in that $U^{[q]}$, when written
simply reduced to a unitary matrix of rank-2, it becomes
indistinguishable from $(U^{[q]})^\dagger = U^{[q^\ast]}$ which,
however, may differ by a sign, e.g., for half-integer spins in the
case of SU(2). Importantly, $1j$-symbols allow to revert arrows in
lines in a tensor network by inserting $\mathbb{I} = U^{[q]\dagger}
U^{[q]}$.

\subsection{Automata approach}
 
Firstly, we briefly recapitulate the automata approach for
constructing MPOs of the Hamiltonian \cite{Frowis.f+:2010:Tensor,
Pirvu.b+:2010:MPO, Crosswhite.g.m+:2008:Automata}.  Consider, for
example, the quantum Ising chain $H=\sum_i S_i^x S_{i+1}^x - h
S_i^z$, which provides a simple example of a Hamiltonian with a
single nearest-neighbor interaction term together with a local term
(here with magnetic field strength $h$).
We need to compute and store the matrix elements of the spin
operators \{$S_x$,$S_z$\}.  Together with the identity operator,
$\mathbb{I}$, these form a basis of local operators that enter a
rank-4 tensor $T_{\alpha, \alpha', \sigma, \sigma'}$, where by rank
we refer to the number of indices (or legs in a graphical depiction)
of a given tensor.  The tensor $T$ is the elementary local tensor of
the MPO, with $\sigma$ the local state space of a given site, and
$\alpha$ the virtual bond states that tie together the MPO. The
tensor $T$ has the same form for every site due to the translational
invariance of the Hamiltonian [assuming open boundary condition, the
open virtual indices of the $T$-tensors for the first and last site
are contracted (``capped'') with a start and a stop state,
respectively; see below].

To be concrete, each tensor $T$  contains $D_H^2 d^2$ matrix
elements, where $d$ is the dimension of the local state space
$\sigma$, and $D_H$ is the bond dimension of the virtual state space
$\alpha$.  Every {\it matrix} element of $T$ in the indices
$(\alpha,\alpha')$ is linked to a local operator with {\it matrix}
indices $(\sigma,\sigma')$.  It is therefore natural to group the
relevant local operators into an (orthogonal) set, that we will also
index below.  For the Ising model above, for example, the relevant
set of local operators is given by \{$\mathbb{I}$,$S_x$,$S_z$\}.

The virtual bond state space is given by a start state [$\alpha=1$,
or equivalently, $(1,0,0,\ldots)^T$], a stop state [$\alpha=2$, or
equivalently, $(0,1,0,\ldots)^T$], followed by $\alpha=3, \ldots,
m_\mathrm{int}+2$ which assigns an index position to every one of
the $m_\mathrm{int}$ interaction terms in the Hamiltonian that
stretches across a given bond in the MPO (strictly speaking,
$m_\mathrm{int}$ corresponds to the number of operators that need to
be stored across a given bond which may be less than the number of
elementary interaction terms in the Hamiltonian if interaction terms
can be grouped by factorizing out specific operators). Hence the
dimension of the virtual bond state space is given by $D_H = 2 +
m_\mathrm{int}$.
For example, for the Ising model above, the $i$-th bond in the
system in between sites $i' \le i$ and $j'\ge i+1$ carries the
single interaction term $S_i^x S_{i+1}^x$, hence $m_\mathrm{int}=1$
and $D_H=3$.

The general strategy then for setting up the MPO w.r.t the specific
example of the Ising model is as follows: starting from the left
end, the bond state space, i.e., the automaton is initialized in the
start state ($\alpha=1$). This is carried through the MPO (therefore
$T_{1,1}=\mathbb{I}$) until an interaction term in the Hamiltonian
occurs, say at site $i$, which brings the automaton into the state
$\alpha=3$ (therefore $T_{1,3}=S^x)$. Having only nearest neighbor
terms, the subsequent $T$ tensor e.g., at site $i+1$ immediately
brings down the automaton to the end state $\alpha=2$ (therefore
$T_{3,2}=S^x$).  By having completed the interaction term, the
automaton stays in that state (hence $T_{2,2}=\mathbb{I}$).
Overall, what has been encoded this way was simple the interaction
term $\mathbb{I}_1 \otimes \mathbb{I}_2 \otimes \ldots \otimes
\mathbb{I}_{i-1} \otimes S^z_{i} \otimes S^z_{i+1} \otimes
\mathbb{I}_{i+2} \otimes \ldots \otimes \mathbb{I}_{L}$.  With the
same line of arguments, the local transverse field term, say at site
$i$, is described by $T_{1,2}=-h S^z$, which directly brings the
automaton from the start into the end state. By translational
invariance, there is nothing special about site $i$, though.
Therefore all of the {\it matrix} elements of the tensor $T$
specified above must hold for every site.  Given these local tensors
in MPO, the summation (trace) over geometric indices $\alpha$ is
equivalent to adding up all the interaction terms in the total
Hamiltonian.

\begin{figure}[tbp]
\includegraphics[angle=0,width=0.95\linewidth]{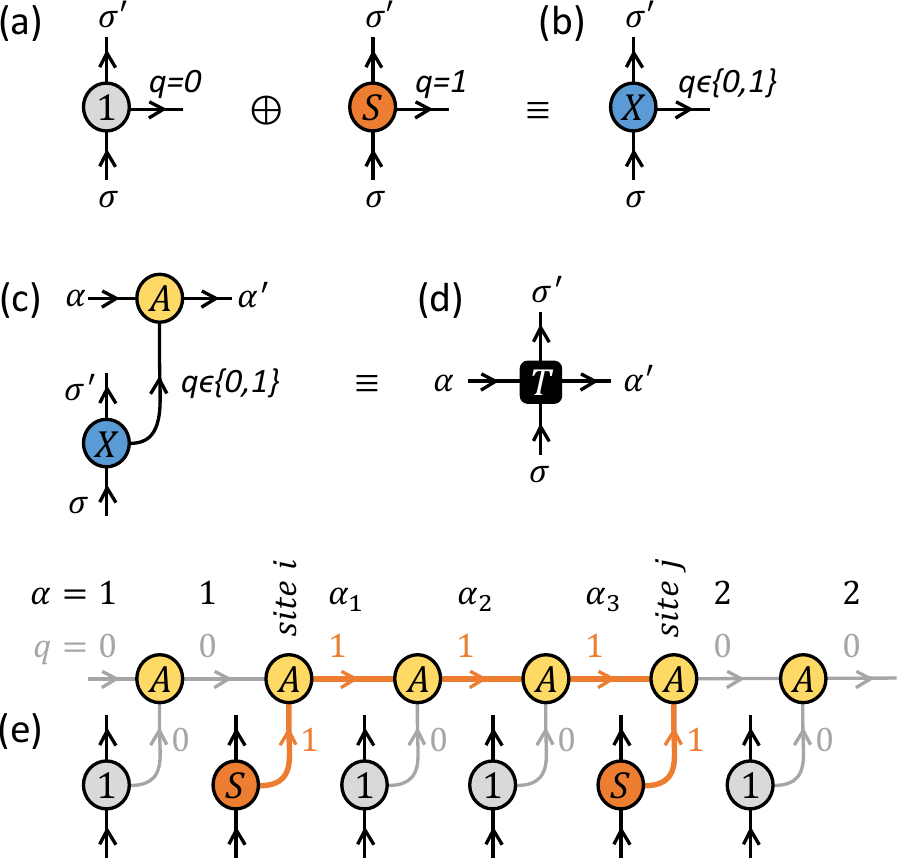}
\caption{(Color online)
(a) The local identity and spin operator
  for the $\mathrm{SU(2)}$ spin symmetric Heisenberg model,
  which transform according to $S \equiv q=0$ and $q=1$,
  respectively. All relevant local operators, including
  the identity operator, can be combined into the 
  rank-3 tensor $X$ [panel (b)].
(c) The super-MPS described by the rank-3 tensor $A$
  can be contracted with the operator index in $X$
  to form the rank-4 local tensor $T$ in the SU(2)
  invariant MPO [panel (d)].
(e) Typical sequence that occurs in the construction
  of the MPO via the super-MPS that describes a specific
  individual interaction term $\hat{S}_i
  \cdot \hat{S}_j^\dagger$.
  This demonstrates how the quantum number of the
  IROP $S$ simply stretches like a string (red line)
  along the virtual bonds in between the two sites
  $i$ and $j$ where the spin operators act.
  The values $\alpha_1,\alpha_2,\ldots$ on given
  $A$-tensors are reserved for this very specific
  interaction term, where in general the indices
  $\alpha_i>2$ are not all the same.
  Note that the arrows on the red line are reversed
  w.r.t. site $j$ which thus need to incorporate the
  Clebsch-Gordan coefficients $(1,1;0)$ that combines
  the $S=1$ multiplet of the spinor $S$ with its dual
  (also $S=1$) into a scalar. It is this CGC on
  the $A$-tensor of site $j$ within the super-MPS that
  takes properly care of the dagger in the scalar
  product $\hat{S}_i \cdot \hat{S}_j^\dagger$.
  Similarly, also the coupling strength $J$ is 
  encoded with the $A$-tensor in the super-MPS.
}
\label{Fig:SMPO}
\end{figure}

\subsection{From super-MPS to MPO}

In the presence of global continuous symmetries, all state spaces
must be organized into symmetry multiplets. Naturally, this also
implies a directedness of lines in a tensor network.  From the point
of view of a given tensor, the direction on its lines indicate bra
or ket nature of these state spaces which, in pictorial language, is
equivalent to legs (lines) entering or leaving a given tensor,
respectively.  For U(1) symmetries it implies that the sum of all
charges that enter a tensor must be exactly equal to the sum of all
charges leaving it.  For SU(2) symmetries, the fusion of all ingoing
lines must result in a symmetry sector that exactly matches a
symmetry sector resulting from the fusion of all outgoing lines.  If
all lines are ingoing, the tensor must be scalar in that the
(skipped) outgoing line transforms as a singleton index that
transforms like the vacuum state (and vice versa, if all lines are
outgoing).

In contrast to the Ising model above which has no simple continuous
symmetry for $h\neq0$, let us continue with the model system of
interest in this work, the (anisotropic) Heisenberg model
[compare \Eq{eq:HeisenbergH}, using $J:=1$]
\begin{eqnarray}
   \hat{H} = \sum_{\langle i,j \rangle}
   \underset{= \tfrac{1}{2}
     \bigl( \hat{S}^+_i \hat{S}^-_{j} + \hat{S}^-_i \hat{S}^+_{j}
   \bigr)}{
   \underbrace{\hat{S}^x_i \hat{S}^x_{j} + \hat{S}^y_i \hat{S}^y_{j}}
   } + 
   \Delta 
   \hat{S}^z_i \hat{S}^z_{j}
\text{ ,}\label{eq:Ham:gen}
\end{eqnarray}
where we temporarily introduce hats on top of operators, in order to
differentiate them from symmetry labels (e.g., $\hat{S}^z$ vs.
$S^z$).  The model in \Eq{eq:Ham:gen} is U(1) symmetric as it
preserves $\Sztot$. In the isotropic case, $\Delta=1$, 
it becomes SU(2)
spin symmetric.  Then the spin operators need to be grouped into a
spinor,
\begin{eqnarray}
   \mathbf{\hat{S}} \equiv \left(
      \begin{array}{c}
         -\tfrac{1}{\sqrt{2}} \hat{S}^+ \\
          \hat{S}^z \\
         +\tfrac{1}{\sqrt{2}} \hat{S}^- \\
      \end{array}
   \right)
\label{eq:Sop}
\end{eqnarray}
such that \Eq{eq:Ham:gen} can be rewritten
in SU(2) invariant form, 
\begin{eqnarray}
   \hat{H} = \sum_{\langle i,j \rangle}
   \mathbf{\hat{S}}_i^\dagger \cdot \mathbf{\hat{S}}_j
\text{ .}\label{eq:Ham:SU2}
\end{eqnarray}
Note that the relative weights and signs in \Eq{eq:Sop} are
important for consistency with standard conventions on SU(2) spin
multiplets.  In particular, the operators in the spinor in
\Eq{eq:Sop} exactly represent, top to bottom, the states
$S^z=(+1,0,-1)$ of an $S=1$ spin multiplet (e.g., see
Ref.~\cite{Weichselbaum.a:2012:QSpace}).  In contrast, the Hermitian
set of operators $(\hat{S}^x,\hat{S}^y, \hat{S}^z)$ does not.
However, using \Eq{eq:Sop}, the dagger on one of the spinors in
\Eq{eq:Ham:SU2} is important.

In general, a local operator acting on some physical site is a
spinor, i.e., a collection of operators that transforms like some
multiplet $q$ [cf. Fig.  \ref{Fig:SMPO}(a)].  This can be written as
the irreducible operator (IROP) $\hat{X}^{[nq;q_z]}$ where the
composite index $(nq;q_z)$ naturally specifies entire state spaces
\cite{Weichselbaum.a:2012:QSpace}, or here an operator space: the
index $n$ differentiates between local IROPs that transform
according to the same irreducible representation $q$. By definition
of an index, $n=1,2,\ldots$, we therefore also introduces an
arbitrary but fixed order to the local operators.  The label $q_z$,
finally, fully differentiates the operators within a given spinor
\cite{Weichselbaum.a:2012:QSpace}.  For example, within SU(2), $q_z$
simply stands for $S_z$.

The matrix elements of IROPs are determined via the Wigner-Eckart
theorem. For a generic spinor, this IROP acquires a third dimension,
which indexes the operators in the irreducible set.  Scalar
operators then are special.  With one in- and one outgoing index,
the third index having $q=0$ is a trivial singleton dimension that
may safely be skipped.  In this sense, scalar operators can be
reduced to rank-2, and are block-diagonal.  For U(1) spin symmetry,
for example, scalar operators are the identity operator
$\hat{\mathbb{I}}$ or the spin projection operator $\hat{S}^z$.  In
contrast, the operators $\hat{S}^\pm$ carry $q=\pm\tfrac{1}{2}$,
hence switch between symmetry sectors, and therefore are not
considered scalar operators.

All local operators eventually can be combined into a single rank-3
tensor $\hat{X}$ [cf. Figs.  \ref{Fig:SMPO}(a,b)].  The third index
then represents the state space $|nq;q_z\rangle$
\cite{Weichselbaum.a:2012:QSpace} of the ``supervectors''
$X^{[nq;q_z]}$. For efficiency, the set of local operators should be
orthogonal in the sense
\begin{eqnarray}
   \mathrm{tr} \Bigl[
   \bigl( X^{[nq;q_z]} \bigr)^\dagger X^{[n'q';q'_z]}\Bigr]
   \propto \delta_{nn'} \delta_{q,q'} \delta_{q_z,q'_z}
\text{ ,}\label{eq:ortho:lops}
\end{eqnarray}
with arbitrary normalization, otherwise.  This is also in the spirit
of an orthogonal local (super-) state space of a \mbox{(super-)}
MPS.  Conversely, assuming that the tensor $T$ of the MPO is given,
note that the intermediate supervector index that connects the
super-MPS with the local operators $\hat{X}$ [cf. \Fig{Fig:SMPO}(c)]
may also be generated by the reverse operation of splitting off the
local state space $(\sigma,\sigma')$ from the tensor $T$ via SVD.
Then by construction, the operators in $\hat{X}$ would be
orthonormal.

For both, conceptual and implementational transparency, we can
construct an MPO as a super-MPS of operators.  By this we mean, that
the local state space of the super-MPS are ``superstates'' that
actually refer to a set of orthogonal local operators [e.g., see
\Fig{Fig:SMPO}(a-b)].  By finally contracting the super-MPS (rank-3
tensors) with the local operators along the intermediate index, this
leads to the final rank-4 tensors $T$ that constitutes the MPO [cf.
Figs.~\ref{Fig:SMPO}(c,d)].  Note that the intermediate index also
specifies an arbitrary but fixed order of the local operators
(`supervectors'').

Now the structure of an interaction term as in \Eq{eq:Ham:SU2} is
generic: a non-scalar irreducible operator $\hat{X}^{[q]}_i$ acting
on site $i$ must be paired up, i.e., contracted on the spinor index
into a scalar term of the Hamiltonian with another operator
$\bigl(\hat{Y}^{[q]}_j\bigr)^\dagger$ acting on site $j$ that
transforms according to exactly the same irreducible representation
(typically $\hat{Y}=\hat{X}$; here we also ignore 3- or more-site
interactions). This observation holds both, for abelian and
non-abelian symmetries.

The construction of the super-MPS that encodes the MPO is greatly
simplified by the simple bilinear structure of 2-site interactions
as in \Eq{eq:Ham:gen} or \Eq{eq:Ham:SU2}. In partiuclar, the
super-MPS can be built completely analogous to the automata approach
above, while paying simple attention to symmetry sectors.  The start
($\alpha=1$) and the stop ($\alpha=2$) state on the virtual bonds
transform like scalars (i.e., have $q=0$), whereas the bond states
$\alpha>2$ directly inherit the symmetry labels from the underlying
IROPs in the 2-site interactions [cf. \Fig{Fig:SMPO}(e)].

For the Heisenberg model in \Eq{eq:Ham:gen}, the set of local
operators is given by $\hat{X}=\{ \hat{\mathbb{I}}, \hat{S}^z,
\hat{S}^+, \hat{S}^- \}$ for the U(1) symmetric setup, and by
$\hat{X}=\{ \hat{\mathbb{I}}, \hat{\mathbf{S}}\}$ for the SU(2)
symmetric setup. In either case, the set of local operators is
orthogonal as in \Eq{eq:ortho:lops}. Note also that while in the
U(1) symmetric case, also the daggered operator $(\hat{S}^+)^\dagger
= \hat{S}^-$ appears in the set, this is not the case for the SU(2)
symmetric case, since SU(2) is self-dual, and therefore
$\mathbf{\hat{S}}_i^\dagger \cdot \mathbf{\hat{S}}_j =
\mathbf{\hat{S}}_i \cdot \mathbf{\hat{S}}_j ^\dagger $.
Specifically, with $U^{[1]} \propto (1,1;0)$ a unitary
transformation that corresponds to the Clebsch-Gordan coefficients
which combine a spin $S=1$ with its dual (again $S=1$) into a
singlet [cf. $1j$-symbol earlier], the spin-spin interaction can be
written as $\mathbf{\hat{S}}_{i}^\dagger \cdot \mathbf{\hat{S}}_j =
\sum_{r,r'=1}^3 \hat{S}^r_{i} \, U^{[1]}_{rr'} \hat{S}^{r'}_j $.
Therefore the action of the dagger on one of the spin operators can
be transferred via the unitary $1j$-symbol into the definition of
the super-MPS itself, proper sign-convention on $U^{[1]}$ implied.

For more complicated cases, like the snake MPO representation of 2D
Heisenberg Hamiltonian, longer range interactions need to be
included.  This is straightforward in the automata construction
above, yet requires that the bond dimension $D_H$ increases [see
also \Fig{Fig:SMPO}(e)].  The period of the translational invariance
of the MPO also increases from $1$ to the width $W$ of the system
and hence requires at least $W$ different $A$-tensors in the
super-MPS [cf.  \Fig{Fig:SMPO}(c-d)].

Once the super-MPS is obtained, one can use standard MPS techniques
to check whether it can be compressed.  An important ingredient here
is that the local supervector space is orthogonal, indeed [cf.
\Eq{eq:ortho:lops}].  If the bond-dimension $D_H$ can be reduced at
no cost, i.e., by discarding singular values that are strictly zero,
the super-MPS and subsequently the MPO contains inefficiencies that
may be simply removed with an improved setup of the super-MPS
itself. On the other hand, for long-ranged systems the bond
dimension $D_H$ may simply become too large, in practice, for an
exact representation of the Hamiltonian. In this case, standard MPS
truncation techniques may be employed on the level of the super-MPS
itself. Here a uniform normalization of the supervector space (i.e.,
the local operators) is advised such that standard MPS techniques
are directly applicable without any further ado.  Alternatively, one
may truncate on the level of the MPO, either by SVD or variational
techniques. The latter is unavoidable for MPO products or sums in
any case, as will be discussed next.

\section{Initialization in the XTRG algorithm}
\label{App:Init}

In this section, we compare three different initializations of
$\rho(\taui)$ in the XTRG algorithm.  The quality of our initial
$\rho(\taui)$ for small $\taui$ is measured by estimating the
relative error of the free energy $|\delta F(\beta)/F(\beta)|$,
starting from exponentially small $\beta=2\taui$ (i.e., the first
data point after initialization at $\beta=\taui$) down to
intermediate temperatures $\beta \gtrsim 10$.

\subsection{Series expansion vs. Trotter-Suzuki initialization}

\begin{figure}[tbp]
\includegraphics[angle=0,width=0.9\linewidth]{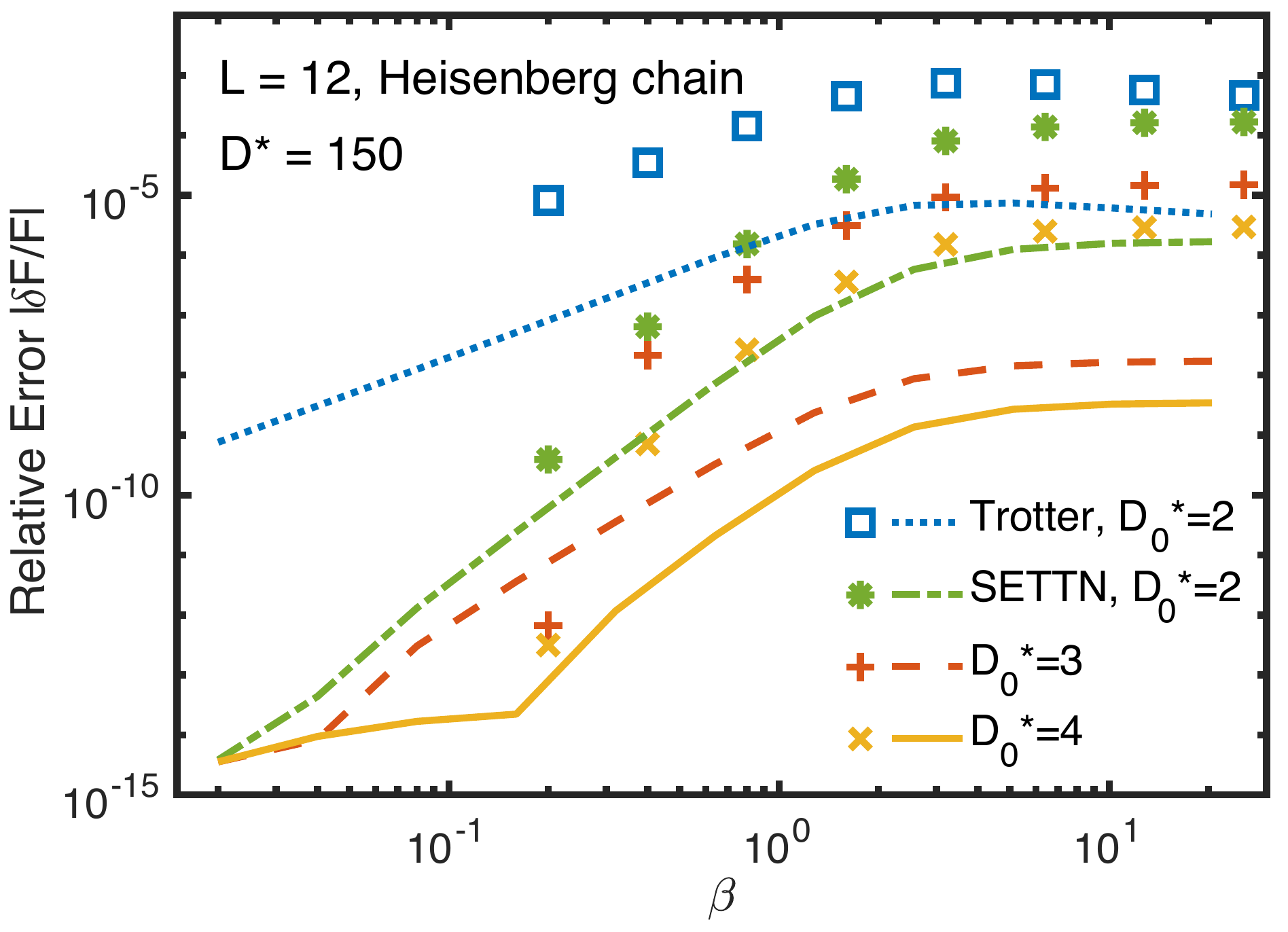}
\caption{(Color online) Comparison between the
   first-order Trotter-Suzuki and the series-expansion
   initialization schemes on an $L=12$ Heisenberg chain. Two
   sets of data are shown from calculations starting with
   $\taui=0.1$ (symbols) and $\taui=0.01$ (lines).  After the
   initialization, the MPO $e^{-\taui H}$ is fed into an XTRG
   evolution, where the number of retained bond states is set
   to $\Dstar=150$.  Same color represents the same type of
   initialization, i.e., blue represents Trotter
   initialization, while the green, red and yellow represent
   initialization by SETTN with bond dimensions $\Dstari =
   2, 3, 4$ ($\Di = 4,5,8$), respectively.
}
\label{Fig:Trotter}
\end{figure}

Firstly, we compare the Trotter-Suzuki initialization with series
expansion [\Eq{Eq:RhoTau} in the main tex] followed by XTRG.
Trotter-Suzuki decomposition breaks $e^{-\taui H}$ into product of
local evolution gates, which for nearest-neighbor spin-1/2 chains
within first-order can be represented as an MPO with bond dimension
$\Dstari=2$ ($\Di=4)$, comprised of $\tfrac{1}{2} \otimes
\tfrac{1}{2}$ $ = \underbar{1}^1 \oplus \underbar{3}^1$.  In
\Fig{Fig:Trotter} we plot the relative errors of the free energy
after Trotter-Suzuki initialization at two values of $\taui=0.1$ and
$\taui=0.01$.  At $\beta=2\taui$, the respective errors $|\delta
F/F|$ are $10^{-5}$ and $10^{-9}$, respectively.

Interestingly, first-order Trotter-Suzuki initialization manages to
arrive at an MPO with bond dimension that is {\it lower} than what
is required for the actual representation of the Hamiltonian itself.
But as a consequence, the overall errors are also larger.  For
comparison, nevertheless, we also show data initialized via SETTN at
the same $\Dstari=2$ ($\Di=4$; green data).  The resulting errors
are much lower $10^{-10}$ and $10^{-15}$ for $\taui=0.1$ and 0.01,
respectively.

\begin{figure}[tbp]
\includegraphics[angle=0,width=0.9\linewidth]{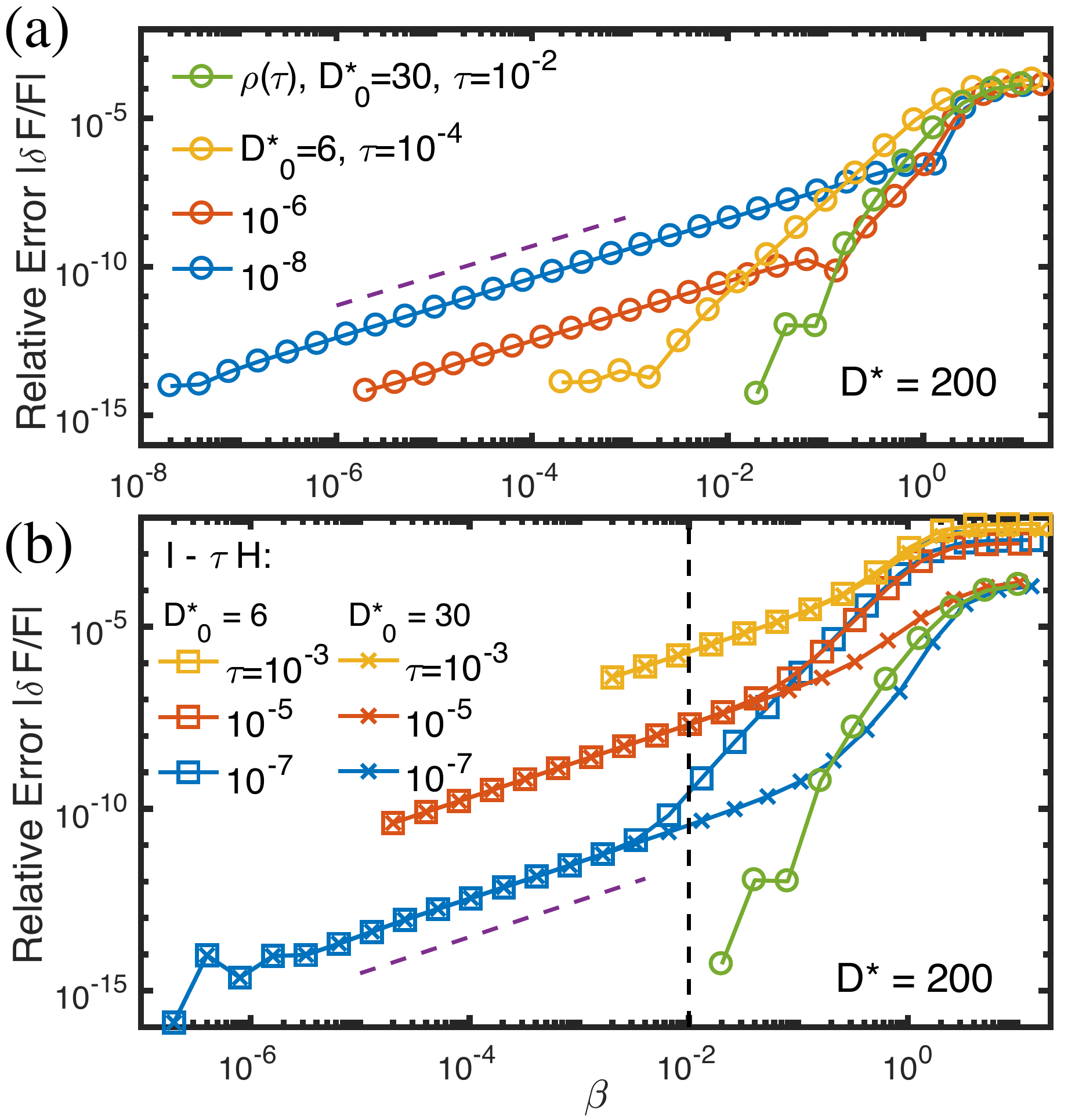}
\caption{(Color online) 
   (a) Relative error of free energy for the Heisenberg model
   on a $4 \times 4$ open square lattice {($\Dstar_H=6$)},
   with various initial $\tau$ {and $\Dstari$} values in the
   XTRG scheme.  Here the $\rho(\beta=\taui)$ is initialized
   via SETTN [cf. \Eq{Eq:RhoTau}].
   (b) Similar to (a) but using lowest order, i.e., linear
   expansion of $\rho(\taui)=\mathbb{I}-\taui H$ for the
   initialization of $\rho(\taui)$ instead of SETTN [only the
   green curve at $\taui=10^{-2}$ uses SETTN, and was copied
   from (a) for direct reference].  Here we use $\Dstar =
   \Dstari$ for all temperatures down to $\beta=10^{-2}$
   (black dashed line), where we switch to $\Dstar=200$.
   Purple dashed lines indicate linear behavior, i.e.,
   $|\delta F/F| \propto \beta$.
}
\label{Fig:VariTau}
\end{figure}

By increasing the initial bond dimension to $\Dstar_H$ and slightly
above, i.e., $\Dstari=3,4$ ($\Di=5,8$), as seen in
\Fig{Fig:Trotter}, SETTN initialization as in \Eq{Eq:RhoTau} can
offer generally better accuracy.  The initial inaccuarcy represents
a systematic error that also propagates along the XTRG procedure
towards lower temperatures.  e.g., for $\beta \sim 10$, the SETTN
initialization with $\Dstari=2$ is still an order of magnitude more
accurate as compared to the Trotter-initialized data, and orders of
magnitude more accurate if $\Di$ is only marginally increased to
$\Dstari=3,4$ relative to the $\Dstar=150$ of the subsequent XTRG.

\subsection{Linear initialization $\rho(\taui) \simeq \mathbb{I}-\taui H$}

By sweeping over several orders of energy scales, the XTRG algorithm
permits a simple linear initialization $\rho(\taui)\simeq
\mathbb{I}-\taui H$ at basically infinitesimal, i.e., exponentially
small $\taui$. In the following, we analyze the effect of the
initial $\taui$ in more detail.

In \Fig{Fig:VariTau}, we compare the initialization of $\rho(\taui)$
with SETTN [panel (a)] and linear expansion [panel (b)],
respectively.  We benchmark the accuracy by analyzing the relative
errors of the free energy $\delta F/F$, with various initial
$\taui=10^{-8}, \ldots,10^{-2}$.  In \Fig{Fig:VariTau}(a) the
initial state $\rho(\beta=\taui)$ is constructed via SETTN [cf.
\Eq{Eq:RhoTau}] and constrained to the bond dimension $\Dstari$ for
$\beta=\taui$ only.  For all other points at lower temperatures
$\beta_n \equiv 2^n \taui$ with $n=1,2,\ldots$ we use $\Dstar=200$.
With this setup, the data in \Fig{Fig:VariTau}(a) at the lowest
temperatures (largest $\beta$)  {exhibits} a similar level of accuracy,
irrespective of the choice of the initial $\taui \lesssim 10^{-4}$
for $\Dstari \ge \Dstar_H = 6$.  Note that $\Dstari = \Dstar_H$ is
the minimal bond dimension to represent the lowest order linear
expansion $\rho(\taui)\simeq \mathbb{I}-\taui H$.  The long
extended straight slopes for $\taui =10^{-6}$ or $\taui=10^{-8}$
are simply $\propto \beta$, as indicated by the guide to the eye
(purple dashed line), which just indicates that the accuracy is
limited by accumulated double precision error of the calculation.
The strong upturn for $\beta \gtrsim 0.1$ then is where truncation
error sets in. Using larger initial $\taui$, e.g., $\taui=10^{-4}$
(yellow curve) error accumulates more strongly, which implies that
$\Dstari=6$ already starts to affect the accuracy at larger
$\beta$. Clearly, for $\tau \gtrsim 10^{-4}$ a larger $\Dstari$,
i.e., higher-order terms in the series expansion are required to
maintain accuracy.  For example, the data for $\taui=10^{-2}$ and
$\Dstari=30$ again shows good accuracy in comparison.

Once $\taui$ is small enough, plain simple lowest-order linear
expansion suffices.  This is analyzed in \Fig{Fig:VariTau}(b) where
we replace series expansion for the initialization of $\rho(\taui)$
by the MPO for $\mathbb{I}-\taui H$.  This is an extremely
convenient initialization that can be simply derived from the MPO
for the Hamiltonian.  In particular, the MPO for this initial
 $\rho(\taui)$ can be exactly represented with bond dimension
 $\Dstari=\Dstar_H$.

Starting with $\beta = \taui \lll 10^{-2}$, XTRG can be used to
exponentially decrease temperature down to values $\beta \approx
10^{-2}$ which still may be considered part of the initialization of
$\rho$ at large temperatures before actual physical energy scales
set in.  Therefore while we always have $\Dstari=\Dstar_H$ for the
very first step, by definition, we can alo constrain $\Dstari$ to
the values specified for the entire range $\beta \le 10^{-2}$
[vertical dashed line in \Fig{Fig:VariTau}(b)].  For $\beta >
10^{-2}$ we allow the MPO to grow up to dimension $\Dstar=200$ to
capture the quickly growing MPO entanglement.

Using $\Dstari = \Dstar_H$ up to $\beta = 10^{-2}$ [squares in
\Fig{Fig:VariTau}(b)], the data already significantly deteriorates
at the largest $\beta \gg 1$ by about two orders of magnitude. This
shows that $\Dstari = \Dstar_H$ up to $\beta=10^{-2}$ is simply too
small, as it introduces systematic errors.  By increasing $\Dstari$
modestly, i.e., $\Dstari=30$ which is still about an order of
magnitude smaller than what is required for large $\beta$, the
systematic errors reduce dramatically.  Starting with $\taui
\lesssim 10^{-4}$, the accuracy at large $\beta$ competes with a
careful higher-order expansion for finite $\taui$ [for reference, we
replotted the data from the $\Dstari=30$ series-expansion
initialization starting with $\taui=10^{-2}$ from panel (a)].
Similar to panel (a), for our data sets with smallest initial
$\taui$, we see wide ranges where the numerical error is simply
$\propto \beta$, and hence given by accumulated double precision
error and, in particular, not truncation error.

The initialization procedure in \Fig{Fig:VariTau}(b) above started
from infinitesimally small $\beta=\tau$ and worked its way up
exponentially using XTRG to $\beta=10^{-2}$.  Using $\Dstar_H <
\Dstari \ll \Dstar_\mathrm{(final)}$, the overall numerical lost of
the entire calculation is strongly dominated by the simulation of
$\beta \gg 1$.  For example, for $\Dstari = 6$ $(30)$ and
$\taui=10^{-7}$, the cost for $\beta \leq 10^{-2}$ is about 7\% and
11\% of the total calculation, respectively.  In this sense the
procedure above is an extremely simple, efficient, and accurate
initialization for finite temperature calculations. The algorithm
works for arbitrary topologies of Hamiltonians, including long-range
interactions or higher-dimensional systems.

Nevertheless, since the series expansion scheme serves as a
systematic way to provide accurate initialization at finite
$\taui<0.1$, in this work, we sticked to initialize of the density
operator $\rho(\taui)$ with our already existing codes on series
expansion.

\section{Compression of matrix product operators}
\label{App:VariComp}

MPO compression is of key importance and frequently used in the XTRG
and SETTN algorithm, to compress the \textit{product} or
\textit{sum} of two MPOs.  The overall procedure follows standard
MPS strategies, where the implications of abelian or non-abelian
symmetries can be largely put aside as an extremely convenient
benefit of using the QSpace tensor library.

Overall, the variational method is preferable due to its higher
efficiency, while the direct SVD compression is also useful as long
as the bond-dimension $D$ is manageable. In the following, we focus
on the variational compression, but also provide details of the SVD
compression along the way. 

\subsection{Compression of MPO product \label{sec:compress:prod}}
  
\begin{figure}[tbp]
\includegraphics[angle=0,width=1\linewidth]{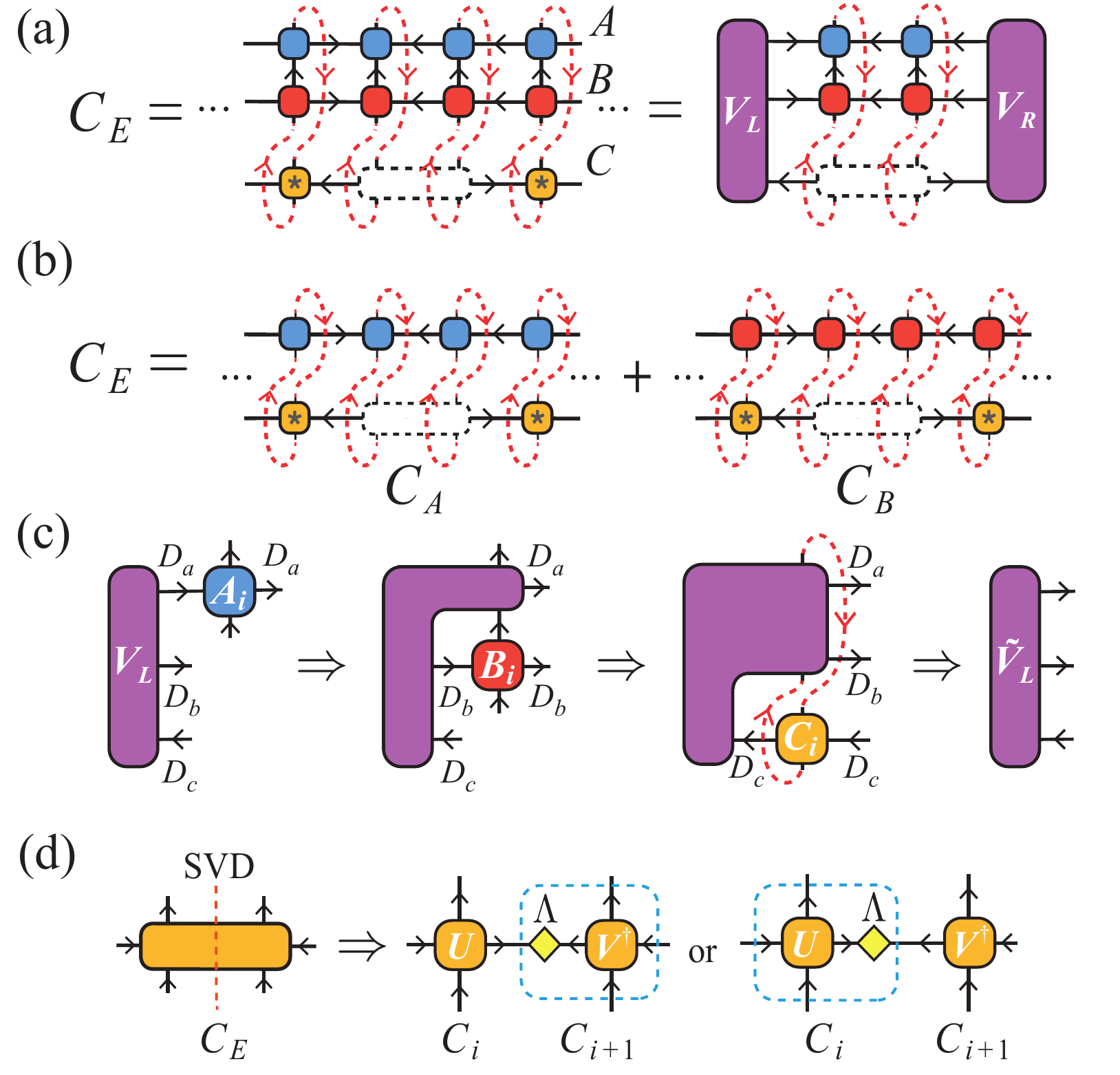}
\caption{(Color online) 
  (a) The environment tensor $C_E$ in two-site variational
  optimization of compressing the product of two MPOs into a single
  MPO, i.e., $C := A\ast B$ [cf.~\Eq{Eq:LSEProd}].  Arrows on the
  horizontal lines demonstrate orthonormalization or, equivalently,
  canonicalization of the MPO. The environment $C_E$ can be
  evaluated by contracting the cluster consists of two environment
  tensors $V_L$ and $V_R$, as well as relevant MPO tensors.
  (b) The environment tensor $C_E$ in two-site variational scheme of
  rewriting the sum of two MPOs into a single MPO, i.e., $C := A+B$
  [cf.~\Eq{Eq:LSESum}].  An asterisk inside a box indicates a
  `daggered' tensor.
  (c) Update the left environment tensor $V_L$ with three local
  tensor $A_i, B_i$ and $C_i$ of MPOs $A, B$ and $C$, respectively.
  The computational costs of the three substeps (from left to right)
  are $\order{D_a^2 D_b D_c}, \order{D_a D_b^2 D_c}$ and $\order{D_a
  D_b D_c^2}$, respectively. $V_R$ can be updated in a similar way.
  (d) Update of the local tensor $T^*_i$ and $T^*_{i+1}$ of MPO $C$,
  by performing SVD on either environment tensor, $C_{E}$ or $(C_{A}
  + C_{B})$.}
\label{Fig:VMPO}
\end{figure}

Consider the product of two MPSs $A \ast B = C$, with bond
dimensions $D_a$, $D_b$, and $D_c$, respectively. The representation
of the product is exact if $D_c = D_a D_b$ which, however, is
typically numerically costly.  Therefore $C$ needs to be compressed
in an efficient and numerically controlled manner.  A typical
example in this paper is the representation of $C:= H^n X = H \ast
(H^{n-1} X)$, with $X=\mathbb{I}$ or $\rho(\beta)$ in SETTN, where
we need to project $A:=H$ to a previously e.g., iteratively obtained
MPO for $B:= H^{n-1} X$ and then compress the ``fat" MPO to bring
down its bond dimensions.  Similarly, in XTRG, one needs to apply
$\rho(\beta)$ onto itself, i.e., $A=B:=\rho(\beta)$, in order to
reach the density matrix $C:=\rho(2\beta)$ at half the temperature. 

The direct SVD compression is straightforward
\cite{Chen.b+:2017:SETTN} but computationally costly. Given the
product of two MPOs, as depicted pictorially in
\Fig{Fig:Diagram}(b), horizontal lines are fused pairwise  {with respect to} 
the same horizontal bond position into a single ``fat'' index of
dimension $D_a D_b$, and then truncated via SVD. Using symmetries,
abelian or non-abelian symmetries alike, the fusion step includes a
simple tensor product of two state spaces. For this it is important,
that the arrows along the virtual bond state spaces (horizontal
lines) are parallel and point all in the same direction
[\Fig{Fig:Diagram}(b)].

The computational cost of SVD compression scales as $\order{D_a^3
D_b^3}$.  For the series expansions, say, when constructing $H^n
\rho$ with $D_a=D_H$ and $D_b=D$, this is still relatively cheap,
$\order{D^3}$, given that typically $D_H \ll D$.  In contrast, the
numerical cost becomes quickly prohibitive for XTRG since with
$D_a=D_b=D$ the cost scales as $\order{D^6}$.
 
The variational method can significantly reduce numerical cost, and
therefore is mainly adopted in the present study.  For this, we use
a two-site update similar to standard DMRG procedures to allow
adaptive adjustment of bond dimensions.  This is particularly
important when exploiting symmetries, abelian and non-abelian alike,
to optimally adapt bond dimensions w.r.t. to each individual
symmetry sector.  This way, irrelevant symmetry sectors drop out
automatically, whereas possibly new relevant symmetry sectors can
emerge or get strengthened.

For the variational approach, we minimize the cost function
(Frobenius norm squared),
\begin{eqnarray}
   && \Vert (A\ast B) - C \Vert_F^2 = \notag \\
   && \ \ \ C^{\dagger}C - (A\ast B)^{\dagger}C-C^{\dagger}(A\ast B) +
   \mathrm{const}
\text{ .}\label{eq:MPO:prod}
\end{eqnarray}
We then take the partial derivative with respect to the product of
two adjacent local tensors $C_i C_{i+1}$ in the full MPO of $C$.
This results in the linear system of equations,
\begin{equation}
   \frac{\partial [C^{\dagger} C]}{\partial (C_i C_{i+1})^\ast} =
   \frac{\partial[C^{\dagger} (A\ast B)]}{\partial (C_i C_{i+1})^\ast}
\text{ ,} 
\label{Eq:LSEProd}
\end{equation} 
to be solved iteratively for $i=1, \ldots, L-1$, with $L$ the length
of the MPO (while all simulations in this work are based on real
numbers, for the simplicity of the derivation, nevertheless, we
assume complex numbers).  Both sides of \Eq{Eq:LSEProd} can be
expressed as fully contracted tensor networks, except for the
missing tensors $C_i^\ast$ and $C_{i+1}^\ast$ (i.e., with ``punched
holes''), as shown in \Fig{Fig:VMPO}(a).  Therefore both sides of
\Eq{Eq:LSEProd} represent tensors of rank-6.  For a canonicalized
MPO $C$, where all lines in $C$ are directed towards the {\it
orthogonality center}, here at sites $(i,i+1)$, the left-hand side
is simply $C_i C_{i+1}$.  The right-hand side defines {the}
{generalized overlap} tensor $C_E$ [cf. \Fig{Fig:VMPO}(a)].
\EQ{Eq:LSEProd} therefore directly states the solution for $(C_i
C_{i+1})$ to the local optimization problem w.r.t.  to sites
$(i,i+1)$.

To obtain $C_E$, one needs to iteratively update the left/right
environment tensors $V_L/V_R$ of the three-MPO product until the
structure in the right-hand-side of \Fig{Fig:VMPO}(a) reached.  As
shown in \Fig{Fig:VMPO}(c), to update the $V_L/V_R$ tensors, we
iteratively contract the local tensors $A_i, B_i$ and $C_i$ of MPO
$A, B$ and $C$ with it. The procedures of updating $V_L/V_R$
constitute the most time-consuming ones, which scale as
$\order{D^4}$ (assuming $D_a=D_b=D_c=:D$), when we square $\rho$ and
compress it in XTRG.  Nevertheless, this is still computationally
much cheaper as compared to the SVD compression of $\order{D^6}$
cost as briefly discussed above. 

Given the optimized product $C_i C_{i+1} = C_E$, we need to split
$C_E$ into the actual product shape $C_i C_{i+1}$ by performing SVD,
$C_E = U \, \Lambda\, V^{\dagger}$ as shown {schematically} in
\Fig{Fig:VMPO}  {(d)}.  To gauge the MPO $C$ in a canonical form,
$C_i$ is updated with $U$, and $C_{i+1}$ with $\Lambda V^{\dagger}$
in a left-to-right sweep, while in a right-to-left sweep, the matrix
$\Lambda$ is contracted with $U$ and thus associated with $C_i$,
instead.  One typically only needs a few full sweeps (left to right
and vice versa) to converge the cost function and obtain the optimal
MPO for $C$. In this work, typically at most $\lesssim 4$
sweeps were sufficient.

\begin{figure}
\includegraphics[angle=0,width=0.85\linewidth]{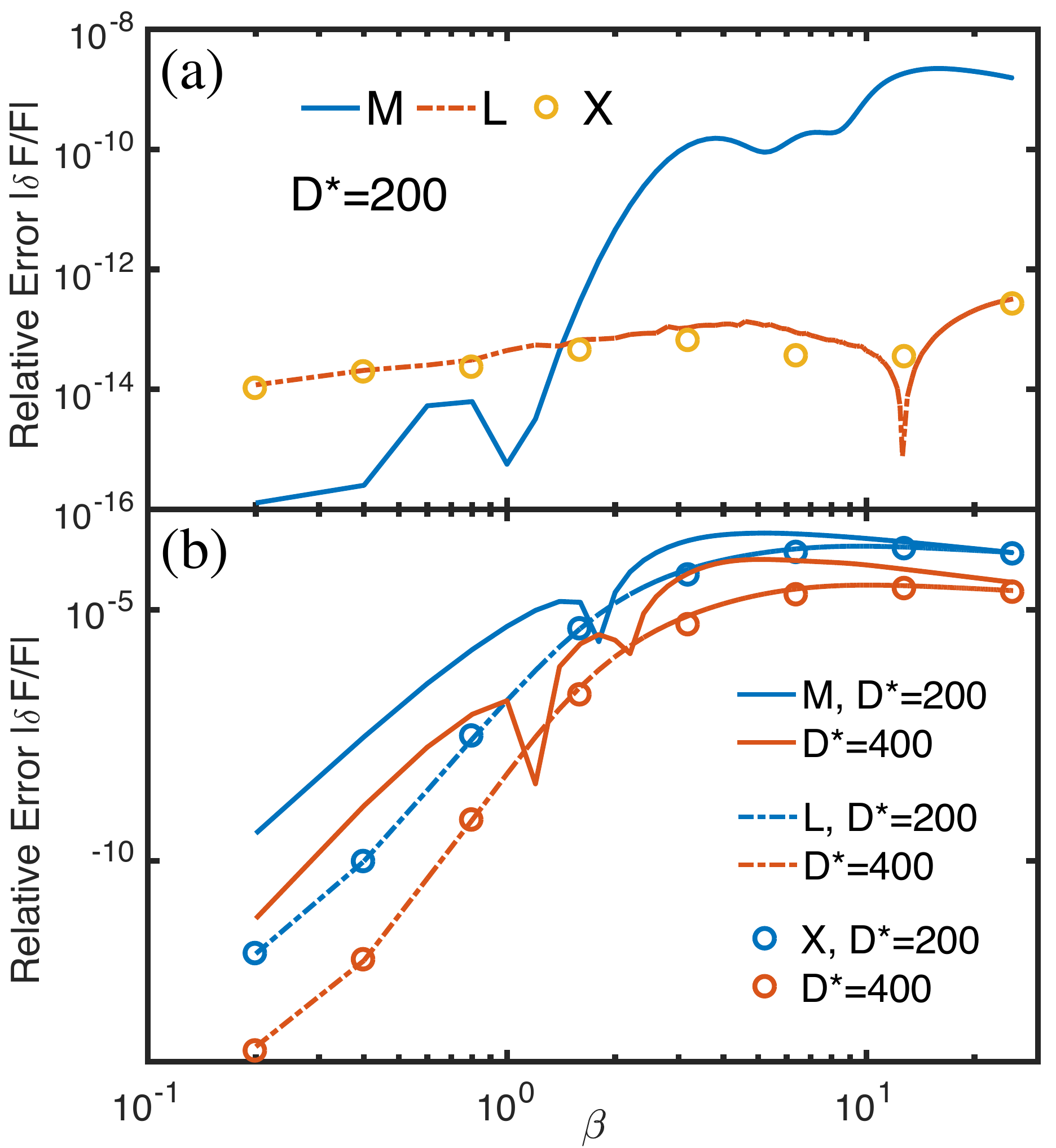}
\caption{ (Color online)
   Relative errors of the free energy fully within SETTN in (a) an
   $L=16$ Heisenberg chain, and (b) a 4$\times$4 Heisenberg model
   on the square lattice (OBC). The data is computed along various
   $\beta$ grids, including Maclaurin (M) expansion around
   $\beta=0$, Taylor expansion in linear (L) and exponential (X)
   $\beta$ scales.  In (a) we used $\Dstar=200$
   resulting in relative CPU run times M:L:X= 2.34:5.25:1.
   In (b) we used $\Dstar=200, 400$.
   For $\Dstar=200$, this resulted in 
   relative CPU run times M : L : X = 2.1 : 4.95 : 1,
   and for $\Dstar=400$, and M : L : X = 1.4 : 4.2 : 1.
}
\label{Fig:CompSETTN}
\end{figure}

\begin{figure*}
\centering
  \includegraphics[width=1\linewidth]{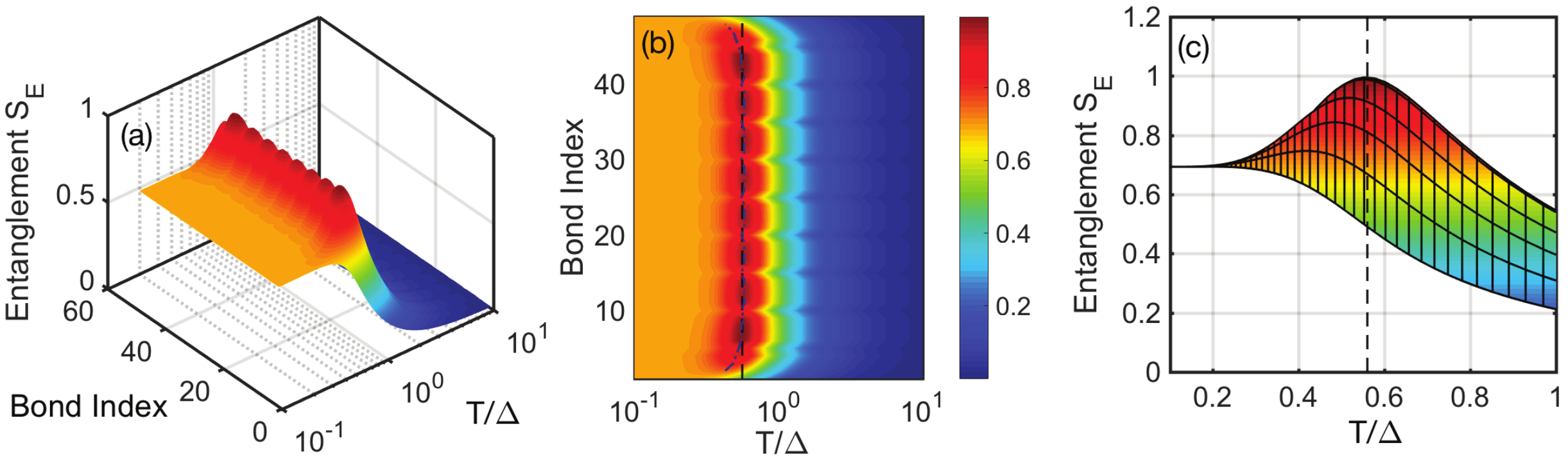}
\caption{(Color online) 
(a) Block entanglement landscapes vs.  temperature $T/\Delta$ and
   bond index for the XXZ model on a square lattice (using
   $\Delta=5$, $L=10, W=5$, keeping upt to $D=500$ states, i.e.,
   similar analysis as in \Fig{Fig:XXZTc} but for smaller system
   size).
(b) Top view, and (c) side view vs. $T/\Delta$ of the
   same data as in (a).  The black dashed line represents the exact
   value $\Tc=0.56$ and the dash-dotted line is tracking the
   ``ridge" of entanglement landscape (i.e., maxima vs. $T$ on each
   bond) in (a), the latter coincides with $\Tc$ in the central
   bonds of the system.  Incidentally, the {\it envelope} of
   entanglement curves also shows a peak around $\Tc$ (with the
   resolution $\delta T/\Delta \simeq 0.02$).
}
\label{Fig:XXZCyl}
\end{figure*}

\subsection{Compression of MPO sum
  \label{sec:compress:sum}}

Summation of MPOs is an essential technique, e.g., in the series
expansion $\rho(\beta) = \sum_{n=0}^{N} \frac{(-\beta)}{n!}H^n$.
Here we generalize standard procedures for the addition of MPS to
MPO. For this, we also resort to a 2-site variational approach as
illustrated in \Fig{Fig:VMPO}(b).  To find a optimal MPO for $C = A
+ B$, where the generalization to more than two vectors is
straightforward, we minimize the cost function, 
\begin{eqnarray}
&& \Vert (A+B)-C \Vert_F^2 = \notag \\
&& \ \ \ C^{\dagger}C - (A+B)^{\dagger}C-C^{\dagger}(A+B) 
+ \mathrm{const}
\text{ .}\label{eq:MPO:plus}
\end{eqnarray}
Again we take the partial derivative with respect to the product of
two adjacent local tensors $C_i C_{i+1}$ of the MPO $C$, resulting
in the linear system of equations,
\begin{equation}
   \frac{\partial(C^{\dagger}C)}{\partial (C_i C_{i+1}){^\ast}} =
   \frac{\partial [C^{\dagger}(A+B)]}{\partial (C_i C_{i+1}){^\ast}}
\text{ ,}
\label{Eq:LSESum}
\end{equation}
to be solved iteratively for $i=1, \ldots, L-1$.  Again, using a
canonicalized MPO for $C$, \Eq{Eq:LSESum} simply reduces to $C_i
C_{i+1} = C_A + C_B$, with $C_A$ etc. generalized overlap matrices
to be computed iteratively [cf. \Fig{Fig:VMPO}(b)].  The remainder
of the algorithm proceeds exactly the same as the compression of MPO
products above.

\section{Series expansion tensor network simulations with linear versus
exponential $\beta$ grid} \label{App:LvsX}

In this appendix, we compare SETTN calculations with three schemes
of selecting expansion point set: (M) Maclaurin scheme which expands
$\rho(\beta)$ around $\beta=0$, i.e., there is only one expansion
point in the set; (L) point-wise Taylor expansion around a linear
$\beta$ set, $\beta = n \taui$ with $n$ an integer; (X) exponential
set $\beta=2^n \taui$.  The results are summarized in
\Fig{Fig:CompSETTN}.

In \Fig{Fig:CompSETTN}(a), we compare the accuracy of the above grid
setups (M,L,X) in the calculation of the free energy for an $L=16$
Heisenberg chain.  Compared to (M), both (L) and (X) are clearly
superior, as they gain four orders of magnitude in accuracy for the
largest $\beta$.

We also compare these three grid setups within SETTN to the more
challenging system of a 4$\times$4 SLH.  Here
due to the significantly larger truncation errors across all
approaches, the gain of (L,X) over (M) is significantly less
pronounced, as seen in \Fig{Fig:CompSETTN}(b).  To reduce the
truncation error, the number of virtual bond states would have to be
increased significantly from the $\Dstar=200$ or even $400$ in the
present calculation.  As for the numerical efficiency, although
having very good accuracy, the computational overhead of the linear
scheme (L) is significant in any case. In contrast, the logarithmic
scheme (X) strongly reduces the computational cost [see caption to
\Fig{Fig:CompSETTN} for explicit numbers] without losing any
accuracy.  By minimizing the number of intermediate and thus also
truncation steps by moving to large $\beta$ in the fastest possible
way, the numerical errors of (X) are on the lower end.

\section{Exact solution of the XY chain at finite temperature}
\label{App:XY-Chain}

For completeness, we provide analytical expression of the partition
function in a 1D XY model. The Hamiltonian is given by
\begin{align*}
 H &= J\sum_i
   (S^x_i S^x_{i+1} + S^y_i S^y_{i+1}), \\
    &\equiv \tfrac{J}{2}\sum_i
    (S^+_i S^-_{i+1} + S^-_i S^+_{i+1}). \numberthis
\end{align*}
Exploiting Jordan-Wigner transformation, the Hamiltonian can be
mapped onto a plain  {fermionic} tight-binding chain,
\begin{align}
  H = \tfrac{J}{2}\sum_i (c^{\dagger}_i c_{i+1} + h.c.)
  =\sum_{k=1}^L \epsilon_k c_k^{\dagger} c_k
\text{ ,}
\end{align}
which for open boundary condition (OBC), is diagonalized by
the 1-particle eigenstates,
\begin{align}
c_k &= \sqrt{\tfrac{2}{L+1}}
\sum_{i=1}^L c_i \sin\bigl(\tfrac{k\pi}{L+1}i\bigr) \notag \\
\epsilon_k &= J\cos\bigl(\tfrac{k\pi}{L+1} \bigr)
\text{ ,}
\end{align}
where $c_i^{(\dagger)}$ [$c_k^{(\dagger)}$] are fermionic
annihilation (creation) operators at site $i$ (``momentum" $k$),
respectively  \cite{Tu.h:2017:Klein, UniEntropy-2017}.

The partition function is fully determined by the dispersion
$\epsilon_k$, \begin{equation} \mathcal{Z}^{\rm{XY}}_{\rm{OBC}} =
\prod_{k=1}^L (1+e^{-\beta \epsilon_k}).  \end{equation} with free
energy $F=-T \, \ln{\mathcal{Z}}$.

\section{Entanglement measurements of XXZ model on the cylinder geometry}
\label{App:EntCyl}

The entanglement landscape, along with its top and side views, for
the 2D XXZ model on a square-lattice cylinder ($L=10, W=5$,
$\Delta=5$) is analyzed in \Fig{Fig:XXZCyl}.  Similar to the
analysis of the larger system in \Fig{Fig:XXZTc}, we can again
observe prominent peaks which can be help to pinpoint the critical
temperature $\Tc$ [vertical dashed lines in \Fig{Fig:XXZCyl}(b-c)].
In the top view \Fig{Fig:XXZCyl}(b) we again see that the peak
position $\Tc^S$  [represented as the dash-dotted lines in
\Fig{Fig:XXZCyl}(b)] in the very center of the system deviates
slightly from $\Tc=0.56$ in that it approaches from above, i.e.
$\Tc^S> \Tc$.  This is different from \Fig{Fig:XXZTc}, in the main
text where $\TcS$ approaches $\Tc$ from the low-temperature side.
 {The underlying reason is the different boundary conditions,} i.e., 
cylindrical BC here vs. fully open BC in \Fig{Fig:XXZTc}. 

Interestingly, we find that $\TcS$ determined from cylinder
geometry again serves as a good estimate for $\Tc$. As shown in
\Fig{Fig:XXZCyl}(c), we can see that the profile of entanglement
curves overall (i.e., selecting the maximal entanglement over bonds
for any given temperature), dubbed entanglement envelope, shows a
 peak in close proximity to  $\Tc=0.56$.

\bibliography{Research}

\end{document}